\documentclass[apj,twocolumn,numberedappendix]{emulateapj}
\usepackage{epsfig}
\usepackage{amsmath}
\usepackage{natbib}
\usepackage{subfigure}
\usepackage{longtable}
\usepackage{color}
\usepackage{epstopdf}

\def\spose#1{\hbox to 0pt{#1\hss}}
\def\simlt{\mathrel{\spose{\lower 3pt\hbox{$\mathchar"218$}}
     \raise 2.0pt\hbox{$\mathchar"13C$}}}
\def\simgt{\mathrel{\spose{\lower 3pt\hbox{$\mathchar"218$}}
     \raise 2.0pt\hbox{$\mathchar"13E$}}}

\usepackage{graphicx}

\begin{document}

\title{Cluster candidates around low power radio-galaxies at \scriptsize{z}\normalsize$\sim1-2$ in COSMOS}
\author{G. Castignani\altaffilmark{1}, M. Chiaberge\altaffilmark{2,3,4}, A. Celotti\altaffilmark{1,5,6}, C. Norman\altaffilmark{2,7},G. De Zotti\altaffilmark{1,8}}
\altaffiltext{1}{SISSA, Via Bonomea 265, 34136, Trieste, Italy \\  castigna@sissa.it}
\altaffiltext{2}{Space Telescope Science Institute, 3700 San Martin Drive, Baltimore, MD 21218}
\altaffiltext{3}{INAF - IRA, Via P. Gobetti 101, Bologna, I-40129}
\altaffiltext{4}{Center for Astrophysical Sciences, Johns Hopkins University, 3400 N. Charles Street, Baltimore, MD 21218, USA}
\altaffiltext{5}{INAF - Oservatorio Astronomico di Brera, via Bianchi 46,  23807, Merate, Italy}
\altaffiltext{6}{INFN- Sezione di Trieste, via Valerio 2, 34127, Trieste, Italy}
\altaffiltext{7}{Department of Physics and Astronomy, Johns Hopkins University, Baltimore, MD 21218, USA}
\altaffiltext{8}{INAF-Osservatorio Astronomico di Padova, Vicolo dell'Osservatorio 5, I-35122 Padova, Italy}


\begin{abstract}
{We search for high redshift ($z\sim$1-2) galaxy clusters using low luminosity radio galaxies (FR~I) as beacons and our  newly developed Poisson Probability Method (PPM) based on photometric redshift 
information and  galaxy number counts.
We use a sample of 32 FR~Is within the Cosmic Evolution Survey (COSMOS) field from \citet{chiaberge2009} catalog.
 We derive a reliable subsample of 21 {\it bona fide} Low Luminosity Radio Galaxies (LLRGs) and a subsample of 
11 High Luminosity Radio Galaxies (HLRGs), on the basis of photometric  redshift information and 
 NRAO VLA Sky Survey (NVSS) radio fluxes.  
The LLRGs are selected to have 1.4~GHz rest frame luminosities lower than the fiducial FR~I/FR~II divide. 
This also allows us to estimate the comoving space density of 
sources with $L_{1.4}\simeq 10^{32.3}\,\hbox{erg}\,\hbox{s}^{-1}\,\hbox{Hz}^{-1}$ at $z\simeq 1.1$, which strengthens the case for a 
strong cosmological evolution of these sources. In the fields of the LLRGs and HLRGs we find evidence that 14 and 8 of them reside 
in rich groups or galaxy clusters, respectively. Thus, overdensities are found around $\sim70\%$ of the FR~Is, independently of the
 considered subsample. This rate is in agreement with the fraction found for low  redshift FR~Is and it is significantly higher than 
that of FR~IIs at all redshifts. Although our method is primarily introduced for the COSMOS survey, it may be applied to
 both present and future wide field surveys such as SDSS Stripe 82, LSST, and Euclid.
Furthermore, cluster candidates found with our method are excellent targets for
next generation space telescopes such as JWST.}

\end{abstract}

\keywords{galaxies: active - galaxies: clusters: general - galaxies:
high redshift}


\section{Introduction}\label{sec:intro}
Cluster of galaxies are among the most massive large scale structures in the Universe.
They form from gravitational collapse of matter concentrations induced by perturbations of the primordial density field
\citep{peebles1993,peacock1999}.
Galaxy clusters have been extensively studied to understand how large scale structures form and evolve during cosmic time,
from galactic to cluster scales \citep[see][for a review]{kravtsov_borgani2012}.

Despite this,
the properties of the cluster galaxy population and their changes with redshift
in terms of galaxy morphologies, types, masses, colors \citep[e.g.][]{bassett2013,mcintosh2013},
and star formation content  \citep[e.g.][]{zeimann2012,santos2013,strazzullo2013,gobat2013,casasola2013,brodwin2013,zeimann2013,
alberts2013}
 are still debated,
especially at redshifts $z\gtrsim1.5$.

It is also unknown when the Intra Cluster Medium (ICM)
virializes and starts emitting in X-rays
and upscattering the CMB through the Sunyaev - Zel'dovich (SZ) effect \citep{sunyaev_zeldovich1972}.
See \citet{rosati2002} for a review.
More in general, the formation history of the large scale structures and the halo assembly
history \citep[e.g.][]{sheth2004,dalal2008,adami2013} are not fully understood.

High redshift cluster counts are used to constrain cosmological parameters \citep[e.g.][]{planckXX_2013}, to test the
validity of the $\Lambda$CDM scenario and quintessence models \citep{jee2011,mortonson2011,benson2013}.
Cluster counts are strongly sensitive to the equation of state of the Universe, especially at $z\gtrsim1$ \citep{mohr2005}, when the
Universe starts accelerating and the dark energy component starts becoming dominant.
The Sunyaev-Zel'dovic (SZ) effect, weak lensing measurements \citep{rozo2010}, X-ray scaling relations and data \citep{vikhlinin2009,mantz2010}
are used to evaluate the mass, the redshift of the clusters, and their mass function.
Moreover, high redshift cluster samples might be  used to test 
the  (non-)Gaussianity of the primordial density field and to test alternative 
theories beyond General Relativity \citep[see][and references therein for a review]{allen2011,weinberg2012}.

Searching for high redshift $z\gtrsim1$ galaxy clusters is therefore a fundamental issue of modern astrophysics
to understand open problems of extra-galactic astrophysics and cosmology
from both observational and theoretical perspectives.

An increasing number of high redshift $z\gtrsim1$ spectroscopic confirmations
of cluster candidates have been obtained in the last years.
 To the best of our knowledge, there are in the literature only 11 spectroscopically confirmed
$z\gtrsim1.5$ clusters
\citep{papovich2010,fassbender2011,nastasi2011,santos2011,gobat2011,brodwin2011,brodwin2012,zeimann2012,stanford2012,muzzin2013,newman2013}.
Only some of them have estimated masses greater than $10^{14}$~M$_\odot$.
In addition to them, \citet{tanaka2013} spectroscopically confirmed a $z=1.6$
X-ray emitting group, whose estimated
mass is $3.2\times10^{13}~M_\odot$. A $z\sim1.7$ group associated with a $z\sim8$ lensed background galaxy was found by \citet{barone_nugent2013}.

Several methods use photometric and/or spectroscopic redshifts
to search for high redshift overdensities 
\citep{eisenhardt2008,knobel2009,knobel2012,adami2010,adami2011,george2011,wen_han2011,jian2013}.
They are generally less effective at $z\gtrsim1.5$.
This is due to the difficulty of obtaining spectroscopic redshift information for a sufficient
number of sources at $z>1$, to the significant photometric redshift uncertainties, and
to the small number density of objects.

High redshift clusters have been searched for by using several other independent techniques; such as e.g.
those that use X-ray emission \citep[e.g.][]{cruddace2002,bohringer2004,henry2006,suhada2012} or
the SZ effect \citep[e.g.][]{planckXXIX_2013,hasselfield2013,reichardt2013}.
However, such methods require a minimum mass and are rapidly insensitive
for detecting $z\gtrsim1.2$ clusters \citep[see  e.g. discussion in][]{zeimann2012}.
This seems to be true also for the SZ effect.

It is commonly accepted that
early-type passively evolving galaxies segregate within the cluster core
and represent the majority among the galaxy population, at least at redshifts $z\lesssim1.4$
 \citep[e.g.][]{menci2008,tozzi2013}.

Various methods search for distant clusters
taking advantage of the segregation of red objects in the cluster core.
Such searches are commonly performed adopting  either optical \citep{gladders2005} or
infrared \citep{papovich2008} color selection criteria.
They find a great number of cluster candidates, even at $z\sim2$ \citep[e.g.][]{spitler2012}. However, all these
methods seem to be
less  effective at redshifts $z\gtrsim1.6$. Moreover, such methods require a significant presence of red galaxies.
There might be a bias in excluding clusters with a significant amount of star forming galaxies  or, at least, in selecting only those
overdensities whose galaxies exhibit specific colors \citep{scoville07b,george2011}.



Powerful radio galaxies \citep[i.e. FR~IIs,][]{fr74} have been  extensively used for high redshift cluster searches 
\citep[e.g.][]{rigby2013,koyama2014}.
High redshift (i.e. $z\gtrsim2$) high power radio galaxies
are frequently hosted in Lyman-$\alpha$ emitting protoclusters \citep[see][for a review]{miley_debreuck2008}.
Recently \citet{galametz2012} and \citet{wylezalek2013} searched for Mpc-scale
structures around high redshift (i.e. $z\gtrsim1.2)$  high power radio galaxies
using an infrared (IR) color selection \citep{papovich2008}.

The radio galaxy population comprises FR~I and FR~II sources \citep{fr74}.
Edge-darkened (FR~I) radio galaxies are those where the surface brightness decreases from the core of the source to the lobes or the plumes
of the jet at larger scales. Conversely, the surface brightness of edge-brightened (FR~II) radio galaxies has its peak at the 
edges of the radio source. 

FR~I radio galaxies are
intrinsically dim and are more difficult to find
at high redshifts than the higher power FR~IIs.
This has so far
limited the environmental study of the high redshift ($z\gtrsim1$) radio galaxy population to the FR~II class only.

However, due to the steepness of the luminosity function, FR~I radio galaxies represent the great majority among
the radio galaxy population.
Furthermore, on the basis of the radio luminosity function,
hints of strong evolution have been observationally suggested by previous work \citep{sadler2007,donoso2009}.
Furthermore, their comoving density is expected to reach a maximum around $z\sim1.0-1.5$
followed by a slow declining at higher redshifts, according to some
theoretical model \citep[e.g.][]{massardi2010}.


At variance with FR~II radio galaxies or other types of active galactic nuclei (AGN),
low-redshift FR~Is are typically hosted by undisturbed ellipticals or giant ellipticals of cD type
\citep{zirbel96}, which
are often associated with the
Brightest Cluster Galaxies \citep[BCGs,][]{vonderlinden2007}.
Furthermore, FR~Is are preferentially found locally in dense environments
\citep{hill1991,zirbel1997,wing2011}.
This suggests that FR~I radio galaxies could be more effective
for high redshift cluster searches than FR~IIs.

\citet[][hereinafter C09]{chiaberge2009} derived the first sample of $z\sim1-2$ FR~Is
within the Cosmic Evolution Survey (COSMOS) field \citep{scoville07}.
\cite{chiaberge2010} suggested the presence of overdensities around three of their
highest redshift sources.
Based on  galaxy number counts, the authors found that the Mpc-scale environments of these sources are 4$\sigma$
denser than the mean COSMOS density.
\citet{tundo2012} searched for X-ray emission in the fields of the radio galaxies of the C09 sample.
They took advantage of the Chandra COSMOS field (C-COSMOS).
They did not find any evidence for clear diffuse X-ray emission from the surroundings of the radio galaxies.
However, their stacking analysis suggests that, if present, any X-ray emitting hot gas would have temperatures lower than $\sim$2-3~keV.
Furthermore, \cite{baldi2013} derived accurate photometric redshifts for each of the sources in the \cite{chiaberge2009} sample.

The goal of this project is to search for high redshift clusters or groups using FR~I radio galaxies  as beacons.
In this paper we apply
the new method we developed to achieve such a goal.
The Poisson Probability Method (PPM) has been introduced in a separate paper \citep{PPMmethod},
it is tailored  to the specific properties of the sample (C09) we consider, and it uses
photometric redshifts.
For comparison, we also apply the \cite{papovich2008} method that was previously used in other work
to search for high redshift $z\gtrsim1.2$ cluster candidates \citep[e.g.][]{galametz2012,mayo2012}.

We firstly redefine the sample by carefully selecting those sources that can be safely considered as low radio power
FR~Is at $z\sim1-2$.
This is done by estimating the luminosity of each radio galaxy in the sample on the
basis of their most accurate photometric redshifts available to date \citep{baldi2013}, and
a careful revision of all the adopted radio fluxes.

The main aim of this work is to confirm  statistically that the great majority of FR~I radio galaxies at $(z\sim1-2)$ reside
in dense Mpc-scale environments,
as found at low redshifts. We also discuss the properties of the detected overdensities in terms of their significance, estimated redshift,
location, richness, and size, as inferred from the PPM.
A careful spectroscopic confirmation of the candidates is however required to have a fully reliable picture of the cluster properties.

In particular, throughout the text we will refer to the Mpc-scale overdensities as clusters, cluster
candidates, and overdensities, with no distinction. However, we keep in mind that these large scale structures could
show different properties and they might be virialized clusters or groups, as well as
still forming clusters or proto-clusters.

We describe the adopted sample in Sect.~\ref{par:sample}, the sample redefinition in Sect.~\ref{par:sample_redef}.
In Sect.~\ref{sec:integrated_LF} we estimate the space density of 1.4~GHz sources at $z\sim1$.
 We apply our newly developed method to search for overdensities and we discuss the results in Sect.~\ref{sec:PPM} 
and Sect.~\ref{sec:PPMresults}, respectively.
In Sect.~\ref{sec:papovich_test} we apply the \citet{papovich2008} method to search for 
overdensities and we discuss the results.
In Sect.~\ref{sec:Discussion} we summarize and discuss our results and the main implications of our findings.
In Sect.~\ref{sec:conclusions} we draw conclusions and we outline possible future applications
of our work.

Throughout this work we adopt a standard flat $\Lambda$CDM cosmology with
matter density $\Omega_m=0.27$ and Hubble constant $H_0=71~\hbox{km}\,\hbox{s}^{-1}\,\hbox{Mpc}^{-1}$ \citep{Hinshaw2009}.

\section{The sample}
\label{par:sample}

The  COSMOS survey \citep{scoville07} is a
1$^\circ$.4$\times$1$^\circ$.4 equatorial survey that includes
multiwavelength imaging and spectroscopy from the radio to the X-ray
band. COSMOS is also entirely covered by the Very Large Array Faint Images of the Radio Sky at Twenty-Centimeters
(VLA FIRST) survey at
1.4~GHz \citep{becker95}, and it includes HST observations,
\citep{koekemoer07}.

Due to its high sensitivity, angular resolution, and wide spectral
coverage, COSMOS is suitable to study large scale structures at high
redshifts, with unprecedented accuracy and low cosmic variance.

Hereafter in this work we will refer to Low (High) Luminosity Radio Galaxies,
i.e. LLRGs (HLRGs).
The LLRGs will denote those radio galaxies with radio power typical of FR~Is, while the HLRGs
will denote radio galaxies with radio powers generally higher than the FR~I/FR~II radio power divide
\citep[L$_{1.4GHz}\sim 4\times10^{32}$~erg~s$^{-1}$~Hz$^{-1}$,][]{fr74}.\footnote{See Sect.~\ref{par:subsample} 
and
 Sect.~\ref{sec:HLRGs} for robust definitions of the two classes, concerning our sample.}
This does not imply that the LLRGs are FR~Is and the HLRGs are FR~IIs, especially at high redshift.
This is because the HLRGs of our sample have radio powers only slightly higher than those typical of local FR~Is.
In fact, all the sources in our sample (including the HLRGs) have radio powers about $\sim2$~orders of magnitude
lower than those typical of high-z radio galaxies \citep[$z\gtrsim2$,][]{miley_debreuck2008}.
Furthermore, both the LLRGs and HLRGs might include radio galaxies of transitional type.
Therefore, despite the radio galaxies in our sample do not clearly exhibit
all the properties typical of local FR~Is we will refer to both the LLRGs and the HLRGs as FR~I radio galaxies,
except where otherwise specified.

C09 searched for FR~Is candidates at $1\lesssim z
\lesssim2$ in the COSMOS field, using multiwavelength selection
criteria. Here, we briefly summarize the main steps of the procedure, while more details are given in C09.

The two basic assumptions are: (i) the FR~I/FR~II divide in radio
power per unit frequency (set at L$_{1.4GHz}\sim 4\times10^{32}$~erg~s$^{-1}$~Hz$^{-1}$)
does not change
with redshift; (ii) the  magnitudes and colors of the FR~I hosts at
$1<z<2$ are similar to those of FR~IIs within the same redshift bin,
as in the case of local radio galaxies \citep[e.g.][]{zirbel96, donzelli07}.
Note that the photometric redshifts are affected by great uncertainties, so they
do not constitute a selection criterion. In the following we summarize
the source selection procedure adopted by C09:

\begin{enumerate}
\item FIRST radio sources in the COSMOS field whose
observed 1.4~GHz fluxes are in the range expected for FR~Is
at $1<z<2$ (1$<$F$_{1.4}<$13 mJy) are considered.
\item Sources with FR~II radio morphology, i.e. showing clear
edge-brightened radio structures, are rejected.
\item Those with bright optical counterparts (m$_{\rm i,Vega}<21$) are
then excluded since they are likely lower redshift galaxies with radio
emission produced by e.g. starbursts. Note also that this constraint
assumes that the magnitude of the FR~Is hosts are similar to those of
FR~IIs.
\item u-band dropouts are rejected as they are likely Lyman-break
galaxies at $z>2.5$ \citep{giavalisco2002}.
\end{enumerate}

The selection of the radio sources is based mainly on a flux
requirement, criterion (1). The following ones (2, 3, 4)
are used only to discard spurious sources from the sample.

The source COSMOS-FR~I~236, tentatively classified in C09 as a QSO, was
later identified with a known QSO at the spectroscopic redshift $z=2.132$
\citep{prescott06}.
Similarly to what done for all sources in our sample (see Sect.~\ref{par:subsample} and Sect.\ref{sec:HLRGs}), we estimate
that the total radio power of this
source is 1.96$\times$10$^{33}$~erg~s$^{-1}$~Hz$^{-1}$, based on its redshift and  FIRST radio
flux of 7.10~mJy \citep[see][]{baldi2013}.
We also assume a radio spectral index $\alpha=0.8$ (see Sect.~\ref{sec:K-correction}).
Therefore, since this is typical of high power FR~IIs and radio
loud QSOs, we do not consider this source in this paper. Steepening the radio spectrum, i.e. increasing the value of the spectral index $\alpha$, would increase the estimated radio power, reinforcing our conclusions.
Hence, our sample comprises 36 sources. Note that the sample, as for
any flux limited one, is affected by the well-known Malmquist bias
and thus includes higher/lower power radio sources at high/low redshifts (see Sect.~\ref{par:subsample}, \ref{sec:HLRGs}).

As the aim of this work is to search for clusters of galaxies in
the fields of the low power radio galaxies of the C09 catalog, in the following section we redefine the sample by selecting only
{\it bona fide} low luminosity objects, based on the latest
photometric (or spectroscopic, when available) redshift
estimates. While we cannot exclude that the remaining (high power)
sources are associated with a dense environment, we will
consider them separately.

Hereinafter, we will refer to our sources using the ID number only, as
opposed to the complete name COSMOS-FR~I~nnn.

\section{Sample redefinition}
\label{par:sample_redef}
The aim of this section is to derive a reliable sample of low
luminosity radio galaxies (LLRGs) that, based on the information
available to date, have ${\rm L}_{\rm 1.4~ GHz}$ lower than the fiducial
separation between FR~Is and FR~IIs. In order to do so we require robust measurements
of the total radio fluxes, accurate photometric redshifts (in
absence of firm spectroscopic redshifts) and assumptions on the
K-correction.

\subsection{Radio fluxes}
\label{par:radio_fluxes}

As discussed above, the C09 sample was selected using the radio fluxes
from the FIRST survey \citep{becker95} which was performed by using
the VLA B-configuration at 1.4~GHz and it covers 10,000 square degrees
of the North and South Galactic Caps. The COSMOS field entirely
resides within the area mapped by FIRST. Post-pipeline radio maps have
a resolution of $\sim 5$~arcsec. The detection limit of the FIRST
catalog is $\sim 1$~mJy with a typical rms of 0.15~mJy. When we
make use of the FIRST survey, we adopt the flux densities from the
catalog as of October 10th, 2011.
However, the FIRST
radio maps may be missing a substantial fraction of any extended low
surface brightness radio emission from the lobes of our radio sources,
which are close to the detection limit.  This is particularly
important because of the relatively high angular resolution provided
by the used VLA configuration, which is more suitable for detecting
compact or unresolved radio sources.

While being slightly shallower than FIRST, the NRAO VLA Sky Survey (NVSS) survey
\citep{condon1998} may be more suitable for our purposes, since it
was obtained by using the VLA-D configuration at 1.4~GHz. The angular
resolution of the NVSS radio maps is 45 arcsec (FWHM). Thus, it is
more suitable for detecting extended emission of the sources in our
sample. Therefore, in order to derive the total radio luminosity of our
sources, we use the NVSS fluxes and upper limits (as of
October 10th, 2011), when possible. In the NVSS
catalog\footnote{\url{http://www.cv.nrao.edu/nvss/}} at the
coordinates of the C09 objects, we find 26 of the 36 sources.

While the FIRST survey is complete down to a flux of 1~mJy, the
completeness of the NVSS catalog is only 50$\%$ at its formal limit of
2.5~mJy, while rises rapidly to 99$\%$ at 3.4~mJy \citep{condon1998}. Thus, the
drawbacks of using NVSS sources are as follows: i) sources with total
radio flux $<3.4$~mJy might not be included. ii) The identification of
the NVSS counterpart of each source is not trivial. Due to the lower
angular resolution rms uncertainties are about 7 arcsec at the NVSS limit, as affected by confusion.
Furthermore, the extended radio morphology of many of the radio sources might be complex. Therefore, since the NVSS is more sensitive
to the extended emission than FIRST, the centroid of the FIRST source could
not coincide with that in the NVSS map.  Also note that, even if
the limit of the NVSS catalog is set at 2.5~mJy, some of our fainter
sources are detected in the radio maps.

To overcome these inconveniences we use FIRST \citep{becker95}  and VLA COSMOS \citep{VLA_COSMOS}.
FIRST has a flux density threshold of 1~mJy and a positional accuracy of $\lesssim$1~arcsec for radio pointlike sources.
VLA COSMOS has a angular resolution of 1.5''$\times$1.4'' and a sensitivity limit of 45~$\mu$Jy/beam. It is therefore deeper and with higher angular resolution than FIRST.
For the majority of the
objects it is straightforward to identify the radio
sources in the above surveys. The few cases in which the
identification is problematic are discussed in the following.

\begin{figure}
\begin{center}
\includegraphics[width=0.4\textwidth, angle=270,natwidth=610,natheight=642]{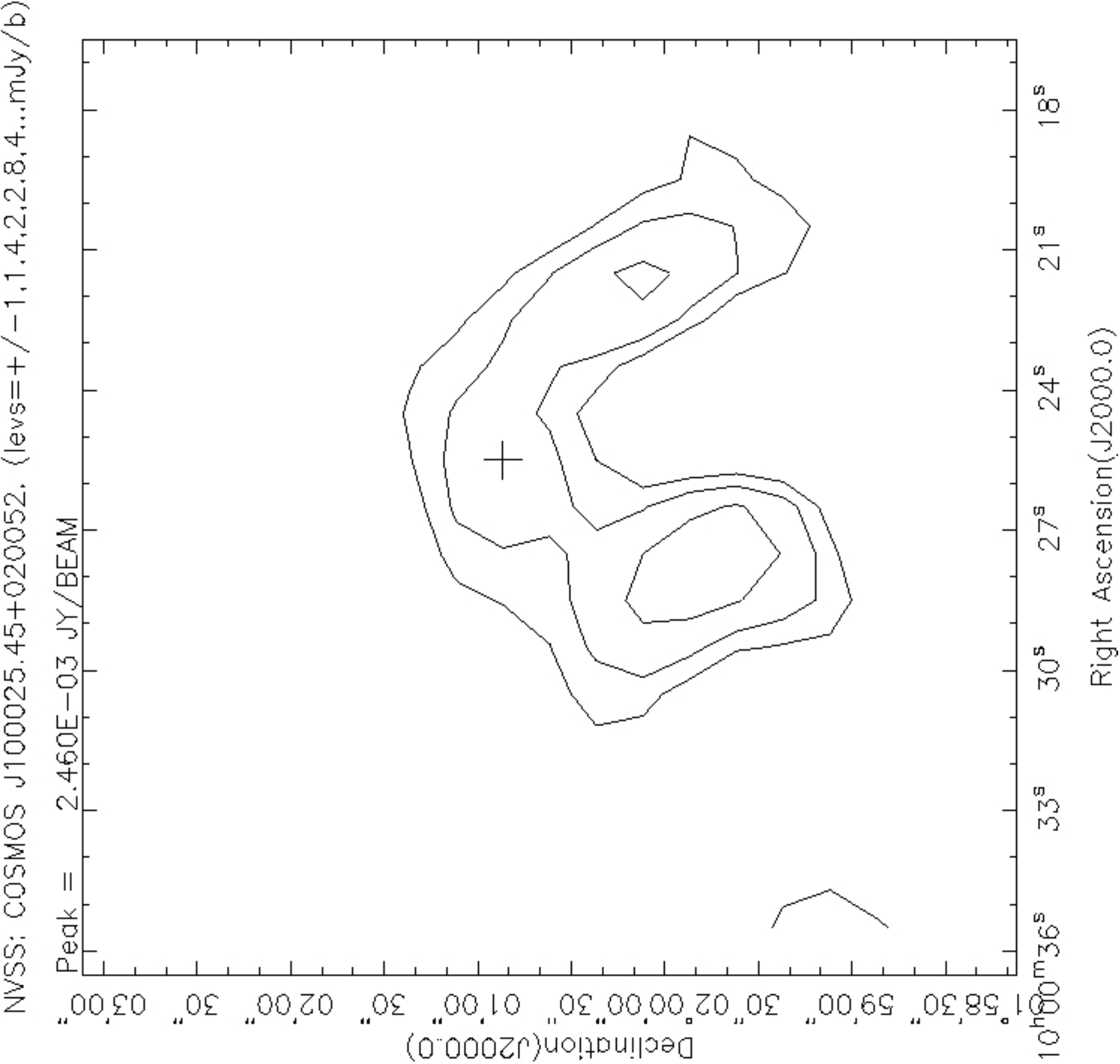}
\end{center}
\caption{NVSS map, field of 05. The cross marks the coordinates of the radio source.}
\label{fig:FRI_005_NVSSmap}
\end{figure}

For these cases we consider the VLA COSMOS maps to clearly identify the radio sources, as described in the following
for source 05.
In Figure~\ref{fig:FRI_005_NVSSmap} we show the NVSS radio map of the field around the object 05.
Visual inspection reveals the
presence of a complex radio morphology, which might be (erroneously)
identified with either the narrow-angle tail \citep[NAT, e.g. NGC~1265,][]{Odea_Owen1986} or the wide-angle tail
\cite[WAT, e.g. 3C465,][]{venturi1995} radio
morphology. The NVSS catalog reports sources at distance of $\sim60$ and $\sim67$ arcsec to the
SW and SE from the VLA-COSMOS coordinates of the source 05, and fluxes of 3.4 and 3.7~mJy, respectively.
A third radio source located at the position
of 05 is visible in the map, but it is below the threshold of the NVSS
catalog.

\begin{figure*} \centering
\subfigure{\includegraphics[width=0.4\textwidth,natwidth=610,natheight=642]{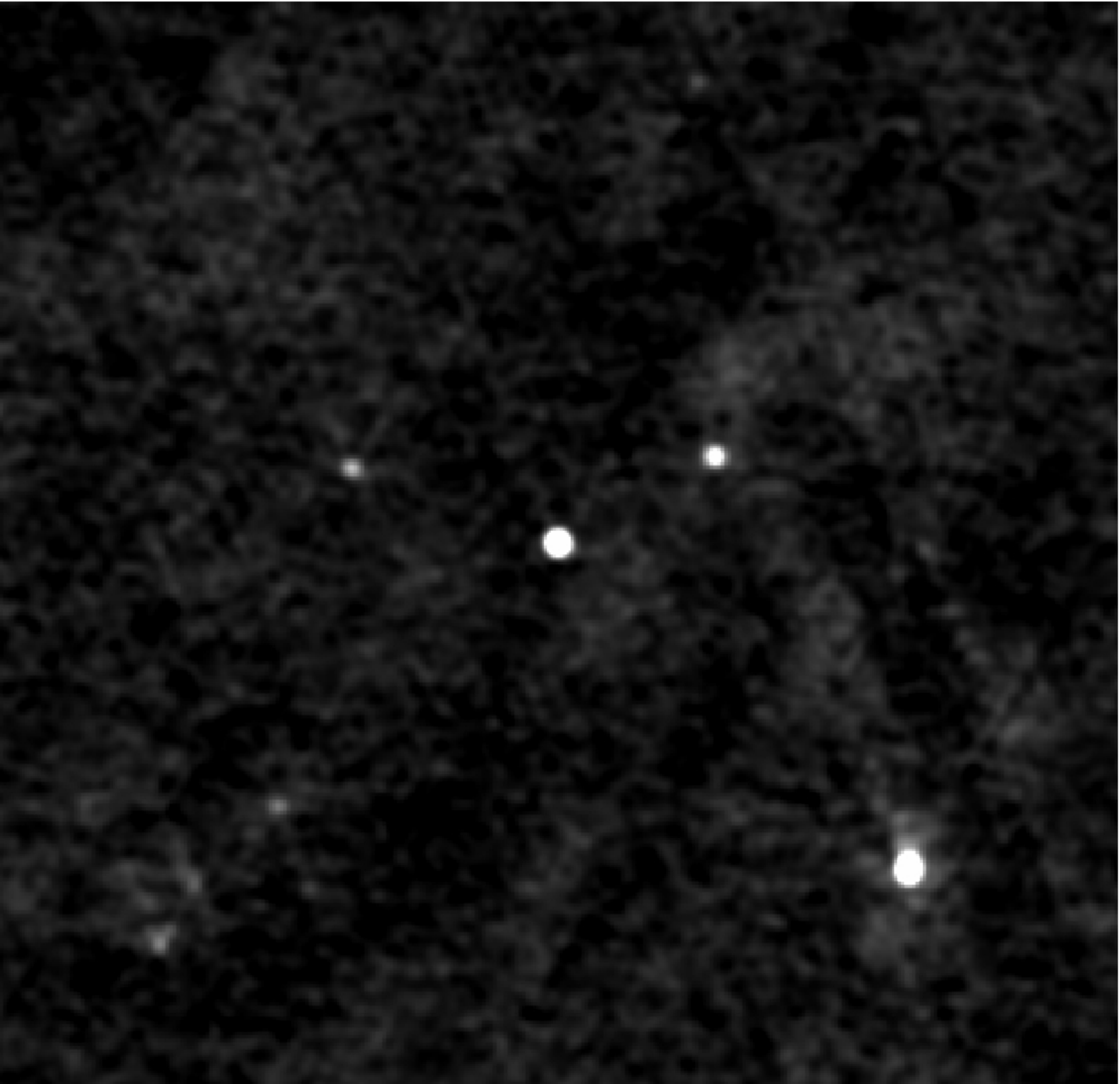}}\qquad
\subfigure{\includegraphics[width=0.4\textwidth,natwidth=610,natheight=642]{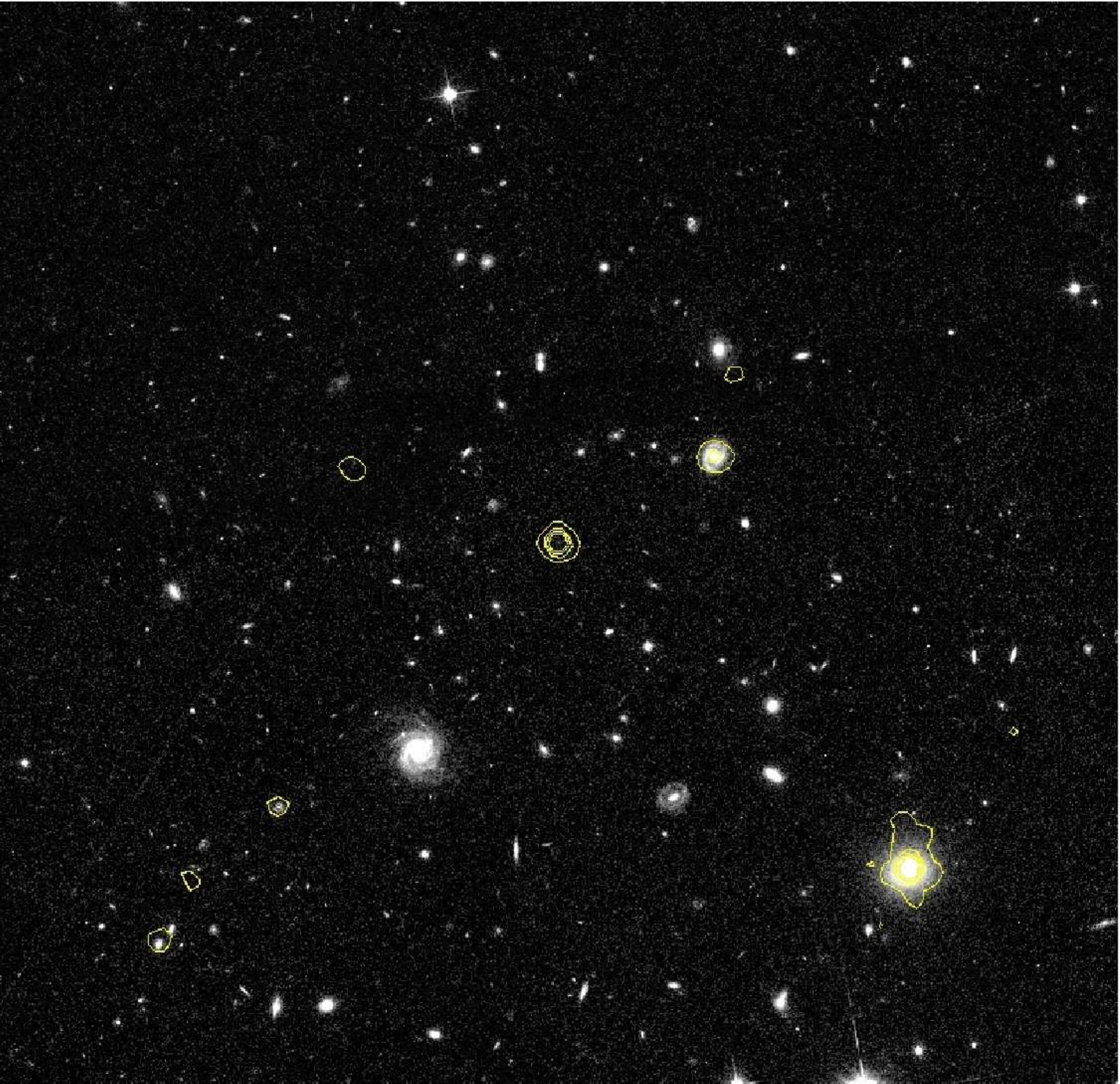}}
\caption{Field ($3'\times3'$ dimensions) of 05. Left: VLA-COSMOS map. Right: HST image taken
from ACS and the F814W filter. Yellow contours are from VLA-COSMOS. The
angular scale is the same for both of the panels.}
\label{fig:FRI_005_VLAplusHSTmap}
\end{figure*}

In Figure \ref{fig:FRI_005_VLAplusHSTmap} (left) we show the
same field as seen with VLA-COSMOS, at much higher angular
resolution. Such image shows the presence of a number of point-like
sources and some extended emission. In the right panel we report
the HST image of the same field, taken with the Advanced Camera for
Surveys (ACS) and the F814W filter, as part of the COSMOS survey. The
radio contours from VLA-COSMOS are over-plotted in yellow. It is clear
that the radio sources seen in VLA-COSMOS overlap with foreground
galaxies. This generates the complex extended emission seen in the
NVSS map. By using higher resolution radio data and the optical image,
we are able to overcome the confusion problem in the NVSS map. The
NVSS catalog misses our source and detects only the two unrelated brighter
radio emitting regions.

\begin{figure*} \centering
\subfigure{\includegraphics[width=0.4\textwidth, angle=270,natwidth=610,natheight=642]{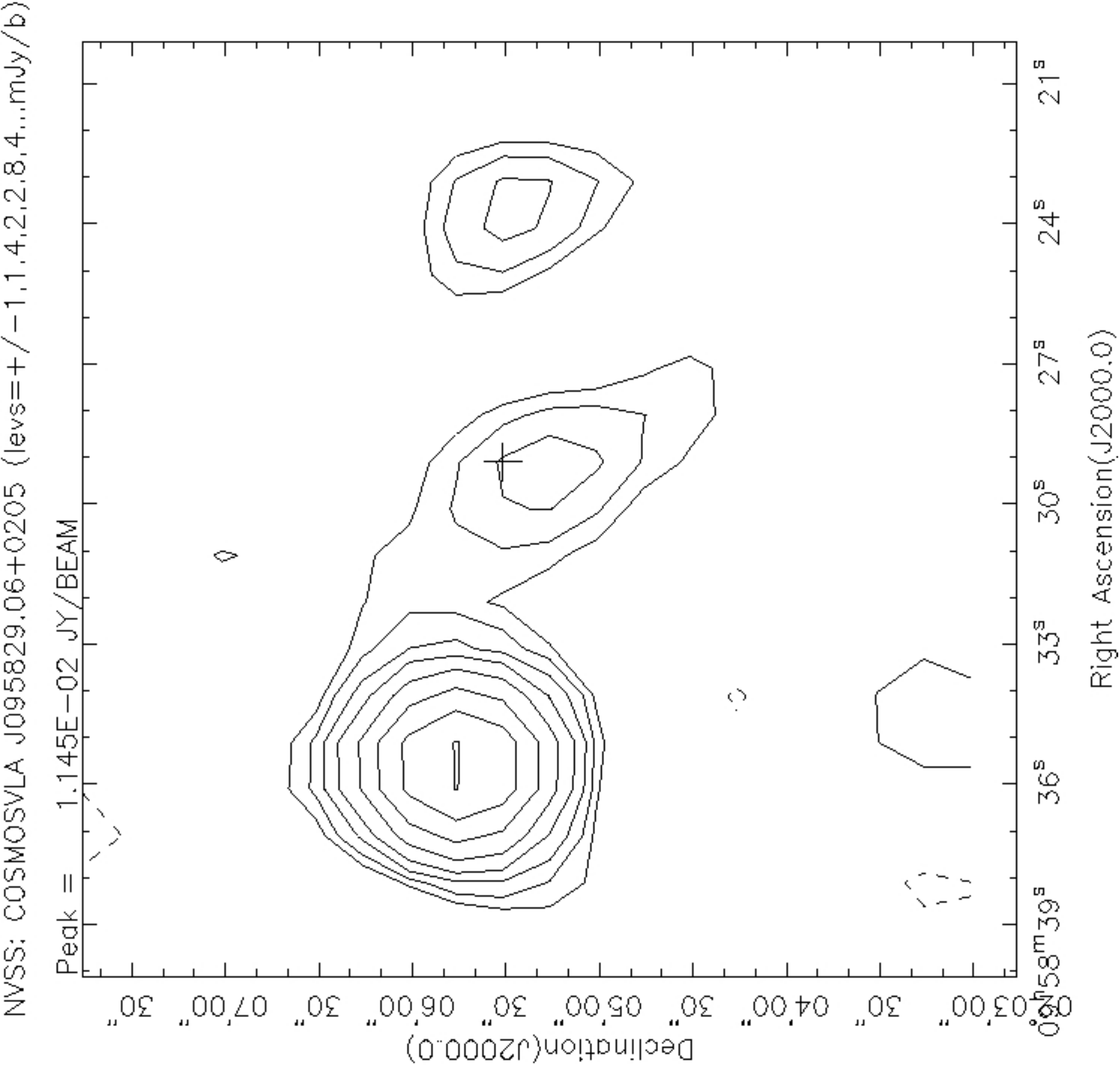}}
\subfigure{\includegraphics[width=0.4\textwidth,angle=270,
natwidth=610,natheight=642]{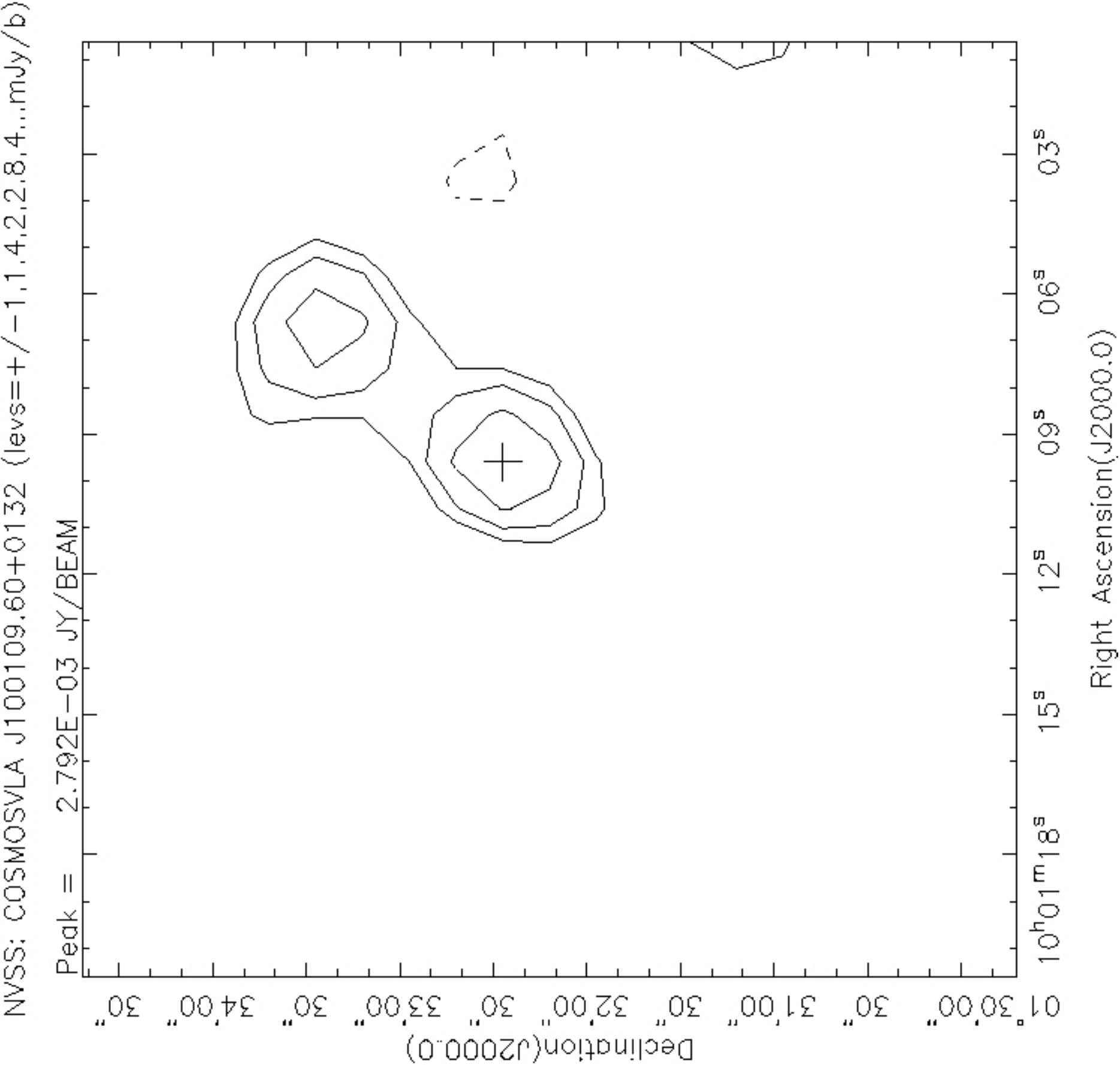}}\qquad
 \caption{NVSS maps, the cross marks the coordinates of the radio
source. Left: field of 26. Right: field of 224.}
\label{fig:NVSSmaps_confusion}
\end{figure*}

Similarly, other sources have extended radio morphology, as clear from visual inspection of the NVSS maps.
The angular separation between the coordinates reported in the NVSS catalog
and those obtained by using VLA-COSMOS are about $\sim$15~arcsec.
This is the case of sources 26, 52, 202, 224, and 228, where such angular separations are
15.37, 16.4, 12.82, 12.43, and 18.52 arcsec, respectively.
In Figure~\ref{fig:NVSSmaps_confusion} we report the NVSS fields of 26 and 224, as examples.
These sources show a radio morphology similar to that of 05. However, a bright source is
clearly present in each of these two fields, very close to the radio
galaxy. They are merged in the NVSS map in a single structure due to
the low NVSS angular resolution.

\begin{figure*} \centering
\subfigure{\includegraphics[width=0.4\textwidth, angle=270,natwidth=610,natheight=642]{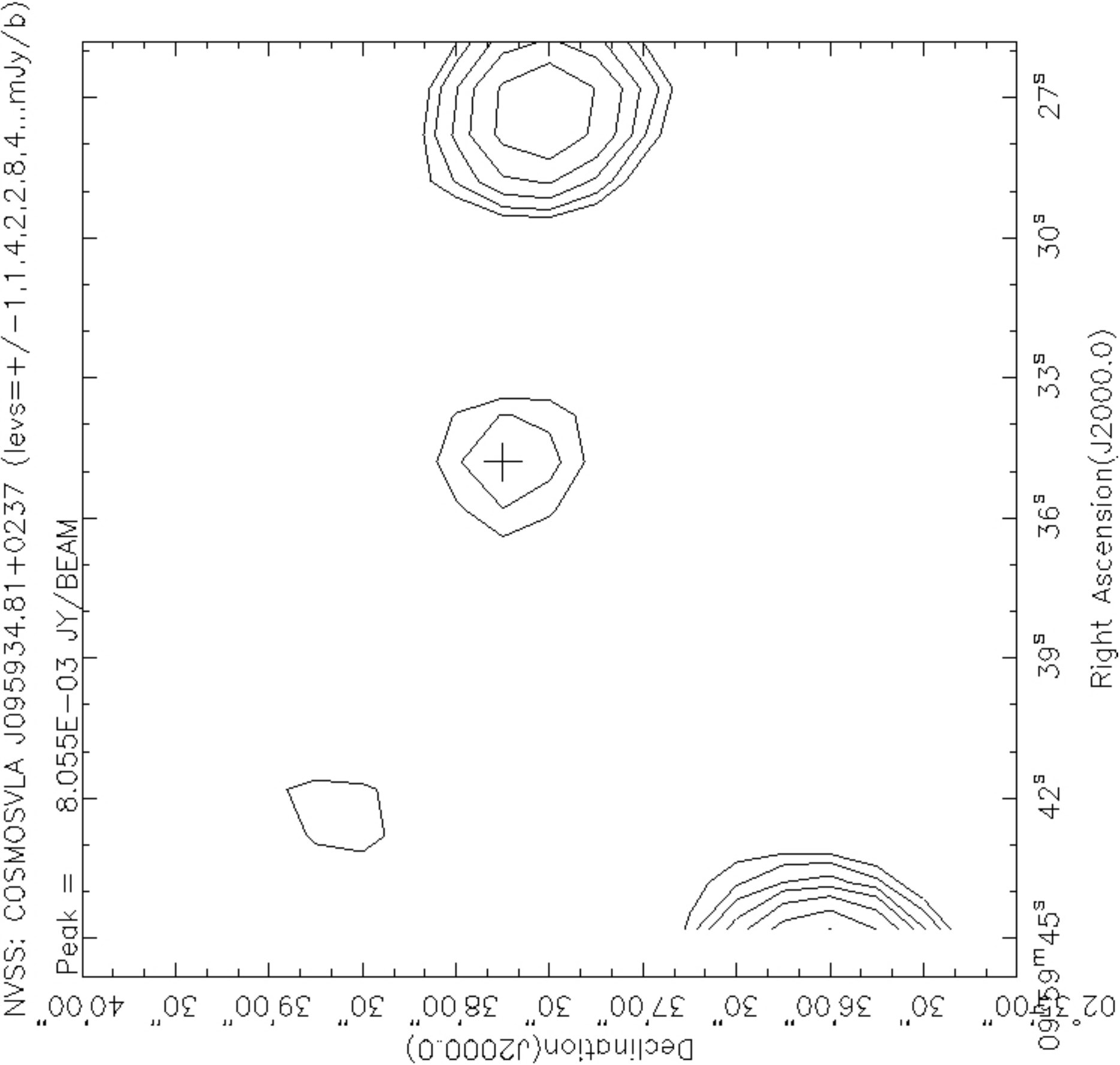}}\qquad
\subfigure{\includegraphics[width=0.4\textwidth, angle=270,natwidth=610,natheight=642]{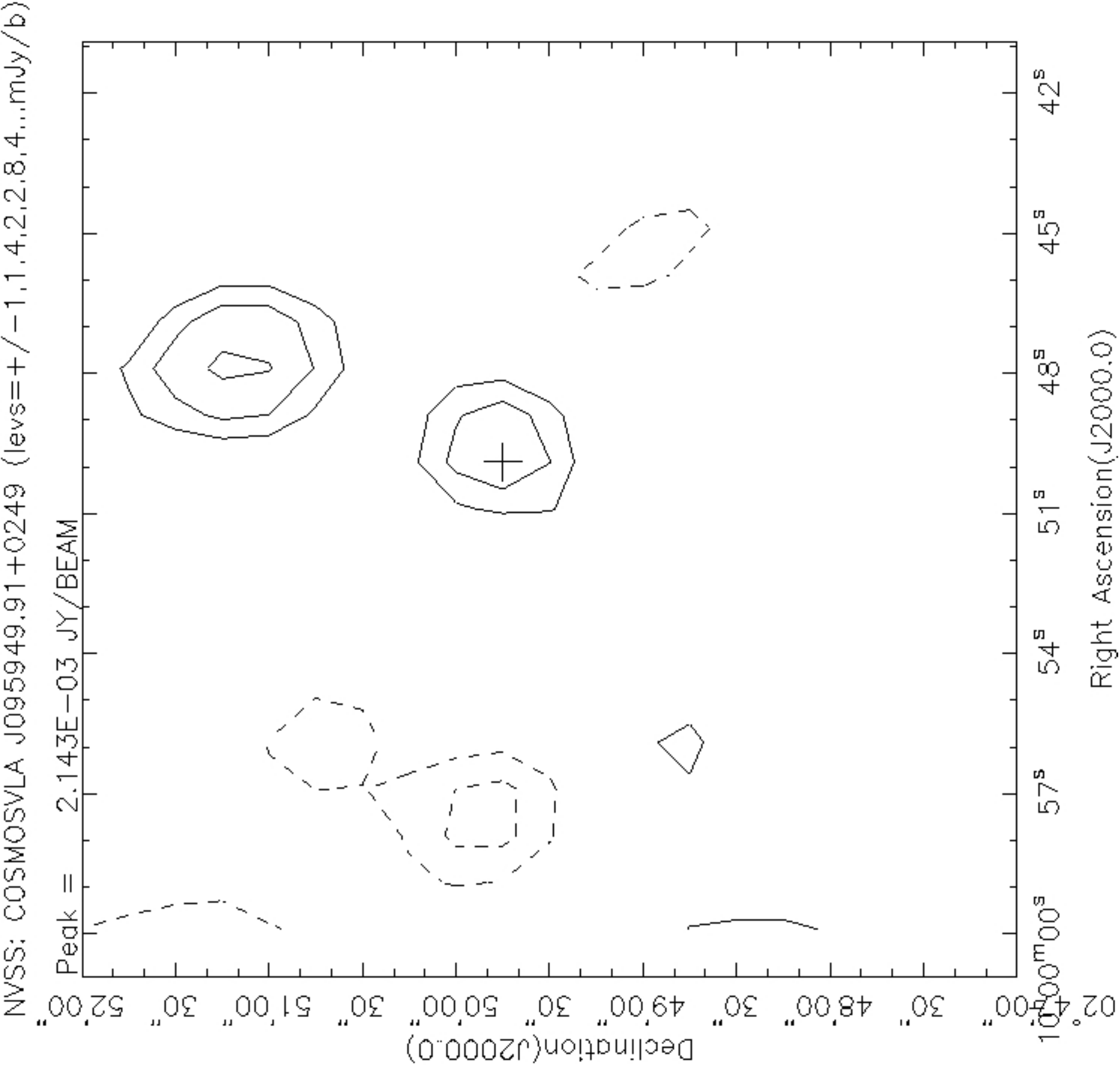}}
\caption{NVSS maps, the cross marks the coordinates of the radio
source. Left: field of 22. Right: field of 39. Examples of sources not included in the NVSS catalog, but clearly present in the NVSS maps. Their 1.4~GHz fluxes are close to the NVSS 2.5~mJy limit.}
\label{fig:NVSSmaps_offset}
\end{figure*}

We  consider the radio NVSS maps of all  of the eight sources that are not present in the NVSS catalog.
We visually inspect each map and search for the presence of radio contours centered around the position of the radio source.
For five out of the eight we find evidence of a radio source located at the coordinates of the radio galaxy.
This is the case of sources 11, 20, 22, 27, and 39, where the radio contours are consistent
with a radio flux close to the NVSS formal limit of 2.5~mJy.
In Figure~\ref{fig:NVSSmaps_offset} we report the fields of 22 and 39, as examples.
Being very close or below the formal completeness limit, we expect that possible systematics might occur in the flux measurements.
Therefore we adopt a fiducial 2.5~mJy upper limit
for all of the eight sources which are not included in the NVSS catalog.

The fiducial FIRST and NVSS flux uncertainties for the sources in our sample are within $\sim$0.1-0.2~mJy and $\sim$0.4-0.6~mJy,
respectively. However, we prefer not to report the flux uncertainty associated with each source. This is because we are
considering fluxes down to the completeness limit of both the FIRST and the NVSS surveys and, therefore,
the flux uncertainties might be underestimated.

\subsection{Redshifts}
\label{sec:redshifts}
We adopt accurate photometric redshifts derived by \citet[][hereinafter B13]{baldi2013} through a careful analysis of
the spectral energy distribution (SED) of the host
galaxies.
As mentioned in Sect.~\ref{par:sample}, throughout this paper we
adopt the photometric redshifts derived in B13, that specifically
focused on the sample considered here. These photometric redshifts
have a great advantage with respect to those in \citet{mobasher2007} and \citet[][hereinafter I09]{ilbert2009},
which were automatically derived by using the COSMOS photometric catalogs.

I09 estimated photometric redshifts by using the photometric data points from 30 bands for those sources with I$<25$ in 
in the deep Subaru area of the COSMOS field \citep{taniguchi2007}.
B13 carefully identified the optical counterparts of the radio sources in all of the
photometric bands.  The authors discovered that, in a few cases, sources in
different bands were misidentified in the COSMOS source list, therefore leading to erroneous photometric redshift estimates.
B13 also performed a more refined SED
modeling, with the inclusion of two stellar populations. At variance
with the I09 catalog, B13 considered only
broad band photometric data and excluded narrow and medium band data,
which can be strongly contaminated by emission lines that are not included in
the stellar templates.

We also search for the spectroscopic redshift of our sources in the zCOSMOS-bright \citep{lilly2007}
and MAGELLAN \citep{trump2007} catalogs. Only 7 out of the 36 sources in our sample are found.

In agreement with B13 we do not use the spectroscopic
redshift for object 25. This is because of its clear misidentification in the
MAGELLAN catalog  (see Sect.~6.1 in B13).  Therefore, for the great majority of the sources we have to rely
on photometric redshifts.

The redshifts of three (namely 27, 52, and 66) out of the 7 sources for which spectroscopic redshifts are available are significantly
outside the $z\sim1-2$ range of C09 selection. Therefore we exclude them from the sample.
Redshifts  $z=0.2847$  and $z=0.7417$ are reported in the MAGELLAN catalog for the sources 27 and 52, respectively.
The redshifts reported for source 66 in the MAGELLAN and the zCOSMOS-bright catalog are consistent with each other
and equal to  $z=0.6838$ and $z=0.6803$, respectively.
Searching for cluster candidates at intermediate or low redshifts (i.e. $z\lesssim0.8$)
 is not the aim of this project.
Therefore, we naturally reject the sources 27, 52, and 66, that are all located at $z\leq0.75$.
We also exclude the source 07 from the sample because it is a peculiar radio source
\citep[as suggested in][]{baldi2013}.
It might be a FR~II radio galaxy at significant high redshift.
It will be studied in a forthcoming paper.
Conversely, we do not exclude those sources (e.g. 28 and 32) that have a photometric redshift formally above $z\sim2$.
This is because, even if they are at redshifts well
outside the fiducial range of our interest, they were not rejected during the C09 selection.
Therefore, they could
comprise similar properties to those of the other galaxies in our sample.
Furthermore, since such sources populate the high redshift tail of our sample,
their Mpc-scale environments are still worth to investigate
(see also Sect.~\ref{sec:papovich_test} for further discussion about source 28).

Summarizing, with respect to the original list given in C09,
we reject sources 07, 27, 52, and 66 (in addition to 236, the QSO we
already discussed above). The sample is thus reduced to 32 objects.

\subsection{ Rest frame radio luminosities}\label{sec:K-correction}

In agreement with C09 we assume that the
radio spectrum in the region around 1.4~GHz is a power-law of the form
$S_\nu\propto\nu^{-\alpha}$,  where $S_\nu$ is the radio flux density
at the observed frequency $\nu$, and $\alpha$ is the spectral index
assumed to be $\alpha=0.8$, accordingly to C09.
Such an assumption
requires that the flat ($\alpha \sim 0$) radio emission of the core is
negligible with respect to the extended emission (jets and lobes) in
the considered spectral range. This is formally correct at the lowest
radio frequencies, but it is less certain at higher
frequencies. However, since the radio data do not allow us to separate
the emission of our sources into different components, we assume that
the measured flux at 1.4~GHz is dominated by the extended emission. If
$\alpha=$~0.3 instead of 0.8, the luminosity would
increase by only a factor of $<1.8$, for the worst case of a
source at $z =2$.

Thus the isotropic  rest frame 1.4~GHz luminosity density is given by:

\begin{equation}
\label{eq:Kcorrection}
L_{1.4} = 4\pi S_{1.4}D_L(z)^2\left(1+z\right)^{\alpha-1},
\end{equation}
where $S_{1.4}$ is the observed flux density at 1.4~GHz, D$_L$ is the
luminosity distance.

\begin{figure*} \centering
\subfigure{\includegraphics[width=0.48\textwidth,natwidth=610,natheight=642]{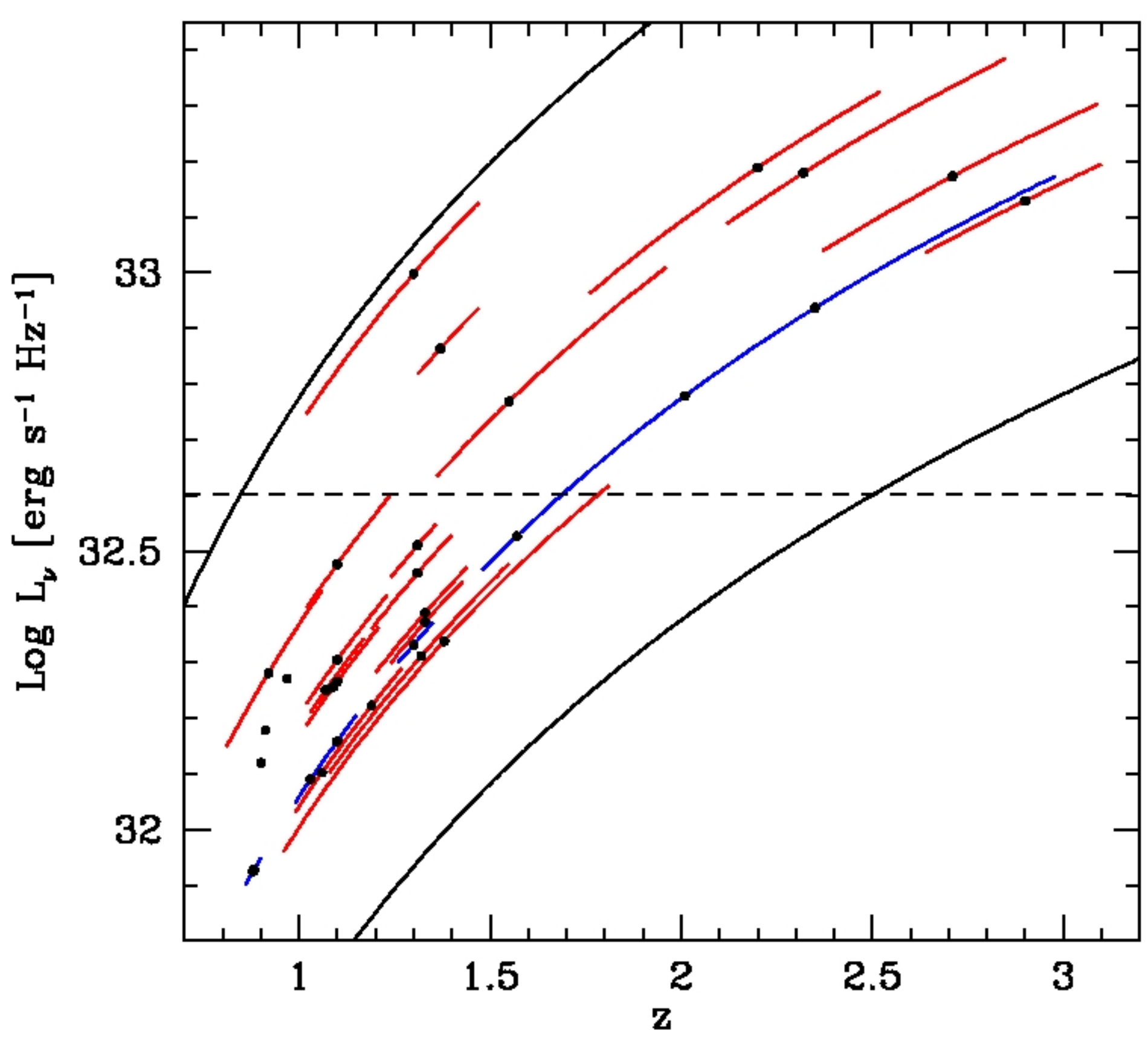}}\qquad
\subfigure{\includegraphics[width=0.48\textwidth,natwidth=610,natheight=642]{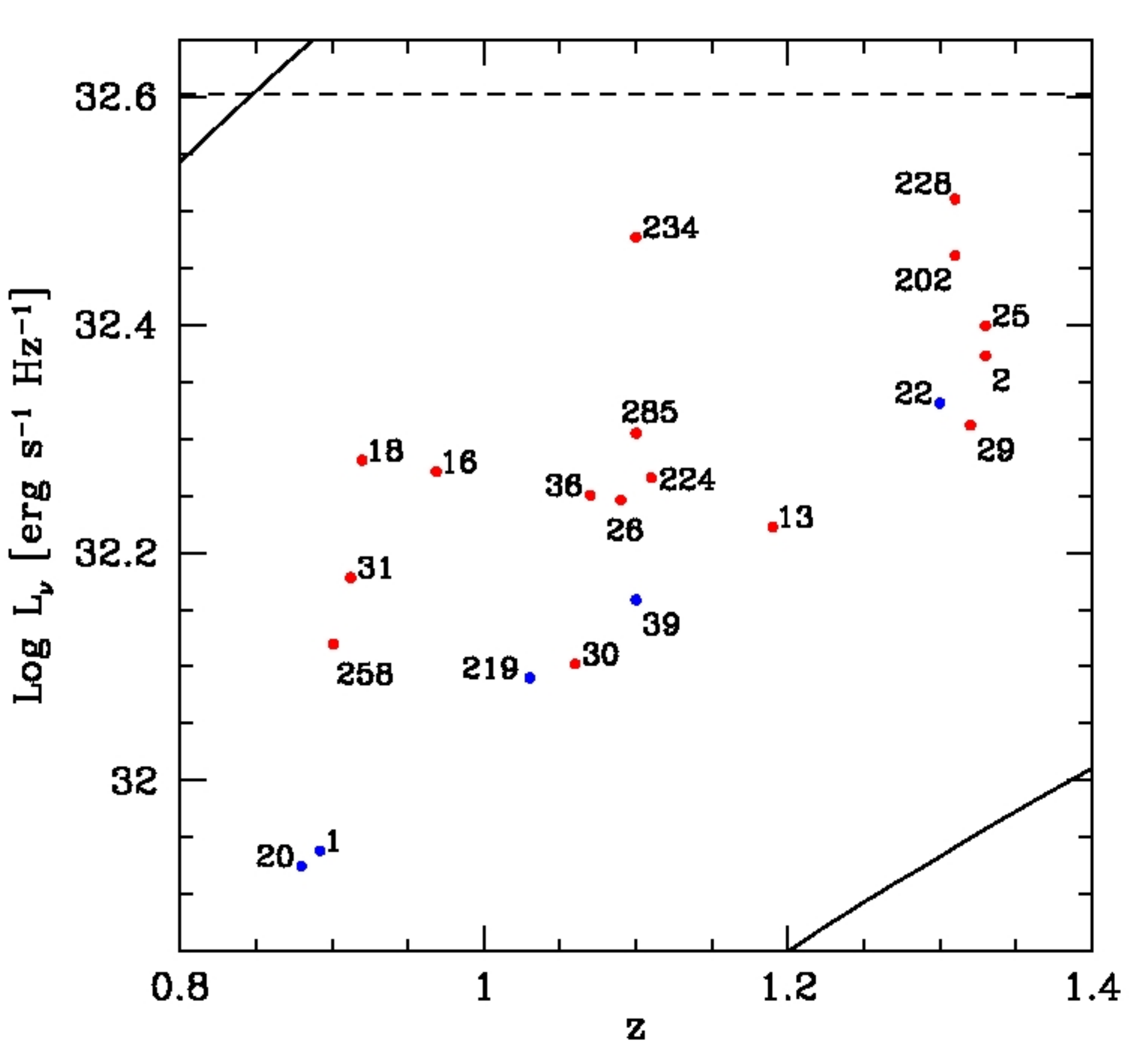}}
\caption{Left: Luminosity vs redshift  scatterplot.
Red lines correspond to sources with  NVSS counterpart and
fluxes. Blue lines correspond to upper limits at 2.5~mJy flux for the sources with no NVSS flux.
 Solid black lines: FIRST cut at 1mJy, 13mJy;
 The blue and the red lines indicate the uncertainties on the photometric redshift.
The x- and y-values of the dots
are the redshift and the luminosity of each source, respectively.
Dots that are not associated with lines show the objects with spectroscopic redshifts.
Horizontal dashed line: FR~I/FR~II
luminosity divide, assumed to be constant with redshift.
Right: LLRGs only. Color legend is the same as for the left panel.
 Each dot represents a source, identified by the corresponding ID
number.
}
\label{fig:scatterplot}
\end{figure*}


\subsection{The Low Luminosity Radio Galaxy subsample}\label{par:subsample}

In Figure~\ref{fig:scatterplot}  (left panel) we report the luminosity vs.
redshift scatterplot. 
The lower/upper thick black lines  in the plot are the FIRST sample selection
lower/upper boundaries adopted in C09 (1.0~mJy and 13.0~mJy,
respectively). Since NVSS fluxes are in general higher than FIRST
fluxes, we expect all the sources to lie above the lower
line. 

Since we are interested in searching for clusters around FR~Is, we consider the 1.4~GHz luminosity intervals
spanned for each source, within the redshift uncertainties,
for an assigned 1.4~GHz radio flux.

Therefore, we conservatively select only those sources whose 1.4~GHz luminosity intervals lie entirely below the
FR~I/FR~II radio luminosity divide of $4\times 10^{32}$~erg~s$^{-1}$~Hz$^{-1}$.
According to this criterion we select 21 {\it bona fide} LLRGs, whose
redshifts span the range z~=~0.88-1.33 and have radio luminosities between
L$_{1.4}$~= (0.84-3.24)$\times$10$^{32}$~erg~s$^{-1}$~Hz$^{-1}$.
In Figure~\ref{fig:scatterplot} (right panel) we plot the scatterplot focused on the LLRGs only.
The median redshift and 1.4~GHz luminosity of the LLRGs are $z_{\rm median}=1.1$ and
L$_{\rm 1.4,~median}=1.84\times10^{32}$~erg~s$^{-1}$~Hz$^{-1}$, respectively. For comparison, radio galaxies of similar power,
selected within the 3C catalog, span
a much smaller redshift range. \cite{chiaberge1999} report a range $z=0.0037-0.29$ and a median value $z=0.03$
for their sample of 33 FR~Is.

The LLRGs span  a limited range of luminosity and slightly
broader of redshift.  However, because of the steepness of the radio luminosity function,
most sources are at $z\sim$1.

Being at relatively low redshifts,
these objects and their Mpc-scale environment can be studied in
greater detail than the whole sample of FR~I candidates considered in this work.
This is mainly because COSMOS field number densities are much higher and statistical photometric redshift uncertainties
are smaller than at higher redshifts \citep{ilbert2009}. Furthermore, spectroscopic redshift information is available for some of the LLRGs only and
photometric redshifts from B13 are more accurate for the LLRGs than for the HLRGs, being the latter, on average, at higher redshifts.

Therefore we separate the LLRGs from the remaining sources, that are generally at  higher luminosities and redshifts
than the  HLRGs.
In particular, the photometric redshifts of the LLRGs are better
constrained, since the typical statistical uncertainty dramatically
increases above $z\sim 1.3$ (see e.g. Figure 9 in I09) and because all of the  sources
in our sample with spectroscopic redshifts belong to the LLRG class.

\subsection{The High Luminosity Radio Galaxy subsample}\label{sec:HLRGs}

We consider in this section the remaining sources of the sample, i.e. the HLRGs, that do not belong to the LLRG subclass.
Note that the radio morphology of both the LLRGs and the HLRGs is not of FR~II type.
In fact, sources with a clear FR~II morphology have been rejected as part of the original sample selection in C09.
Furthermore, the cosmological evolution of the FR~I/FR~II radio divide is
still unknown, i.e. high-z FR~I sources might have higher
radio power than those of local FR~Is, as suggested by  \citet{heywood2007}.

This makes the nature of these HLRGs very unclear and suggestive to investigate.
In the following, we consider the HLRGs separately from the rest of the sample (i.e. the LLRGs)
in order to avoid any bias due to  possible differences in the Mpc-scale environments of low and high luminosity sources.

We find 11 HLRGs. Their redshifts and radio luminosities span the intervals $z= 1.30-2.90$ and
L$_{1.4}=(2.18-15.44)\times10^{32}$~erg~s$^{-1}$~Hz$^{-1}$, respectively. The
median redshift and luminosity are $z_{\rm median}= 2.01$,
L$_{\rm 1.4,~median}= 8.64\times10^{32}$~erg~s$^{-1}$~Hz$^{-1}$,
respectively.

\begin{table*}
\caption{Sample properties.}
\label{table:sampleprop} \centering

\begin{tabular}{c}
{\it The Low Luminosity Radio Galaxy subsample}
\end{tabular}

\begin{tabular}{cccccccc}
\hline\hline
ID  & RA  &  DEC  &  redshift  &  FIRST flux &  NVSS flux & L$_{1.4~{\rm GHz}}$  & radio morphology \\
    & [deg] & [deg]   &            &  [mJy]      &  [mJy]     &  [$10^{32}$~erg~s$^{-1}$~Hz$^{-1}$]     &    \\
\hline\hline

COSMOS FR~I 01  & 150.20744     & 2.2818749     & 0.8823$^a$-0.8827$^b$      &       1.06 & --- & 0.85  & compact     \\
COSMOS FR~I 02  & 150.46751     & 2.7598829     & 1.33$\pm^{0.10}_{0.09}$     & 2.25 & 2.6      &     2.36	        & extended     \\
COSMOS FR~I 13  & 149.97784     & 2.5042069     & 1.19$\pm^{0.08}_{0.11}$     &     1.50     &	     	2.4 &     1.67	         & compact  \\
COSMOS FR~I 16  & 150.53772     & 2.2673550     & 0.9687$^a$               & 5.70     &    4.4 & 1.87          & unresolved \\
COSMOS FR~I 18  & 149.69325     & 2.2674670     & 0.92$\pm^{0.14}_{0.11}$	     &	     4.39     &        5.1 &     1.91	         &	     extended     \\
COSMOS FR~I 20  & 149.83209     & 2.5695460     & 0.88$\pm^{0.02}_{0.02}$	     &       1.33     &	     	--- &     0.84	          & extended     \\
COSMOS FR~I 22  & 149.89508     & 2.6292144     & 1.30$\pm^{0.05}_{0.04}$  & 2.74    & --- &     2.14   &  compact     \\
COSMOS FR~I 25  & 150.45673     & 2.5597000     & 1.33$\pm^{0.11}_{0.13}$	     &	     2.18     &        2.7 &     2.45	         &	     compact     \\
COSMOS FR~I 26  & 149.62114     & 2.0919881     & 1.09$\pm^{0.12}_{0.07}$	     &        1.88     &	     	3.2 &     1.80	         & extended     \\
COSMOS FR~I 29  & 149.64587     & 1.9529760     & 1.32$\pm^{0.23}_{0.24}$     & 2.13     &        2.3 &     2.05      &	     compact     \\
COSMOS FR~I 30  & 149.61542     & 1.9910541     & 1.06$\pm^{0.11}_{0.07}$	     &	     1.26     &        2.4 &     1.27	          &	     compact     \\
COSMOS FR~I 31  & 149.61916     & 1.9163600     & 0.9123$^a$-0.9132$^b$        &  3.71     &	     	4.1 	     	     &     1.51	       & compact     \\
COSMOS FR~I 36  & 150.55662     & 1.7913361     & 1.07$\pm^{0.10}_{0.04}$	     &	     3.19 &        3.3 	     	     &     1.78	         & unresolved \\
COSMOS FR~I 39  & 149.95804     & 2.8288901     & 1.10$\pm^{0.05}_{0.05}$        &	     1.37     &	     	--- &     1.44	         & compact     \\
COSMOS FR~I 202 & 149.99506     & 1.6324950     & 1.31$\pm^{0.09}_{0.12}$	     & 1.08        &	     	3.3 	     	     &     2.89	        & extended     \\
COSMOS FR~I 219 & 150.06444     & 2.8754051     & 1.03$\pm^{0.02}_{0.04}$	     &	     1.85 &	        --- &     1.23	          &	     compact     \\
COSMOS FR~I 224 & 150.28999     & 1.5408180     & 1.10$\pm^{0.10}_{0.04}$	     &        3.31 &	     	3.2 	     	     &     1.84	        & extended     \\
COSMOS FR~I 228 & 149.49455     & 2.5052481     & 1.31$\pm^{0.05}_{0.07}$   &  2.04     &        3.7 &     3.24      &	     compact     \\
COSMOS FR~I 234 & 150.78925     & 2.4539680     & 1.10$\pm^{0.14}_{0.08}$	     &	     4.43     &        5.2 &     3.00	          &	     extended     \\
COSMOS FR~I 258 & 149.55934     & 1.6310670     & 0.9009$^b$              &        2.24     & 3.7 	     	     &     1.32	    &    compact     \\
COSMOS FR~I 285 & 150.72131     & 1.5823840     & 1.10$\pm^{0.13}_{0.08}$	     &	     2.95 &	        3.5 &     2.02	          & extended     \\
\hline
\vspace{0.1cm}
\end{tabular}
\begin{tabular}{c}
{\it The High Luminosity Radio Galaxy subsample}
\end{tabular}
\begin{tabular}{cccccccc}
\hline\hline
ID  & RA  &  DEC  &  redshift  &  FIRST flux &  NVSS flux & L$_{1.4~{\rm GHz}}$  & radio morphology \\
    & [deg] & [deg]   &            &  [mJy]      &  [mJy]     &  [$10^{32}$~erg~s$^{-1}$~Hz$^{-1}$]                    &    \\
\hline\hline
COSMOS FR~I 03     & 150.00253     & 2.2586310     & 2.20$\pm^{0.32}_{0.44}$     &	     4.21 &        5.2 & 15.44	          & unresolved \\
COSMOS FR~I 04     & 149.99153     & 2.3027799     & 1.37$\pm^{0.10}_{0.06}$     &	     5.99     &       7.5 &     7.30	         &	     extended     \\
COSMOS FR~I 05     & 150.10612     & 2.0144780     & 2.01$\pm^{0.22}_{0.35}$     &        1.30     &	     	--- &     6.01	         & compact     \\
COSMOS FR~I 11     & 150.07816     & 1.8985500     & 1.57$\pm^{0.14}_{0.09}$	     &       1.13     &	     	--- &     3.36     &	     compact     \\
COSMOS FR~I 28     & 149.60064     & 2.0918673     & 2.90$\pm^{0.20}_{0.26}$     &	     1.77     &        2.4 & 13.46	          &	     compact     \\
COSMOS FR~I 32     & 149.66830     & 1.8379777     & 2.71$\pm^{0.38}_{0.34}$	     &        1.39     &	     	3.1 & 14.88	          & compact     \\
COSMOS FR~I 34     & 150.56023     & 2.5861051     & 1.55$\pm^{0.41}_{0.19}$	     &	     5.25     &	     	4.5 &     5.87     &	     unresolved \\
COSMOS FR~I 37     & 150.74336     & 2.1705379     & 1.38$\pm^{0.43}_{0.42}$	     &	     1.87     &        2.2 &     2.18	          &	     compact     \\
COSMOS FR~I 38     & 150.53645     & 2.6842549     & 1.30$\pm^{0.17}_{0.28}$	     &        10.01     &	     	11.6 &     9.95	          & compact     \\
COSMOS FR~I 70     & 150.61987     & 2.2894360     & 2.32$\pm^{0.53}_{0.20}$	     &	         3.90     &	     	4.5 & 15.10      &	     compact     \\
COSMOS FR~I 226    & 150.43864     & 1.5934480     & 2.35$\pm^{0.63}_{0.31}$	     &     1.25     &        --- &     8.64         &     compact     \\

\hline

\end{tabular}
\tablecomments{Column description: (1) source ID number; (2) RAJ2000 [degree]; (3) DECJ2000 [degree]; 
(4) Redshifts. Photometric from B13 and spectroscopic from either  MAGELLAN \citep[][]{trump2007} or 
zCOSMOS-bright \citep[][]{lilly2007} catalogs
are denoted with the superscript $^a$ or $^b$, respectively;
(5) 1.4~GHz FIRST fluxes [mJy]; (6) 1.4~GHz NVSS fluxes [mJy].
We assume 2.5~mJy flux (reported as --- in the table) for those sources that are not in the NVSS catalog; (7) 1.4~GHz
radio power [$10^{32}$~erg~s$^{-1}$~Hz$^{-1}$].  NVSS flux or 2.5~mJy upper limit adopted. Radio spectrum
assumed: $L_\nu\propto\nu^{-\alpha}$, $\alpha=0.8$; (8) radio morphology as in C09.}
\end{table*}

\subsection{Statistical properties}\label{sec:stat_prop}
\begin{figure*} \centering
\subfigure{\includegraphics[width=0.4\textwidth,natwidth=610,natheight=642]{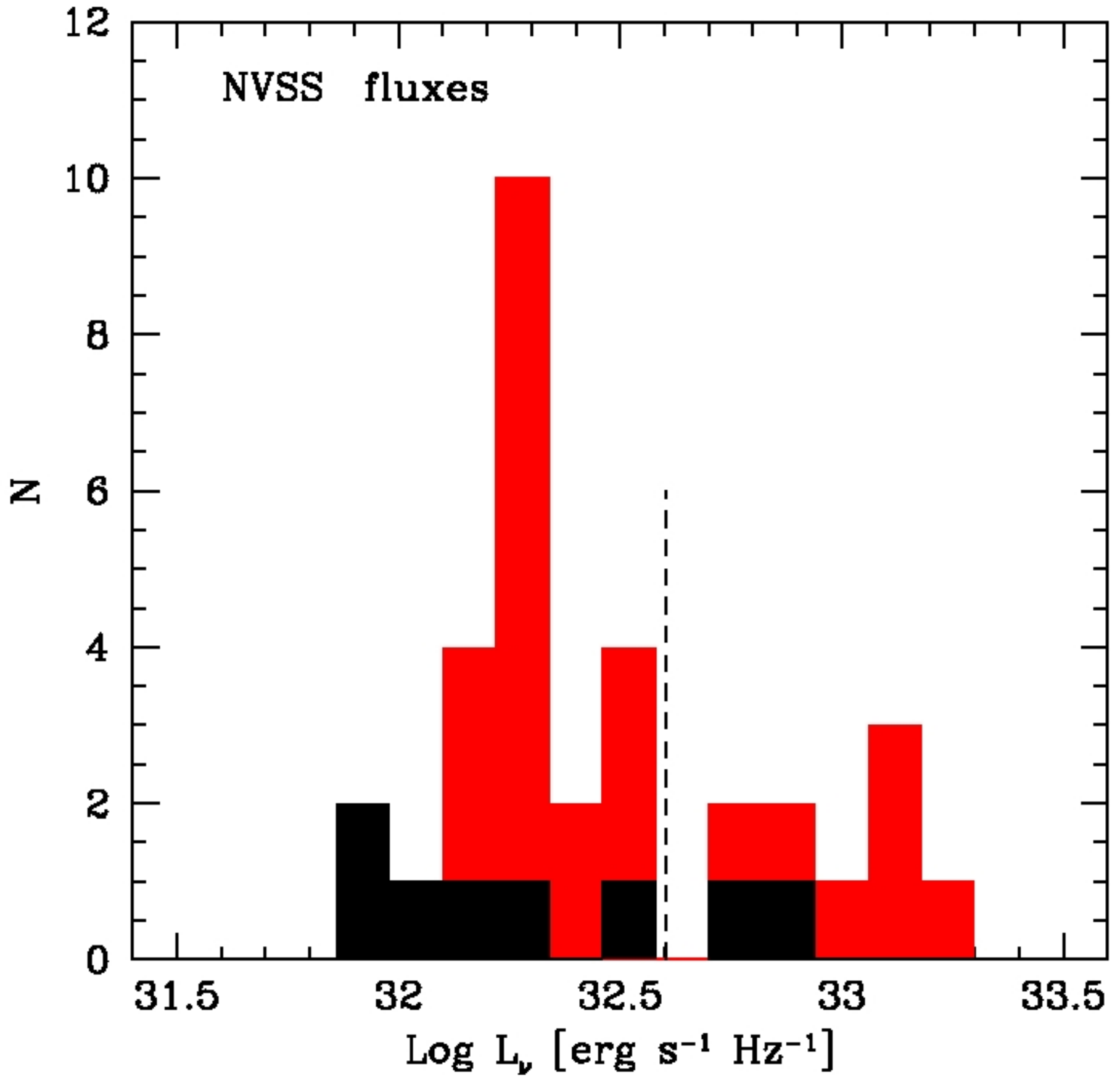}}\qquad
\subfigure{\includegraphics[width=0.4\textwidth,natwidth=610,natheight=642]{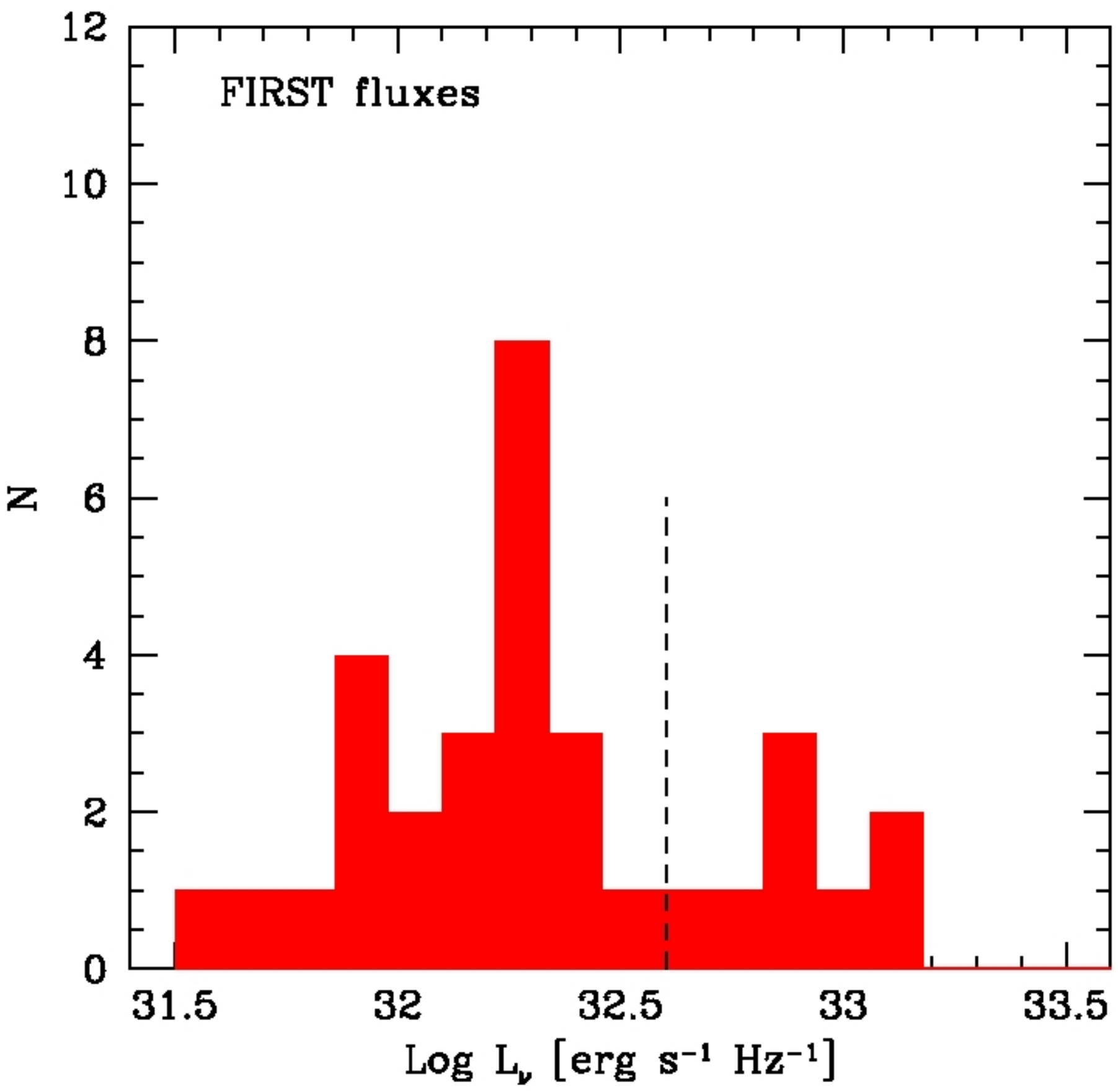}}
\caption{1.4~GHz luminosity histograms for the whole sample. The
vertical dashed line is the FR~I/FR~II radio power divide. Left: NVSS
fluxes adopted. The black regions refer to sources with no NVSS
flux, for which a fiducial 2.5~mJy upper limit is assumed. Right:
FIRST fluxes adopted.}
\label{fig:radio_power_histo}
\end{figure*}

In Table~\ref{table:sampleprop} we summarize the properties of
the sources in our sample, separating them between the LLRGs (top) and
the HLRGs (bottom).
We refer to C09 and their Table~1 for more details about the
sample.
In Figure~ \ref{fig:radio_power_histo} we report
the radio power distribution for our sample obtained by considering NVSS fluxes (left panel) and FIRST fluxes (right panel).
Limited to this section only, we consider also the FIRST instead of the NVSS radio powers only. This is because
FIRST fluxes are available for all the sources in our sample, while this is not the case for NVSS.

The averages of the logarithmic FIRST and NVSS luminosities of the sources in our sample are
$\log[{\rm L}_{\rm 1.4,~FIRST}/({\rm erg~s^{-1}~Hz^{-1}})]=32.32\pm0.41$ and
$\log[{\rm L}_{\rm 1.4,~NVSS}/({\rm erg~s^{-1}~Hz^{-1}})]=32.47\pm0.37$, respectively, where the reported uncertainties are
the rms dispersions around the averages.
This shows that the sources in our sample have, on average, 1.4~GHz radio luminosities slightly below the FR~I/FR~II radio luminosity
divide and that this result is independent of  the two different sets of radio fluxes adopted (i.e. FIRST or NVSS).
However, the logarithmic difference between the FIRST and NVSS luminosities for the sources in our sample
is, on average, $\langle\log({\rm L}_{\rm 1.4,~NVSS}/{\rm L}_{\rm 1.4,~FIRST})\rangle=0.15$ and the rms dispersion around
the average is 0.14~dex.
This can be translated into the fact that, on average,
the 1.4~GHz luminosities estimated from  the NVSS fluxes are
1.5 times than those estimated by adopting
FIRST fluxes.

Therefore, NVSS  are slightly higher than FIRST luminosities for the FR~Is in our sample.
This suggests that the NVSS survey is more sensitive
to the extended emission and it might be more effective than FIRST in order to estimate the
true radio luminosity of our sources.

We test the presence of bimodality in both the FIRST and NVSS
radio power distributions by applying the KMM algorithm described in \citet{ashman1994}.
 The KMM test assumes that the considered distributions are Gaussian functions or a sum of them.
We find that the luminosity distribution is strongly inconsistent with being unimodal at $99.75\%$ confidence level
(i.e. more than 3$\sigma$) if the NVSS fluxes  (or upper limits)  are adopted.
If we adopt the FIRST fluxes for those sources for which we have the NVSS upper limits we find that
the unimodality is rejected at $70.10\%$ confidence level (i.e. just above 1-$\sigma$).
The unimodality is rejected at a level less than 1-$\sigma$ (i.e. $63.88\%$) if the FIRST fluxes are instead considered
for all sources.

The presence of bimodality in the NVSS radio power distribution of the
FR~Is in our sample suggests that the HLRGs might be drawn from a different parent population.
However, the bimodality disappears when the FIRST fluxes are included.
 Futhermore, the Gaussian approximation is a strong assumption and it might not correspond to our case.
Therefore, even if we find evidence of bimodality in the radio power distribution, we cannot draw firm conclusions.



\section{Source space density}\label{sec:integrated_LF}

The careful selection of our sample and the accurate photometric redshifts make possible a reliable estimate of the space 
density of 1.4\,GHz sources at $z\simeq1$, albeit in a narrow luminosity range. 
For this purpose we consider a flux limited sample with NVSS flux density brighter than 2.5 mJy. 
Most (13 out of 19) sources are in the redshift and luminosity ranges $0.9 \le z \le 1.4$ and 
 $ 10^{32.11} \le L_{1.4}/\hbox{erg}\,\hbox{s}^{-1}\hbox{Hz}^{-1} \le 10^{32.51}$. Their median redshift
 and radio luminosity are $z_{\rm median}=1.1$ and $L_{1.4\,\rm median} = 10^{32.30}\,\hbox{erg}\,\hbox{s}^{-1}\,\hbox{Hz}^{-1}$, 
respectively. Only for these there is sufficient statistics to get a meaningful estimate of the space density.

The NVSS catalogue is 50\% complete for unresolved sources with corrected flux density of 2.5 mJy, although its completeness
 rises rapidly to 99\% at 3.4 mJy \citep{condon1998}. To correct for the incompleteness of our sample we have exploited the FIRST survey,
 estimated to be 95\% complete down to 2\,mJy. In our field there are three FIRST sources within the considered luminosity and redshift ranges,
 not present in the NVSS catalog. Only one of them (i.e. source 22) has a FIRST flux density $\ge 2.5\,$mJy. We have added it to sample.

Using the classical $1/V_{\rm max}$ estimator \citep{schmidt1968} we get a comoving density of
 $(6.09^{+1.97}_{-1.77})\,10^{-6}\hbox{Mpc}^{-3}\,(\hbox{d}\log L)^{-1}$. 
The positive error takes into account the possibility that also the other two FIRST 
sources not present in the NVSS catalog are above the 2.5\,mJy limit if observed with the larger NVSS beam. 
Then, the fractional positive error due to incompleteness would be $2/14\simeq 0.14$; we have added it in quadrature to the Poisson error.



A further uncertainty is due to errors on photometric redshifts that may have moved some sources unduly 
in or out of the chosen redshift range. To estimate this uncertainty we have generated $N=1,000$ 
simulated samples randomly assigning to each of the 20 sources in the flux limited sample (including the FIRST source) 
a redshift  randomly drawn from a distribution made of two half-Gaussians with mean equal to the estimated photometric redshift
 and dispersions equal to the positive and negative  1-$\sigma$ redshift errors. 
For each simulated sample we have derived the comoving space 
density with the $1/V_{\rm max}$ estimator, finding $(5.4\pm0.4)\,10^{-6}\hbox{Mpc}^{-3}\,(\hbox{d}\log L)^{-1}$, 
where the errors correspond to the range encompassing 68\% of the distribution. 
 Then, these errors have been added in quadrature to those estimated above.
 This leads  to our final estimate for the comoving space density:
$(5.4^{+2.0}_{-1.8})\,10^{-6}\hbox{Mpc}^{-3}\,(\hbox{d}\log L)^{-1}$.
 

\begin{figure}
\begin{center}
\includegraphics[width=0.45\textwidth,natwidth=610,natheight=642]{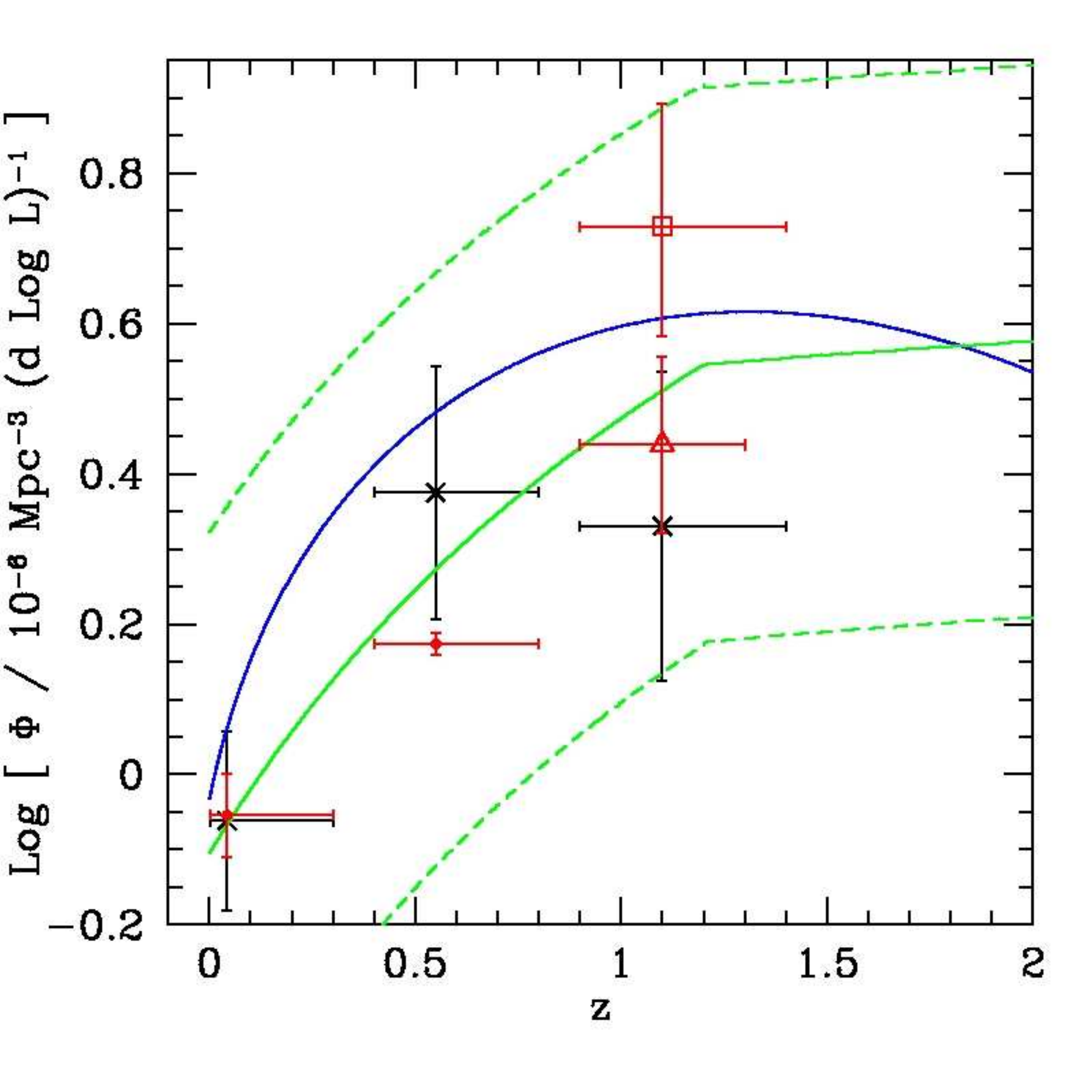}
\end{center}
\caption{Redshift dependence of the comoving space density of 1.4~GHz radio sources 
with $L_{1.4} \simeq 10^{32.3}\,\hbox{erg}\,\hbox{s}^{-1}\,\hbox{Hz}^{-1}$. 
The red points are observational estimates by \citet{mauch_sadler2007} at $z\simeq0.043$,
 \citet{donoso2009} at $z\simeq0.55$, \citet{smolcic2009} at $z\sim1$  (open triangle), and this 
work (open square). The black points are from the \citet{willot2001} model, corrected to the cosmology used in this paper. 
The solid blue line shows the predictions by \citet{massardi2010} for steep-spectrum radio sources.
 The green lines refer to the pure luminosity evolution model by \citet[][model 3 in their Table~3]{mcalpine2013}, 
with its errors. The uncertainties are at 1-$\sigma$ level.}
\label{fig:integrated_LF}
\end{figure}

In Figure~\ref{fig:integrated_LF} we compare our estimate (open square) of the comoving space density 
of 1.4~GHz radio sources with $L_{1.4} \simeq 10^{32.3}\,\hbox{erg}\,\hbox{s}^{-1}\,\hbox{Hz}^{-1}$ and $z\simeq 1.1$ 
with results found in literature for different redshifts. 
Our result is somewhat higher than that by \citet[][see their Table~2]{smolcic2009} at a similar redshift. 
It is also higher than expected from the model by \citet{willot2001}, but consistent with predictions by \citet{massardi2010} 
and \citet[][]{mcalpine2013}.

A comparison with comoving space densities of sources with similar 
luminosities at lower redshifts confirms that they are strongly 
evolving. We find an enhancement of the density by a factor $6.1^{+2.4}_{-2.2}$ 
compared with the \citet{mauch_sadler2007} estimate at $z\sim0$, 
consistent with \citet{rigby2008} who reported an increase by a factor of 5--9 from  
$z\sim 0$ to $z\sim 1$ for FR~I radio galaxies with $L_{1.4} > 10^{32}\,\hbox{erg}\,\hbox{s}^{-1}\,\hbox{Hz}^{-1}$.




\section{The Poisson Probability Method (PPM)}\label{sec:PPM}
Our method to search for overdensities at $z\sim1-2$ has been  introduced and extensively discussed in \cite{PPMmethod}.
The method is based on  galaxy number counts and photometric redshifts.

The Poisson Probability Method (PPM), is adapted from that
proposed by \citet[][see their Appendix A]{gomez1997} to search for X-ray emitting substructures within clusters. The
authors note how their method naturally overcomes the inconvenience of
dealing with low number counts per pixel ($\gtrsim4$), which prevent them from
applying the standard methods based on $\chi^2$-fitting \citep[e.g.][]{davis1993}.
Here we are dealing with a similar problem, since the number counts in the fields of the
radio galaxies are small (see also {Sect.~\ref{sec:results_richness}}).
In fact the COSMOS field survey has, on average,
number densities per unit redshift  ${\rm dn}/{\rm d}z/{\rm d}\Omega \simeq $25, 10, and 3~arcmin$^{-2}$
 at redshift $z\sim$1, 1.5, and 2.0, respectively \citep[see][]{ilbert2009}.
We refer to \citet[][hereinafter Paper~I]{PPMmethod} for a further discussion and a comprehensive description of the PPM.
Here we briefly summarize the basic steps of the procedure:

\begin{itemize}
\item We tessellate the projected space with
a circle centered at the coordinates of the beacon
(in our specific case this is the location of the FR~I radio  galaxy)
and a number of consecutive adjacent annuli.
The regions are concentric and have the same area (2.18~armin$^2$).
 In Figure~\ref{fig:bersaglio} we show the  RGB image of the field of 01. 
The first three regions of the tessellation are shown.

\begin{figure} \centering
\includegraphics[width=0.4\textwidth,natwidth=610,natheight=642]{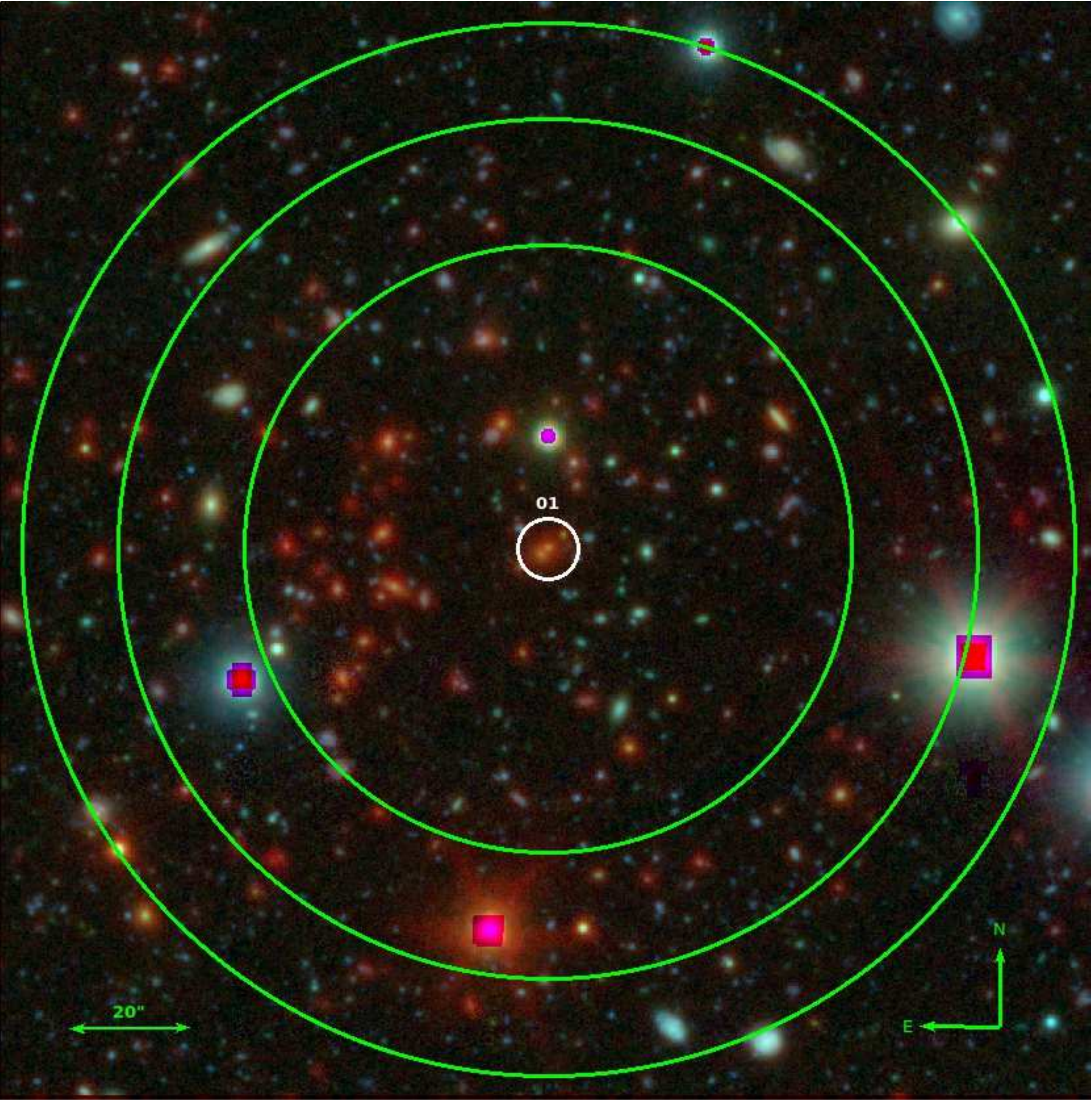}\qquad
\caption{RGB image of the field of 01. The image is obtained using Spitzer 3.6$\mu$m,
 Subaru r$^+$- and Subaru B-band images for the R, G, and B channels, respectively.
Green circles  show the first three regions of the PPM tessellation. 
The white circle is centered at the position of the coordinates of the radio source 01.}
\label{fig:bersaglio}
\end{figure}

\item For each region, we count galaxies with photometric redshifts from the I09 catalog
within a given interval $\Delta z$ centered at the centroid redshift $z_{\rm centroid}$,
for different values of $\Delta z$ and $z_{\rm centroid}$. The values of $\Delta z$ and  $z_{\rm centroid}$
densely span between $0.02-0.4$ and $0.4-4.0$, respectively.

\item  For each area and for a given redshift bin
we calculate the probability of the null hypothesis (i.e. no clustering) to have the observed  or a higher number of galaxies, assuming
Poisson statistics and the average number count density estimated from the COSMOS
field.\footnote{We test
if cosmic variance affects our analysis selecting
four disjoint quadrants in the COSMOS survey to estimate the field density separately from each quadrant.
We verify that the results are independent of the particular choice of the field.
We also note that the beacon is not excluded in estimating the number count density.}
Starting from the coordinates of the beacon we select only
the first consecutive overdense regions
for which  the probability of the null hypothesis
is $\leq30\%$.
We merge the selected regions and we compute the probability, separately, as done for each of them.
Then, we estimate the detection significance of the number count excess as the complementary probability.
We set it equal to zero, if the annulus closest to the radio galaxy has an innermost radius ${\rm r}\gtrsim132$~arcsec,
i.e. we do not consider overdensities that start to be detected at a significant angular separation from the location of the source.
This projected distance corresponds to 0.8~$h^{-1}$~Mpc ($h=0.71$), that is the scale where
the amplitude of the correlation function between Radio Loud AGN (RLAGN) and Luminous Red Galaxies (LRGs)
is reduced to a few percent ($\sim4\%$) of the value at its maximum, up to $z\simeq0.8$ \citep[e.g.,][]{donoso2010,worpel2013}
\begin{figure*} \centering
\subfigure[]{\includegraphics[width=0.45\textwidth,natwidth=610,natheight=642]{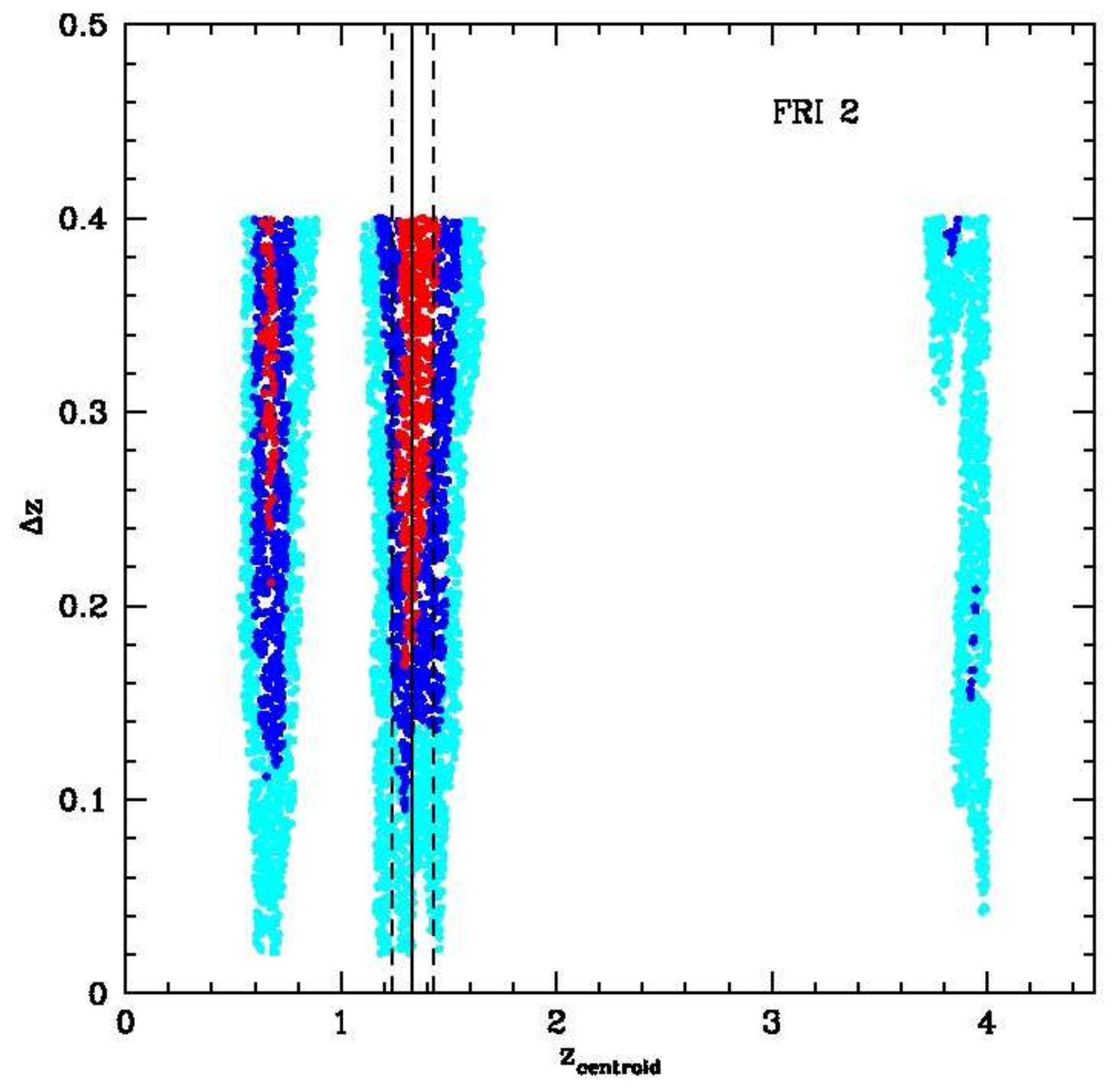}}\qquad
\subfigure[]{\includegraphics[width=0.45\textwidth,natwidth=610,natheight=642]{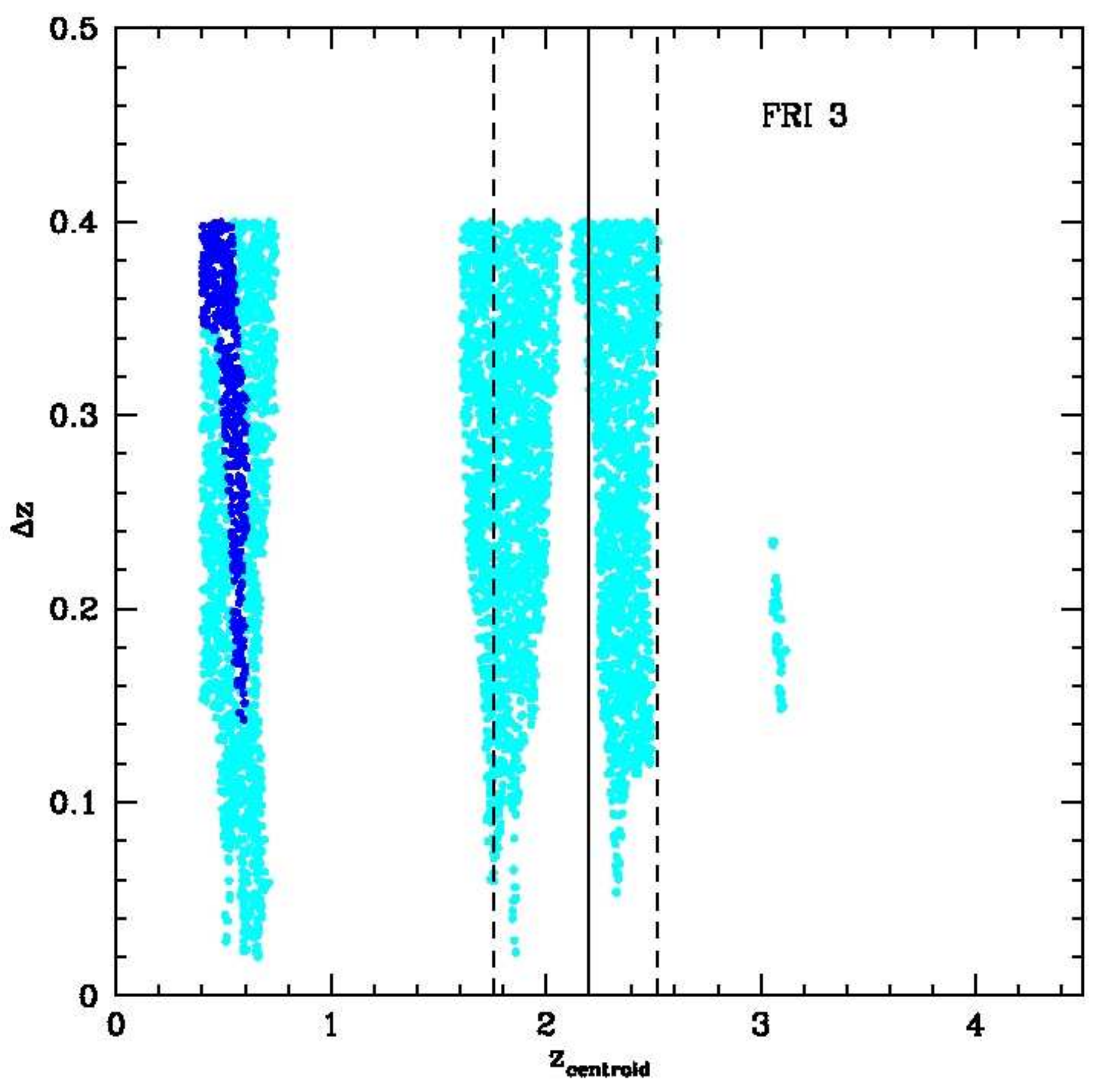}}\qquad
\subfigure[]{\includegraphics[width=0.45\textwidth,natwidth=610,natheight=642]{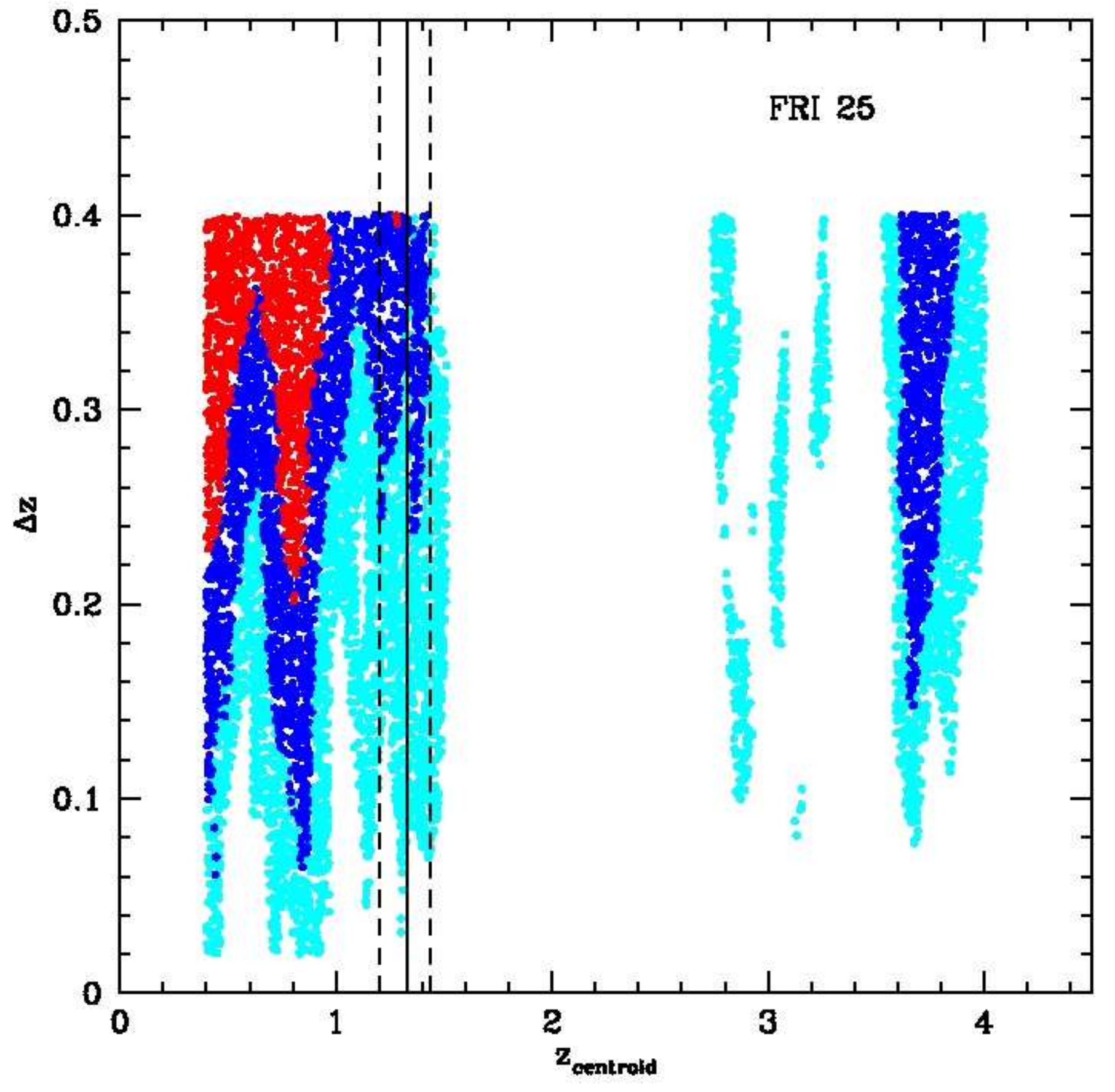}}\qquad
\subfigure[]{\includegraphics[width=0.45\textwidth,natwidth=610,natheight=642]{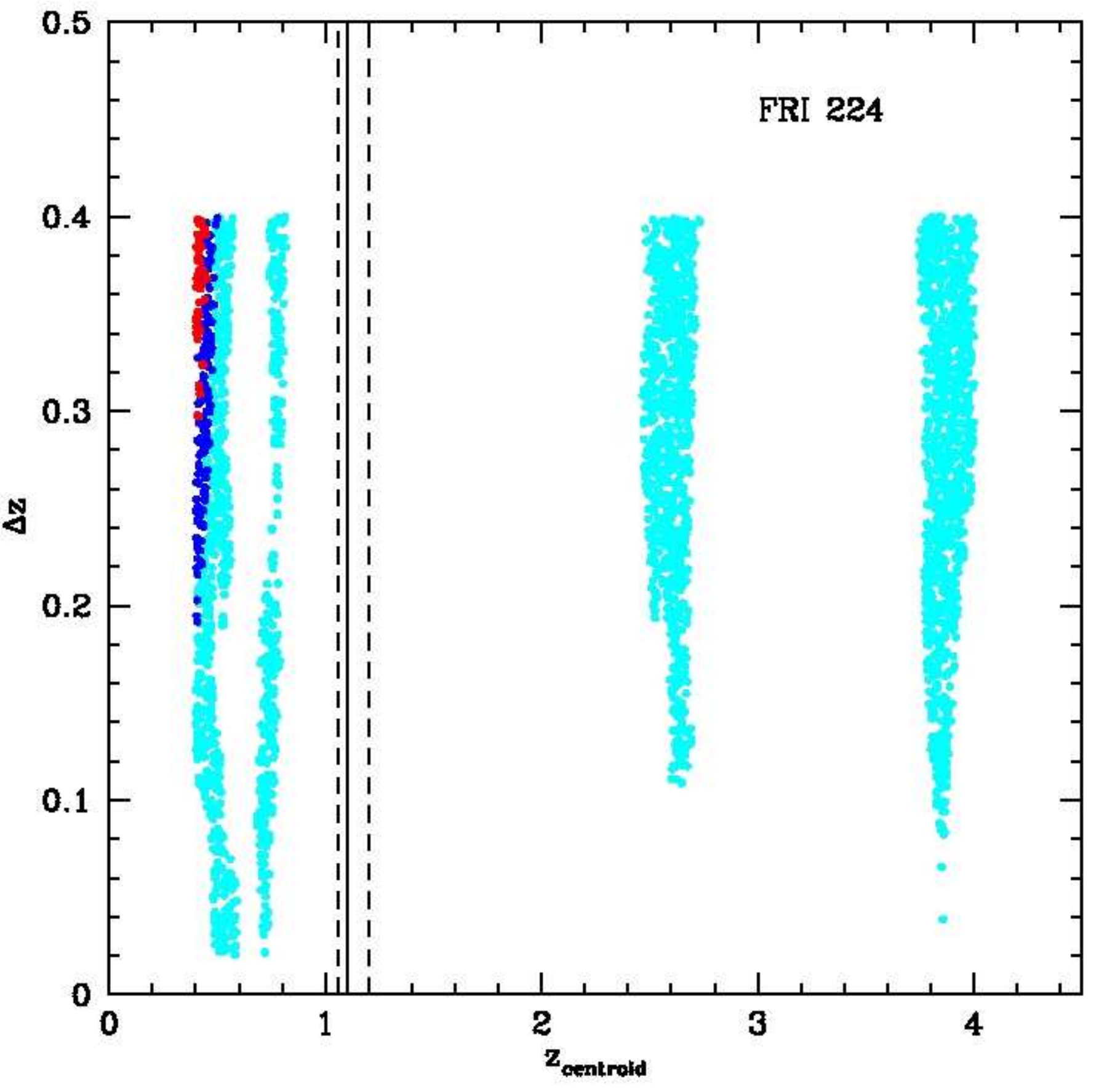}}\qquad

\caption{PPM plots for sources 02 (a), 03 (b), 25 (c), and  224 (d).
The abscissa of the vertical solid line is at the redshift of the source. The vertical dashed
lines show its uncertainties as given in \cite{baldi2013}. We plot only the points corresponding to detected overdensities for
different values of $\Delta z$ and $z_{\rm centroid}$. Color code: $\geq2\sigma$ (cyan points), $\geq3\sigma$ (blue points),
$\geq4\sigma$ (red points). The Gaussian filter which eliminates high frequency noisy patterns has been applied.}
\label{fig:ppm_plots}
\end{figure*}

\item In Figure~\ref{fig:ppm_plots} we show  the resulting plots for some of the sources in our sample.
The points  in each panel represent the probability
estimated for a given choice of the parameters $z_{\rm centroid}$ and $\Delta z$.
We apply a Gaussian filter to eliminate high frequency noisy patterns.
Figure~\ref{fig:ppm_plots} shows the plot where the filter has been applied.

\item We define as overdensities only those regions for which consecutive $\geq2\sigma$ points are present
in a region of the PPM plot at least  $\delta z_{\rm centroid} =0.1$ long on the redshift axis $z_{\rm centroid}$
and defined within a tiny $\delta (\Delta z) =0.01$ wide interval centered at $\Delta z=0.28$.
These values are chosen because of the properties of the errors of the photometric redshifts of our sample and of
the size of the Gaussian filter we apply.
In particular the redshift bin corresponds to the estimated statistical 2-$\sigma$ photometric redshift uncertainty
at $z\sim1.5$ for dim galaxies \citep[i.e. with AB magnitude ${\rm i}^{+}\sim24$,][]{ilbert2009}.
These magnitudes are typical of the galaxies we expect to find in clusters in the redshift range of our interest.
We verified that the results are stable with respect to a sightly different choice of the redshift bin $\Delta z$.
The $2\sigma$ threshold is low, but it is equal to that adopted by previous work that searched for high redshift galaxy clusters
\citep[e.g.][]{durret2011,galametz2012}.

\item In order to estimate the significance of each Mpc-scale overdensity
we apply the same procedure outlined in the previous step, but progressively increasing
the significance threshold until no overdensity is found.
We assign to each overdensity a significance equal to the maximum significance threshold
at which the overdensity is still detected.
Note that in case the overdensity displays multiple
local peaks we do not exclude the lower significance ones.

\item We estimate the redshift of each overdensity as the centroid redshift $z_{\rm centroid}$
at which the overdensity is selected in the PPM plot.

\item We also estimate the size of each overdensity in terms of the minimum and maximum distances from the FR~I beacon at
which the overdensity is detected. In order to do so  we consider all points in the PPM plot within
the region centered around $\Delta z=0.28$ and at least $\delta z_{\rm centroid} =0.1$ long on the redshift axis $z_{\rm centroid}$
which defines the overdensity.
For each of these points the overdensity is detected within certain minimum and maximum distances.
We estimate the minimum and maximum distances of the overdensity as the average (and the median) of the minimum and maximum distances
associated with all of these points, respectively.
We also compute the rms dispersion of the distances as an estimate for the uncertainty.

\item In order to estimate the fiducial uncertainty for the redshift of the overdensity we consider all sources located
within the median minimum and maximum distances from the coordinates of the source within which the overdensity is detected in
the projected space.
We also limit to the sources that have photometric redshifts within a redshift bin $\Delta z=0.28$ centered at the
estimated redshift of the overdensity. This value is chosen to ensure consistency with the value
used for our detection procedure (see above).
We estimate the overdensity redshift uncertainty as
the rms dispersion of the photometric redshifts of
the sources that are selected in the field of the radio galaxy.
In particular, if $N\gg1$ sources were uniformly distributed within the redshift bin $\Delta z=0.28$
we would obtain a rms dispersion of 0.08.
We expect the estimated redshift uncertainty to be around this value.

\item We associate with each radio galaxy any overdensity in its field that is located
at a redshift compatible to that of the radio source itself
(i.e. when the interval centered at the redshift estimated for the overdensity and
with a half-width equal to 2 times the fiducial redshift
error intersects the redshift range defined within the radio galaxy redshift uncertainties).
Note that multiple overdensity associations are not excluded.

\end{itemize}

\section{Results}\label{sec:PPMresults}

In this section we discuss the results of the PPM.
In Figure~\ref{fig:ppm_plots} we show four examples of typical PPM results for fields of the FR~Is. In Panel~(a) 
we report the PPM plot
for the LLRG 02. The photometric redshift of source 02 and that estimated for the overdensity perfectly match. Other two overdensities
are detected in the field of 02 at redshifts $z=0.66~{\rm and}~3.94$, respectively.
They are clearly identified at their estimated redshift by visual inspection of the PPM plot.

Interestingly, the lower redshift cluster is present ($\sim$20~arcsec far from the location of our FR~I)
in both  the $z\lesssim1$ group catalogs of \cite{knobel2009,knobel2012},
who estimated a redshift of $z=0.69$ for the overdensity.

In Panel~(b) we report a similar example for the $z\sim2$ HLRG 03.
Despite the high photometric uncertainties for this source two distinct overdensities are clearly detected within the redshift
 uncertainties of the source 03 at $z=1.82~{\rm and}~2.39$, respectively.
Another overdensity is detected at $z=0.56$, as also clearly identified by visual inspection of the plot.
Interestingly, it is present (with an angular offset of $\sim$20~arcsec from the coordinates of our FR~I)
in the $z\lesssim1$ group catalogs of \cite{knobel2009,knobel2012},
who estimated a redshift of $z=0.66$ for the overdensity.

In Panel (c) we report the PPM plot for the LLRG 25.
A clear overdense (i.e. $\geq2\sigma$) region extends in the PPM plot from $z_{\rm centroid}= 0.40$ to $z_{\rm centroid}=1.51$.
Due to such a large redshift range we interpret the overdense region in the plot as due to a projection effect,
where multiple overdensities are present in the field of 25 at different redshifts.
Our peak finding procedure detects in fact four overdensities within such a redshift interval,
at $z=0.46$, 0.80, 1.23 and 1.37, respectively. Only the last two redshifts agree with the redshift of the radio galaxy,
consistently with our association criterion.
The significances of the two overdensities are similar and equal to 2.7$\sigma$ and 2.8$\sigma$, respectively.
Therefore, we are confident that these two peaks are associated with the same overdensity.
On the contrary, the first two lower redshift overdensities are detected with higher significances of
$3.8\sigma$ and $4.2\sigma$, respectively. Moreover, since they are detected at redshifts significantly below that of the radio galaxy,
we suggest that they are overdensities which are in the field of 25 but they are not associated with the source.
In fact, two overdensities are found in the \cite{knobel2012} group catalog at redshifts of
$z=$0.35 and 0.82 and at angular separations of 8~arcsec and 46~arcsec from the coordinates of the source 25, respectively.
The fact that the redshifts of the $z\sim0.4$ overdensity estimated by \cite{knobel2012} and in this work marginally agree
with each other might be due to the fact that, according to our procedure, we consider sources down to $z_{\rm centroid} = 0.4$.
Therefore, the inconsistency might be due to a boundary effect that would disappear if we considered lower redshift sources.
Note also that we find another clear overdensity in the field of 25 at an estimated redshift of $z=3.72$.
High significance (i.e. $\gtrsim2\sigma$) patterns are also clearly visible in the PPM plot around $z_{\rm centroid}\sim3$.
According to our selection criteria, they are not detected as overdensities but interpreted as noisy features.
These is because they are spiky features that are not stable with respect  to different values for
 the $\Delta z$ and $z_{\rm centroid}$ parameters.

In Panel~(d) we show a clear example where no overdensity is found to be associated with the radio galaxy 224,
altough other three overdensities are detected at redshifts $z=$~0.46, 2.58, and 3.88, well outside the redshift range of our interest.
No group associated with this field is found within the \cite{knobel2009,george2011,knobel2012} catalogs.

In the following sections we will show our results. In Sect.~\ref{sec:cluster_candidates}
we will describe our cluster candidate catalog, in Sect.~\ref{sec:other_cluster_candidates} we will discuss the
presence of other cluster candidates in the fields of our sample of FR~Is that are not associated with our sources.
In Sect~\ref{sec:the_remaining_fields} and Sect.~\ref{sec:results_multiple_associations} 
we will discuss the Mpc-scale environments of the remaining fields and the multiple Mpc-scale
overdensity detections that occur for some of the sources in our sample,  respectively.
In Sect.~\ref{sec:the_clean_catalog} we reconsider our work by rejecting those sources that were masked, classified
as stars, or identified as X-ray AGN in the I09 catalog.
In Sect.~\ref{sec:PPMresults_sizes} and \ref{sec:the_minimum_distance} we will discuss
the projected space information obtained with the PPM, focusing on our cluster size estimates.
 In Sect.~\ref{sec:papovich_test} and Sect.~\ref{sec:results_papovich_test} we will apply the \citet{papovich2008} 
method to our sample and 
compare the results with those obtafined independently by using the PPM, respectively.

\subsection{Cluster candidates}\label{sec:cluster_candidates}

\begin{table*}
\caption{Cluster candidates and their properties as inferred with the PPM.}
\label{tab:PPM_results} \centering
\begin{tabular}{c}
 The Low Luminosity Radio Galaxy subsample
\end{tabular}

\begin{tabular}{cccccccc}
\hline\hline
ID & $z_{\rm source}$ & $z_{\rm overdensity}$ & significance & r$_{\rm min}$ (arcsec) & r$_{\rm max}$ (arcsec) & r$_{\rm max,phys.}$ (kpc) & r$_{\rm max,comov.}$ (kpc) \\
\hline
01     &   0.8823$^a$-0.8827$^b$     &   0.84$\pm$0.07  &   3.5    &   0.0 \hspace{0.3cm}------      &         70.7 \hspace{0.3cm}------       &     536\hspace{0.3cm}-----     &       987\hspace{0.3cm}-----  \\
02     &   1.33$\pm^{0.10}_{0.09}$    &    1.33$\pm$0.09   &  4.3     &   19.2$^{+24.3}_{-19.2}$ (0.0)    &      119.3$\pm$16.2      (122.5)        &    1008$\pm$136                &      2349$\pm$319             \\
16     &   0.9687$^a$             &    1.12$\pm$0.06  &   3.5    &    0.0 \hspace{0.3cm}------     &        100.5$\pm$3.3     (100.0)        &     830$\pm$27                 &      1760$\pm$57              \\
18     &   0.92$\pm^{0.14}_{0.11}$    &    0.80$\pm$0.08   &  5.6      &    0.0 \hspace{0.3cm}------     &       110.4$\pm$25.4     (132.3)        &    834$\pm$191                 &     1501$\pm$345              \\
20     &   0.88$\pm^{0.02}_{0.02}$    &    0.96$\pm$0.06   &  3.9     &   0.0 \hspace{0.3cm}------      &        80.4$\pm$8.3      (86.6)         &    637$\pm$65                  &     1249$\pm$129              \\
22     &   1.30$\pm^{0.05}_{0.04}$    &    1.41$\pm$0.09   &  3.3     &   20.8$^{+24.6}_{-20.8}$ (0.0)    &       94.2$\pm$11.6          (86.6)     &    800$\pm$98                  &    1929$\pm$237               \\
25\tablenotemark{c}     &   1.33$\pm^{0.11}_{0.13}$    &    1.23$\pm$0.07   &  2.8     &   57.2$\pm$9.9   (50.0)         &       120.9$\pm$19.5          (132.3)   &    1009$\pm$162                &     2250$\pm$363              \\
                        &                              &    1.37$\pm$0.08   &  2.7     &   51.1$\pm$4.6   (50.0)         &        86.6 \hspace{0.3cm}------        &    736\hspace{0.3cm}-----      &     1745\hspace{0.3cm}-----   \\
26     &   1.09$\pm^{0.12}_{0.07}$    &    1.15$\pm$0.07  &  3.9     &   42.6$\pm$27.4   (50.0)        &       149.0 $\pm$12.3          (158.1)   &    1237$\pm$102               &     2659$\pm$219              \\
29     &   1.32$\pm^{0.23}_{0.24}$    &    1.34$\pm$0.09  &  2.1     &   77.5$\pm$7.9    (70.7)        &       120.5$\pm$11.2          (122.5)   &    1020$\pm$94                 &     2387$\pm$221              \\
36     &   1.07$\pm^{0.10}_{0.04}$    &    1.18$\pm$0.07  &  3.0     &     0.0 \hspace{0.3cm}------    &        82.6$\pm$6.9           (86.6)    &    685$\pm$57                  &     1494$\pm$124              \\
39     &   1.10$\pm^{0.05}_{0.05}$    &    1.27$\pm$0.06  &  3.5     &     0.0 \hspace{0.3cm}------    &        70.7 \hspace{0.3cm}------        &    597\hspace{0.3cm}-----      &     1356\hspace{0.3cm}-----   \\
228    &   1.31$\pm^{0.05}_{0.07}$    &    1.17$\pm$0.06  &  3.2     &     0.0 \hspace{0.3cm}------    &        70.7 \hspace{0.3cm}------        &    588\hspace{0.3cm}-----      &     1276\hspace{0.3cm}-----   \\
234\tablenotemark{d}      &   1.10$\pm^{0.14}_{0.08}$    &    0.93$\pm$0.08   &  2.5     &     0.0 \hspace{0.3cm}------    &       108.7$\pm$8.3          (111.8)    &    854$\pm$65                  &     1649 $\pm$125             \\
285    &   1.10$\pm^{0.13}_{0.08}$    &    1.01$\pm$0.07   &  2.1     &   50.0 \hspace{0.3cm}------     &        70.7  \hspace{0.3cm}------       &    568\hspace{0.3cm}-----      &     1143\hspace{0.3cm}-----   \\
\hline\hline
\vspace{0.01cm}

\end{tabular}

\begin{tabular}{c}
 The High Luminosity Radio Galaxy subsample
\end{tabular}

\begin{tabular}{cccccccc}
\hline\hline
ID & $z_{\rm source}$ & $z_{\rm overdensity}$ & significance & r$_{\rm min}$ (arcsec)& r$_{\rm max}$ (arcsec)& r$_{\rm max,phys.}$ (kpc) & r$_{\rm max,comov.}$ (kpc)\\
\hline
03\tablenotemark{c}     &  2.20$\pm^{0.32}_{0.44}$       &     1.82$\pm$0.08    &           2.6   &          0.0 \hspace{0.3cm}-----            &        58.7$\pm$11.4          (50.0)      &         502$\pm$97                &       1416$\pm$275           \\
                        &                                &     2.39$\pm$0.09    &           2.5   &         15.8$^{+23.2}_{-15.8}$    (0.0)       &        74.9$\pm$7.0           (70.7)      &         617$\pm$57                &        2093$\pm$195          \\
04                      &  1.37$\pm^{0.10}_{0.06}$       &     1.57$\pm$0.09    &           2.0   &          0.0 \hspace{0.3cm}-----            &        62.4$\pm$10.1           (70.7)     &         532$\pm$86                &       1368$\pm$221             \\
05                      &  2.01$\pm^{0.22}_{0.35}$       &     1.97$\pm$0.07    &           2.2   &          0.0 \hspace{0.3cm}-----            &        50.0\hspace{0.3cm}-----            &         424\hspace{0.3cm}-----    &       1261\hspace{0.3cm}---     \\
28\tablenotemark{c}     &  2.90$\pm^{0.20}_{0.26}$       &     2.71$\pm$0.07    &           2.0   &         86.3$\pm$11.0           (86.6)      &       129.9$\pm$4.2          (132.3)      &         1044$\pm$33               &        3876$\pm$125            \\
                        &                                &     2.98$\pm$0.09    &           2.5   &          0.0 \hspace{0.3cm}-----            &       101.0$\pm$11.7          (100.0)     &         793$\pm$91                &          3159$\pm$366           \\
34                      &  1.55$\pm^{0.41}_{0.19}$       &     1.31$\pm$0.07    &           2.7   &         45.7$\pm$27.9           (50.0)      &       103.6$\pm$8.3          (100.0)      &         871$\pm$69                &          2012$\pm$161           \\
37                      &  1.38$\pm^{0.43}_{0.42}$       &     1.95$\pm$0.07    &           3.0   &         86.6 \hspace{0.3cm}-----            &       121.6$\pm$2.9          (122.5)      &         1035$\pm$24               &         3054$\pm$72               \\
38                      &  1.30$\pm^{0.17}_{0.28}$       &     0.88$\pm$0.07    &           3.7   &          0.0 \hspace{0.3cm}-----            &        90.0$\pm$9.4           (86.6)      &         698$\pm$72                &         1312$\pm$137            \\
226                     & 2.35$\pm^{0.63}_{0.31}$        &     1.99$\pm$0.06    &           2.5   &         70.5$\pm$4.9           (70.7)       &       107.2$\pm$5.8          (111.8)      &         910$\pm$49                &          2723$\pm$147           \\

\hline\hline


\end{tabular}
\\
\tablecomments{Cluster candidates in the fields of the LLRGs (top) and HLRGs (bottom) associated with the corresponding source.
\\
Column description: (1) source ID number; (2)  photometric redshift of the source along with uncertainties  from B13.
  Spectroscopic redshifts from either MAGELLAN \citep[][]{trump2007} or 
zCOSMOS-bright \citep[][]{lilly2007} catalogs
are denoted with the superscript $^a$ or $^b$, respectively;
(3) redshift of the overdensity and corresponding rms dispersion, both estimated with the PPM;
(4) significance of the overdensity estimated by the PPM in terms of $\sigma$;
(5) average minimum radius [arcsec] of the overdensity along with the rms dispersion around the average (both estimated with the PPM). The median value [arcsec] is written between the parenthesis;
(6) average maximum radius [arcsec] of the overdensity along with its rms dispersion around the average (both estimated with the PPM). The median value [arcsec] is written between the parenthesis;
(7) average physical size [kpc] of the overdensity along with the rms dispersion;
(8) average comoving size [kpc] of the overdensity along with the rms dispersion;
The rms dispersions and the median values in columns 5, 6, 7,  and 8 are not reported in those cases where the rms dispersion is null.}
\begin{flushleft}
\tablenotemark{c}{ Sources number 03, 25, 28 are counted twice because multiple peaks are found
to be associated with the corresponding radiogalaxies within the photometric redshift uncertainties.\\}
\tablenotemark{d}{Photometric 
redshifts from \cite{ilbert2009} denoted as {\it zpbest} are adopted. They do not include masked sources, stars,  and X-ray AGN}.
\end{flushleft}
\end{table*}

In Table~\ref{tab:PPM_results} we report the overdensities found in the fields of our sample
that are associated with the corresponding sources, according to the PPM procedure.
We distinguish between the LLRGs (top table) and the HLRGs (bottom table). 
We discuss the estimated sizes in Sect.~\ref{sec:PPMresults_sizes}.
All of the overdensities are robustly detected with respect to slightly different choices of the involved parameters (e.g. a
different choice of the redshift bin $\Delta z$, a different selection threshold, a different choice in the parameters of the
tessellation of the projected space).

According to the overdensity selection procedure outlined in Sect.~\ref{sec:PPM} we find that
22 out of the 32 sources in our sample are hosted in a dense Mpc-scale environment.
The cluster candidates associated with the sources in the sample have an average redshift of $z_{\rm avg}=1.41$
with a rms dispersion around the average of 0.55.
The median redshift is $z_{\rm median}=1.31$.
When calculating these quantities for the fields in which multiple associations between distinct overdensities and the
beacon radio galaxy are identified we only consider the overdensity whose estimated redshift is the closest to that of the radio galaxy.

In particular, we find that 14 radio galaxies out of the 21 LLRGs and 8 out of the 11 HLRGs are associated with overdensities.
This corresponds to a percentage of 67$\%\pm$10$\%$ and 73$\%\pm$13$\%$, for the two subsamples, respectively, where the 1-$\sigma$
uncertainties are estimated according to binomial statistics.
These percentages fully agree within the reported errors.
Therefore the environments of the two subsamples are statistically indistinguishable.
Thus, if we do not distinguish between the two different classes (i.e. the LLRGs and the HLRGs) we find that 22 out of the 32
radio galaxies in our sample (i.e. 69$\%\pm$8$\%$) are found in dense Mpc-scale environments.

The overdensity in the field of 16 is formally not associated with the radio galaxy,
according to the outlined procedure. However, we do not reject it from Table~\ref{tab:PPM_results} because it would be
included if the photometric redshift of the radio source
($z=0.97^{+0.12}_{-0.07}$, see Table~6 in B13)  would be considered instead of the spectroscopic redshift.

Note that, {\it a posteriori}, the redshift estimated for each overdensity in the sample is remarkably
consistent with that of the source estimated in B13.
The overdensity redshift uncertainties are generally small and comparable to typical statistical
photometric redshift uncertainties in I09.

As expected, the overdensities associated with the LLRGs are generally at lower redshifts than those of the HLRGs.
These lower redshift overdensities are also detected, on average, with higher significances ($\sigma_{\rm avg}=3.36$)
than those associated with the HLRGs ($\sigma_{\rm avg}=2.64$).
This effect is in agreement with what pointed out in  Paper~I and it
is mainly due to both increasing photometric redshift errors and to the smaller number counts that occur for increasing redshifts.
If we focus on the overdensities found among the two different subsamples, separately (i.e. the LLRGs and the HLRGs)
we find that the average, the rms dispersion around the average and the median values of the redshifts
of the overdensities associated with the LLRGs are $z_{\rm avg}=1.13$,  ${\rm rms} = 0.20$, and  $z_{\rm median}=1.17$,
respectively.
The average, the rms dispersion around the average and the median values of the redshifts of the overdensities associated with
the HLRGs are $z_{\rm avg}=1.88$ ,  ${\rm rms} = 0.65$, and  $z_{\rm median}=1.97$, respectively.

C10 suggested the
presence of overdensities around three of our highest
redshift sources, namely sources 03, 05, 226. Based on  galaxy number counts, the authors
found that the Mpc-scale environments of these source are 1.7 times denser
with respect to the mean COSMOS density. They translated this into a
4-$\sigma$ overdensity significance. Interestingly, we find this is in full agreement with our results, since we find that
all of  the  three sources reside in high significance ($\sim2.5\sigma$) and high redshift ($z\simeq2$) Mpc-scale overdensities.
The cluster candidate associated with our source 03 is also present in the proto-cluster and group catalog of \cite{diener2013}.
They estimated a redshift of 2.44, that is in good agreement with our estimate ($z=2.39$) for one of the two Mpc-scale
overdensities associated with the source 03. \citet{spitler2012} found a cluster candidate that is about $\sim$3.8-5.4~arcmin
from the source 03. The authors estimated a redshift of $z = 2.2$, on the basis of photometric redshift information. Even if both
the redshift and the projected coordinates are only marginally consistent with those of our cluster candidate,
 it might be possible that the source 03 belongs to the same large scale cluster structure
presented in \citet{spitler2012}.
We also report the PPM plot for the field of this source in Figure~\ref{fig:ppm_plots}, panel (b).
Interestingly, whereas the  independent \citet[][see Sect.~\ref{sec:papovich_test}]{papovich2008}
 method suggests that the source 03 is in a  $\sim3.3\sigma$ overdensity,
it does not detect any overdensity in the fields of sources 05 and 226. We will discuss this in
details in Sect.~\ref{sec:results_papovich_test} and Sect.~\ref{sec:discussion_papovich_test}.

We searched for cluster candidates in  catalogs of  $z\lesssim1$ groups in the COSMOS field
that were obtained by using spectroscopic redshift information  \citep{knobel2009,knobel2012} or photometric redshifts
combined with previous X-ray selected cluster samples \citep{george2011}.
Interestingly, five groups in the fields and redshifts of
our FR~Is are present in these catalogs.
These five source are 01, 16, 18, 20, and 31.
However, we note that the coordinates reported  in \citet{knobel2012}
for the groups and in the fields of 16, 18 and 20 and those of the FR~Is are separated by
$\sim$63, 40, and 42~arcsec, respectively. Therefore, these three associations are only marginally consistent.
Conversely, the offsets for the other two FR~Is (i.e. 01 and 31) are $\lesssim14~\arcsec$; hence the associations are more robust.
The source 258 is the only FR~I in our sample
with a photometric or spectroscopic redshift  less than $z=1$ for which no group was found in these catalogs.
Similarly, the PPM does not find any Mpc-scale overdensity associated with that source.
We also note that the cluster candidate in the field of 01 was previously suggested in \cite{finoguenov2007}.

Redshifts $z=0.88$, 0.92, 0.79, and 0.96 are reported for  the groups associated with  the sources 01, 16, 18, and 20, 
 respectively \citep{finoguenov2007,knobel2009,george2011,knobel2012}.
 The redshifts fully agree with our estimates obtained with the PPM method (see Table~\ref{tab:PPM_results})
for all these overdensities.
A group is also present in the field of our source 31 at an estimated redshift $z=0.91$ in \cite{knobel2009}.
This is exactly the spectroscopic redshift of the FR~I.
Based on spectroscopic redshifts, \cite{knobel2009} associated only two members with this group.
They also estimated a relatively low mass of ${\rm M}=8.9\times10^{12}~{\rm M}_\odot$.
The PPM does not find this group. It might be explained by the fact that the PPM is more effective to find more massive structures,
as discussed in Sect.~\ref{sec:Discussion} and tested in Paper~I.

\subsection{Other cluster candidates}\label{sec:other_cluster_candidates}
\begin{table*}
\caption{Cluster candidates not associated with the radio galaxies as inferred with the PPM.}
\label{tab:PPM_results_abovephotozerr} \centering
\begin{tabular}{c}
 The Low Luminosity Radio Galaxy subsample
\end{tabular}

\begin{tabular}{cccccccc}
\hline\hline
ID & $z_{\rm source}$ & $z_{\rm overdensity}$ & significance & r$_{\rm min}$ (arcsec) & r$_{\rm max}$ (arcsec) & r$_{\rm max,phys.}$ (kpc) & r$_{\rm max,comov.}$ (kpc) \\
\hline
13      & 1.19$\pm^{0.08}_{0.11}$  &   1.42$\pm$0.06  &    3.50     &        0.0 \hspace{0.3cm}------     &     84.4$\pm$8.0   (86.6)         &      719$\pm$68                &         1741$\pm$165 \\
202     & 1.31$\pm^{0.09}_{0.12}$  &   0.91$\pm$0.08  &    2.30     &        7.6$\pm^{17.9}_{7.6}$  (0.0)  &    114.3$\pm$16.6    (122.5)       &     899$\pm$130               &         1718$\pm$249  \\
219     & 1.03$\pm^{0.02}_{0.04}$  &   1.20$\pm$0.06  &    2.60     &        0.0 \hspace{0.3cm}------     &     70.7\hspace{0.3cm}------      &      589\hspace{0.3cm}------   &         1297\hspace{0.3cm}------ \\      	
\hline\hline
\vspace{0.01cm}

\end{tabular}

\begin{tabular}{c}
 The High Luminosity Radio Galaxy subsample
\end{tabular}

\begin{tabular}{cccccccc}
\hline\hline
ID & $z_{\rm source}$ & $z_{\rm overdensity}$ & significance & r$_{\rm min}$ (arcsec)& r$_{\rm max}$ (arcsec)& r$_{\rm max,phys.}$ (kpc) & r$_{\rm max,comov.}$ (kpc)\\
\hline
32      & 2.71$\pm^{0.38}_{0.34}$	&   2.22$\pm$0.07  &    2.20     &        0.0 \hspace{0.3cm}------     &     67.1$\pm$7.8     (70.7)       &      561$\pm$65                &         1808$\pm$210   \\

\hline\hline


\end{tabular}
\tablecomments{Cluster candidates in the fields of the LLRGs (top table) and HLRGs (bottom table) not associated with the radio galaxies.\\
Column description: (1) source ID number; (2)  photometric redshift of the source along with uncertainties  from B13.
(3) redshift of the overdensity and corresponding rms dispersion, both estimated with the PPM;
(4) significance of the overdensity estimated by the PPM in terms of $\sigma$;
(5) average minimum radius [arcsec] of the overdensity along with the rms dispersion around the average (both estimated with the PPM). The median value [arcsec] is written between the parenthesis;
(6) average maximum radius [arcsec] of the overdensity along with its rms dispersion around the average (both estimated with the PPM). The median value [arcsec] is written between the parenthesis;
(7) average physical size [kpc] of the overdensity along with the rms dispersion;
(8) average comoving size [kpc] of the overdensity along with the rms dispersion;
The rms dispersions and the median values in columns 5, 6, 7, 8 are not reported in those cases where the rms dispersion is null.
}
\end{table*}

We now consider
those fields in which no overdensity associated with the radio source is found.
In Table~\ref{tab:PPM_results_abovephotozerr} we report for such fields
the overdensities that would be associated with the radio galaxies if their photometric redshifts, as estimated in B13,
had significantly higher photometric redshift errors.
We adopt the same column description as in Table~\ref{tab:PPM_results}.
We do not consider source number 31, for which a spectroscopic redshift is available. We also report only those overdensities which are
still detected if a smaller redshift bin $\Delta z$
is chosen throughout the PPM procedure.
Interestingly, among these other overdensities, there is a high significance 3.5$\sigma$ overdensity
which is detected in the field of  13 at a redshift $z=1.42\pm0.06$.
\cite{zatloukal2007} also found the presence of a cluster candidate
(i.e. their cluster candidate number 13) in the same field at the redshift $z=1.45$.
We suggest that the two overdensities correspond in fact to the same cluster.

\subsection{The remaining fields}\label{sec:the_remaining_fields}
We discuss in this section the remaining cases for which the difference between the
redshift of the source and the redshift of any overdensity detected in the field is too
 large to make the association plausible. This is the case for the sources 11, 30, 31, 70, 224, and 258.

Source 11 is a HLRG with a photometric redshift $z=1.57^{+0.14}_{-0.09}$. No overdensity is found in its field
within the redshift range $z_{\rm centroid}= 0.4 - 4.0$ considered by the PPM.

Source 30 is a LLRG with a photometric redshift $z=1.06^{+0.11}_{-0.07}$. Three overdensities are found in its fields.
Their estimated redshifts are $z=1.36$, 1.82, and 2.30, respectively.
Their detection significances are 2.0$\sigma$, 2.0$\sigma$, 2.7$\sigma$.

Source 31 is a LLRG at $z_{\rm spec}=0.91$. Four overdensities are detected in its field at redshifts
$z=0.70$, 1.91, 2.27, and 3.62, respectively.
They are detected at a significance level of 3.6$\sigma$, 2.1$\sigma$, 3.1$\sigma$, and 2.7$\sigma$.
Note that none of these overdensities would be associated with the radio galaxy if the photometric redshift
$z=0.88^{+0.03}_{-0.05}$ were adopted from B13, instead of the spectroscopic redshift.
As outlined in Sect.~\ref{sec:cluster_candidates}, a group was found by previous work in the field of 31.
The estimated redshift and mass are $z=0.91$
and ${\rm M}=8.9\times10^{12}{\rm M}_\odot$, respectively \citep{knobel2009}.
As discussed in Sect.~\ref{sec:Discussion} and tested in Paper~I the PPM is more effective to
find richer groups and clusters.  Therefore, it is not surprisingly that our method does not detect this relatively low mass group.

Source 70 is a HLRG with a photometric redshift $z=2.32^{+0.53}_{-0.20}$.
One single overdensity at $z=0.49$ is detected in its field, with a significance of  2.0$\sigma$.

Source 224 is a LLRG with a photometric redshift $z=1.10^{+0.10}_{-0.04}$.
In Figure~\ref{fig:ppm_plots} (panel d), we report the corresponding PPM plot.
Three overdensities are detected in its field at redshifts $z=0.46$,
2.58, and 3.88, respectively.
There high significance patterns are in fact clearly visible in the PPM plot.
Their significance levels are
2.3$\sigma$, 2.5$\sigma$, and 2.6$\sigma$.

Source 258 is a LLRG with at $z_{\rm spec}=0.9009$.
Four overdensities are detected in this field at redshifts $z=2.07$, 2.40, 3.03, and 3.24, respectively.
They are detected with significances of 3.4$\sigma$, 2.4$\sigma$, 2.5$\sigma$,  and 2.3$\sigma$.

\subsection{Multiple associations}\label{sec:results_multiple_associations}
As clear from Table~\ref{tab:PPM_results}, multiple associations are found
in the case of sources 03, 25, and 28, only.
As outlined in Paper~I multiple overdensities might be detected
(i) in presence of projection effects;
(ii) because of incorrect photometric redshift estimates that might be affected by systematics, especially
in the case of the dimmer cluster members (e.g. those with AB magnitude ${\rm i}^{+}\sim24$ in the I09 catalog);
(iii) as a result of multiple local maxima that characterize the patterns of the PPM plot around a given redshift $z_{\rm centroid}$.

We here reconsider in  detail all cases where we find multiple overdensities associated with a single galaxy.
As mentioned above, two overdensities are associated with the source 25
(see also  Figure~\ref{fig:ppm_plots}, panel c).
They have similar significances ($\sim2.5\sigma$) and they are also both detected starting from 50~arcsec from the location of the FR~I.
Such an angular separation corresponds to $\sim400$~kpc at the redshift of the LLRG.
Similar sizes of $\sim$0.7-1.0~Mpc are estimated for the two overdensities (see Table~\ref{tab:PPM_results}).

We visually inspected the field of this source and we did not find any evidence that the non-null offset and the
multiple association are present because of
an artificiality or a technical bias of the I09 catalog occur at the redshift of the radio galaxy (e.g. that some sources at the redshift
of the cluster candidate and in the field of
the FR~I are not included in the I09 catalog  or that their redshifts are erroneously estimated).
Since we do not find any clear discrepancy between the two overdensities and, furthermore, we estimate similar properties for
these two Mpc-scale structures,
we suggest that both the detections are real and they could also correspond to a single cluster candidate associated with source 25.

As mentioned above, two $\sim2.5\sigma$ overdensities are associated with the HLRG 03 (see also Figure~\ref{fig:ppm_plots}, panel b).
They are both detected starting from the coordinates of the radio galaxy (i.e. ${\rm r_{min}}\sim0$~arcsec) and
their estimated sizes are similar (i.e. $\sim$500-600~kpc, see Table~\ref{tab:PPM_results}).
However, they are detected at significantly different redshifts $z=1.82$ and 2.39,
respectively.
Analogously to the case of source 25, we visually inspected the field of 03 and we did not find any evidence that the
multiple association is present because of a technical bias.
Therefore, both the overdensities are equally considered as good, but distinct, cluster candidates, since they are found
at different redshifts.

Two overdensities are associated with the source 28.
They are detected at similar (but different) redshifts $z=2.71$ and 2.98, and with similar significances ($\sim$2.0-2.5$\sigma$).
We also estimate similar sizes for both of them (i.e. $\sim0.8-1.0$~Mpc, see Table~\ref{tab:PPM_results})
However, we find that the overdensity at the lower redshift starts to be detected from 87~arcsec from the radio galaxy.
This corresponds to $\sim700$~kpc at the redshift of the overdensity.
Analogously to the case of sources 03 and 25, we visually inspected the field of 28 and we did not find any evidence that the
non-null offset and the multiple association are present because of a technical bias.
Since we do not find any clear discrepancy between the two overdensities, but nevertheless we estimate different redshifts,
we are not able to conclude if the associations
correspond either to two separate Mpc-scale overdensities at different redshifts or to
a single Mpc-scale structure that is identified as a double pattern in the PPM plot.

\subsection{The clean catalog}\label{sec:the_clean_catalog}

We repeat all the analysis not considering sources that are classified as stars, X-ray AGN, or that
are in masked areas in the I09 list. Hereinafter we refer to this as the clean catalog.
Stars and X-ray AGN are about $\sim4\%$ of the sources in the catalog, while masked sources are about
$\sim13\%-18\%$ (in the redshift range of our interest).
The fields of 36 and 285 were almost completely masked-out most likely because the seeing in the {\it Subaru} optical images
\citep{taniguchi2007} was poor.
We visually inspect the HST image of these fields and we find that all the masked-out objects are in fact galaxies. Therefore,
in these cases we include these masked out objects in our analysis. If the full I09 catalog is adopted we
 find evidence of overdensities in both of these
fields.

Interestingly, we find evidence for a $2.5\sigma$ overdense region associated with the radio galaxy 234 only if the clean
 catalog is adopted,
while no overdensity is found if the complete I09 catalog is adopted.
We visually inspect the HST image of that field and verify that some sources have been masked
southern of the location of source 234 because they
are most likely foreground bright sources.
We also find evidence for a segregation of $z\sim0.93$ sources in the proximity of the radio galaxy 234.
We believe that the discrepancy in adopting the two I09 catalogs is due to the fact that the estimated mean number density of the
COSMOS field is lower if the clean catalog is adopted rather than if the full catalog is considered, while the number of masked
sources in the field of 234 is low enough to detect the overdensity only if the clean catalog is used.
For the sake of completeness, we report the overdensity associated with source 234 in Table~\ref{tab:PPM_results}.
The fields of  36, 234, and 285 are the only cases for which we find a significant difference adopting the two I09 catalogs.

\subsection{Inferred cluster size}\label{sec:PPMresults_sizes}
\begin{figure*} \centering
\subfigure{\includegraphics[width=0.48\textwidth,natwidth=610,natheight=642]{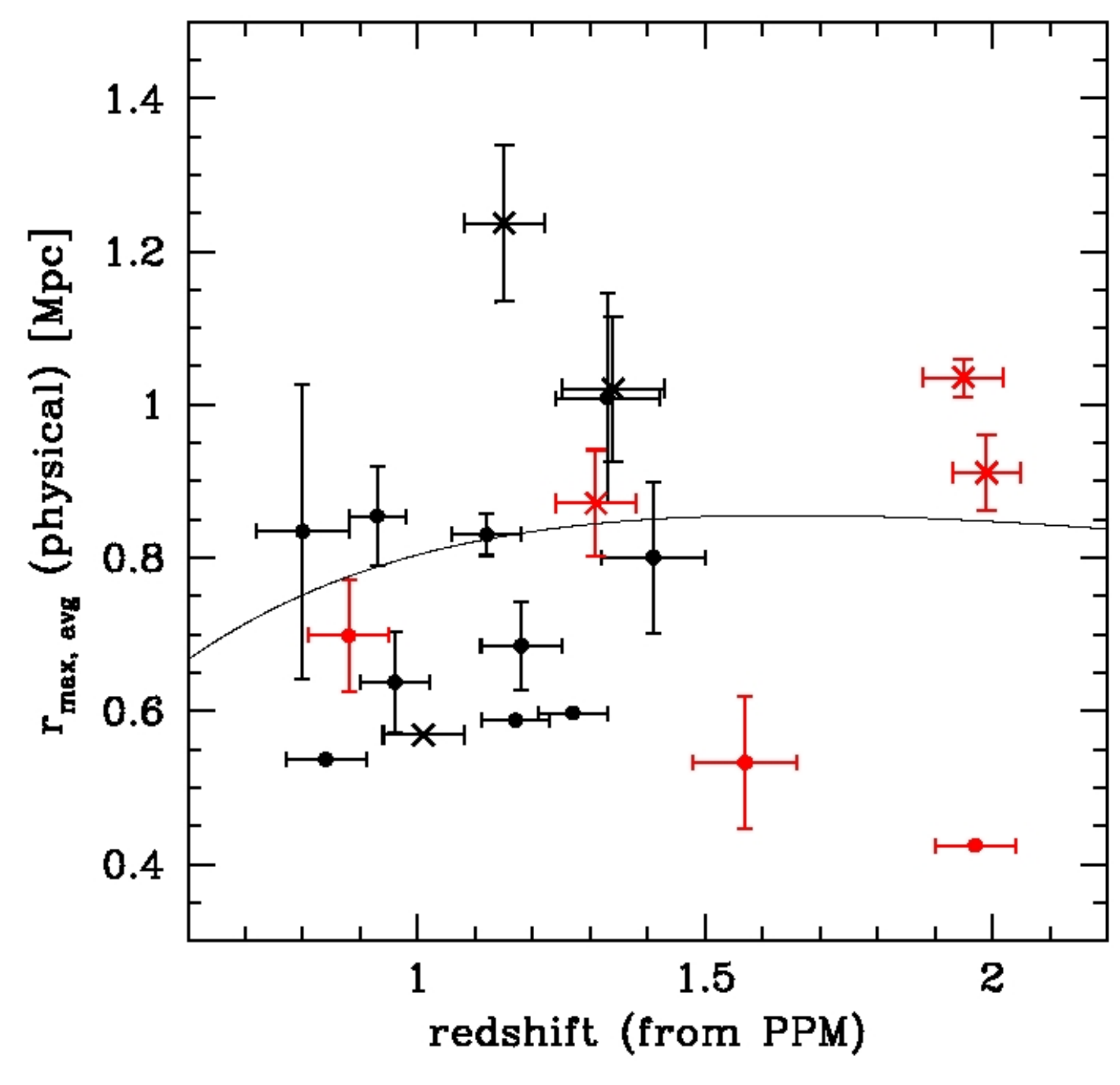}}\qquad
\subfigure{\includegraphics[width=0.48\textwidth,natwidth=610,natheight=642]{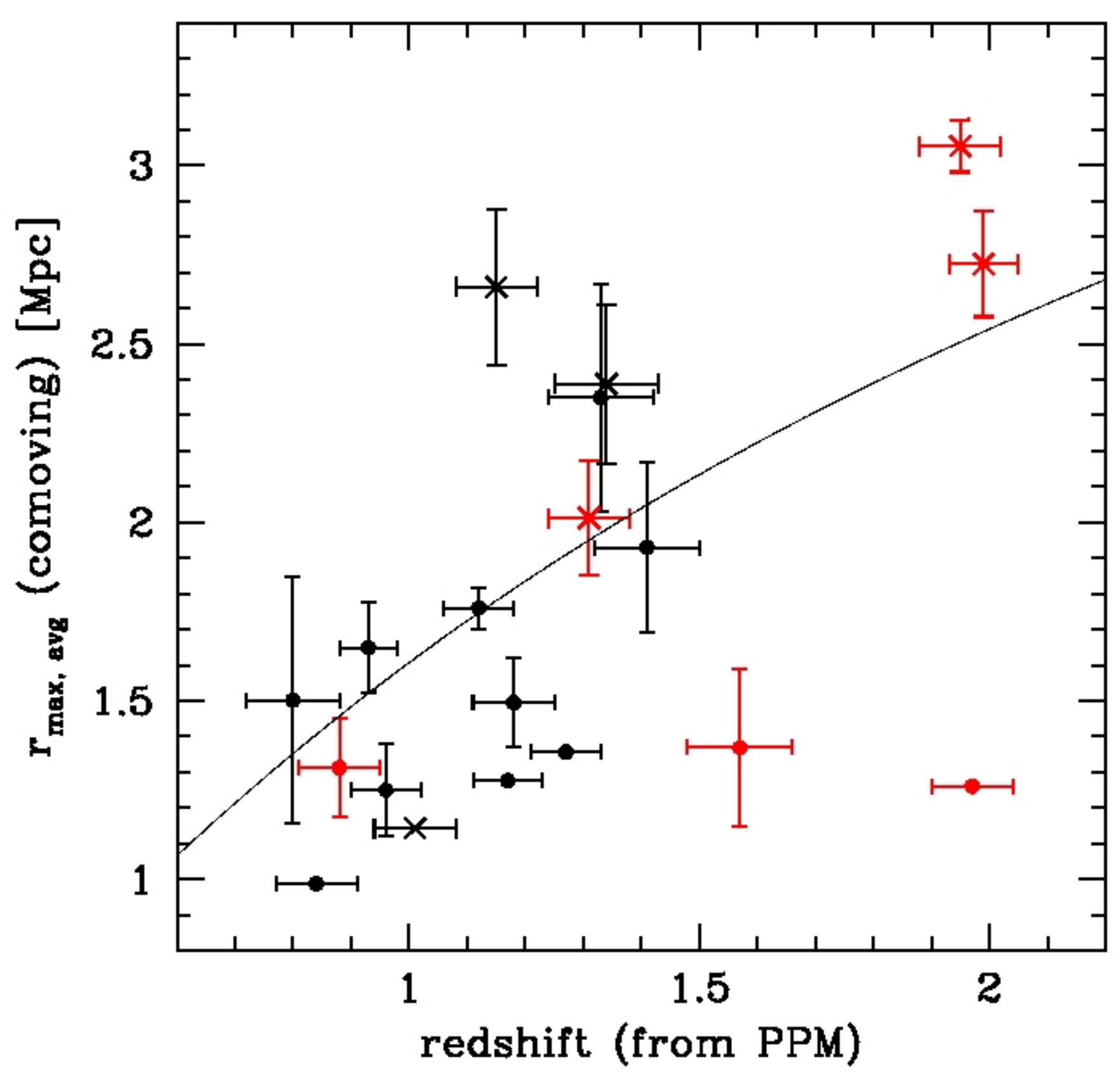}}
\caption{Cluster sizes estimated by the PPM as a function of
their estimated redshifts. The reported uncertainties are the 1-$\sigma$ rms dispersions around the average.
No error is reported in those cases where the rms is null.
Cluster candidates around the HLRGs (red points) and the LLRGs (black points).
Overdensities detected starting from a non null angular separation from the locations of the radio galaxies
are plotted as crosses. The remaining overdensities are plotted as full points.
Sizes are plotted in physical units (left) and in comoving units (right).
Sources with multiple overdensity detections have been conservatively rejected.
The solid black line shows the physical (left panel)
and comoving (right panel) sizes that correspond to 100~arcsec, at each redshift.}
\label{fig:PPMsizes}
\end{figure*}

In this section we limit our discussion to the cluster core sizes estimated by the PPM.
The PPM detects all  of the overdensities within given areas in the projected sky around the location of each radio galaxy.
The procedure is fully described in Paper~I and summarized in Sect.~\ref{sec:PPM}.
The PPM infers the minimum and maximum distances from the coordinates of the radio galaxy at which the overdensity is detected.

The distances are estimated by averaging over all the points of the PPM plot having the significance of the overdensity
and located around the redshift of the overdensity at the fixed bin ($\Delta  z=0.28$).
Such estimates are shown in Table~\ref{tab:PPM_results} for our cluster candidates. Both the average and 
median values are reported.
The median values are less affected by the outliers and are always nevertheless consistent with the corresponding averages within the
rms uncertainties.
These aspects suggest that the overdensities are detected in the projected space with good accuracy and that these detections
are stable with respect  to a different choice of the parameters (i.e. a different centroid of the redshift bin adopted).

 In Figure~\ref{fig:PPMsizes} we plot the comoving (right panel) and physical (left panel) average maximum radii for each overdensity,
along with the corresponding rms dispersions as a function of the estimated redshift of the overdensity along with its formal uncertainty.
We conservatively reject all the sources with multiple overdensity detections.

The cluster candidates around the LLRGs have, on average, comoving (physical) estimated sizes
 of ${\rm r_{\rm avg}}=1672~(784)$~kpc, with a rms dispersion
around the average of 522~(211)~kpc and a median value  ${\rm r_{\rm median}}=1501~(800)$~kpc.
The overdensities around the HLRGs have an estimated average comoving (physical) size
of  ${\rm r_{\rm avg}}=1955~(745)$~kpc, a rms dispersion
around the average of 780~(236)~kpc and a median value  ${\rm r_{\rm median}}=2012~(871)$~kpc.
If we do not distinguish between the two different classes we have an average comoving (physical) value of
 ${\rm r_{\rm avg}}=1762~(772)$~kpc, a rms dispersion
around the average of 607~(213)~kpc and a median value  ${\rm r_{\rm median}}=1501~(800)$~kpc.

Note that these are only rough estimates of the core size of our cluster candidates. However, concerning our project,
we can use them to infer interesting considerations (see also Sect.~\ref{sec:cluster_prop} and \ref{sec:disc_FRIlocation}).
In general, these results suggest that the overdensities in our sample have similar core sizes, independently of the
class considered (i.e. the LLRGs or the LHRGs).

More in general, there seems to be a trend where high redshift sources are also found in overdensities with higher comoving sizes.
We do not find any statistical significance by performing the Spearman test.
Nevertheless, we cannot exclude that less dense overdensities occur at high redshifts.
Diffuse protoclusters with star-forming galaxies have been in fact
found at redshifts higher than $z\sim2.0$ \citep{steidel2000,venemans2007,capak2011,noble2013}.
However, we suspect that this trend is artificial and due to the dependence of the estimated size with redshift or by the 
low number count statistics.
Another possibility is that the cluster size could be overestimated at most by a factor of $\sim2$
if (i) the radio galaxy were not located in the central regions of the cluster core (as tested in Paper~I);
(ii) in the cases when $r_{\rm min}$ is not null (the crosses in Figure~\ref{fig:PPMsizes}), where
${\rm r_{\rm max}}$ might not be a good cluster size estimator (see also discussion in Sect.~\ref{sec:disc_FRIlocation}).

\subsection{The minimum distances}\label{sec:the_minimum_distance}

The cases where the minimum distances are estimated to be small or null
likely correspond to those where the coordinates of the radio galaxy fairly coincide with the center
of the associated overdensity.

However, some of the overdensities are detected starting from a positive angular separation of $\gtrsim$50~arcsec from the
coordinates of the radio galaxy.
Such an offset corresponds to a physical scale of 422~kpc at the median redshift estimated for our cluster candidates (i.e. $z=1.3$).

These cases are controversial and are further discussed in Sect.~\ref{sec:disc_FRIlocation}.
They might be Mpc-scale overdensities where the radio galaxy  is in the outskirts of the overdensity.
This has been investigated in Paper~I through the help of simulations.
We have found that the method is able to detect cluster candidates even if the coordinates of the
cluster  are known with an accuracy of $\sim$100~arcsec and that the inferred minimum radii are
only in some cases greater than zero.
Alternatively, in these cases the radio galaxies might be hosted in underdense regions within their cluster environment.

As outlined above we also visual inspected the fields of some sources (namely 25 and 28) for which the overdensity starts to be detected
from a non null separation from the location of radio galaxy.
Even if we find a depletion in the number of photometric redshifts around the radio galaxy around its assumed redshift,
we are confident that no technical bias occurred, concerning the estimation of photometric redshifts  in the I09 catalog.


\section{The Papovich method}\label{sec:papovich_test}
 In this section we adopt a method \citep[][hereinafter P08]{papovich2008}
based on an IR color selection to search for cluster candidates in the field of the galaxies of our sample.
The P08 method has been widely used in the literature  \citep{mayo2012, galametz2012, wylezalek2013} to 
search for clusters at $z\gtrsim1.3$;
it utilizes the 1.6~$\mu$m bump in the SED of red galaxies, due to a minimum in the opacity of
the H$^{-}$ ion, present in the atmospheres of cool stars \citep[and references therein]{john1988,galametz2012}.
We apply such a method to our sample to see how many objects we can positively detect.
In Sect.~\ref{sec:results_papovich_test} we compare these results with those
obtained by adopting our newly developed  PPM.

The P08 method requires wide field observations at both 3.6 and 4.5~$\mu$m.  We use the  Spitzer-COSMOS (S-COSMOS) archive
catalog\footnote{\url{http://irsa.ipac.caltech.edu/data/SPITZER/S-COSMOS/}}.
S-COSMOS covers the entire COSMOS field.
It is a deep infrared imaging survey carried out with the {\it Spitzer} Telescope.
Mpc-scale overdensities are identified as regions of higher concentration of red sources with respect to the average density,
which is derived as follows, similarly to what done in previous work \citep{mayo2012,galametz2012}.

We choose $\sim$300 randomly selected non overlapping circular fields of 1~arcmin radius each.
The number of the fields is limited and cannot be increased indefinitely because we require the fields to be non overlapping
and to lie within the COSMOS area.

We  conservatively consider the objects
in the S-COSMOS catalog  that are detected at both 3.6 and 4.5~$\mu$m with a signal to noise ratio $S/N>10$.
This criterion is equivalent to that applied by P08 and similar to what done in previous work \citep{galametz2012,wylezalek2013}.
The S/N limit ensures that only well-detected objects enter the sample \citep{papovich2008}.
We also limit our analysis to those sources that are brighter than 1~$\mu$Jy,
which is the confusion limit of the S-COSMOS survey 
 at both 3.6 and 4.5~$\mu$m  \citep{sanders2007}.

 Then, we select all the sources satisfying $([3.6]-[4.5])_{AB}>$-0.1~mag. 
Hereafter we denote as $[3.6]$ and $[4.5]$ the apparent  AB magnitudes
 at the (observer frame) wavelength equal to 3.6 and 4.5~$\mu$m, respectively.

\begin{figure}
\begin{center}
\includegraphics[width=0.4\textwidth,natwidth=610,natheight=642]{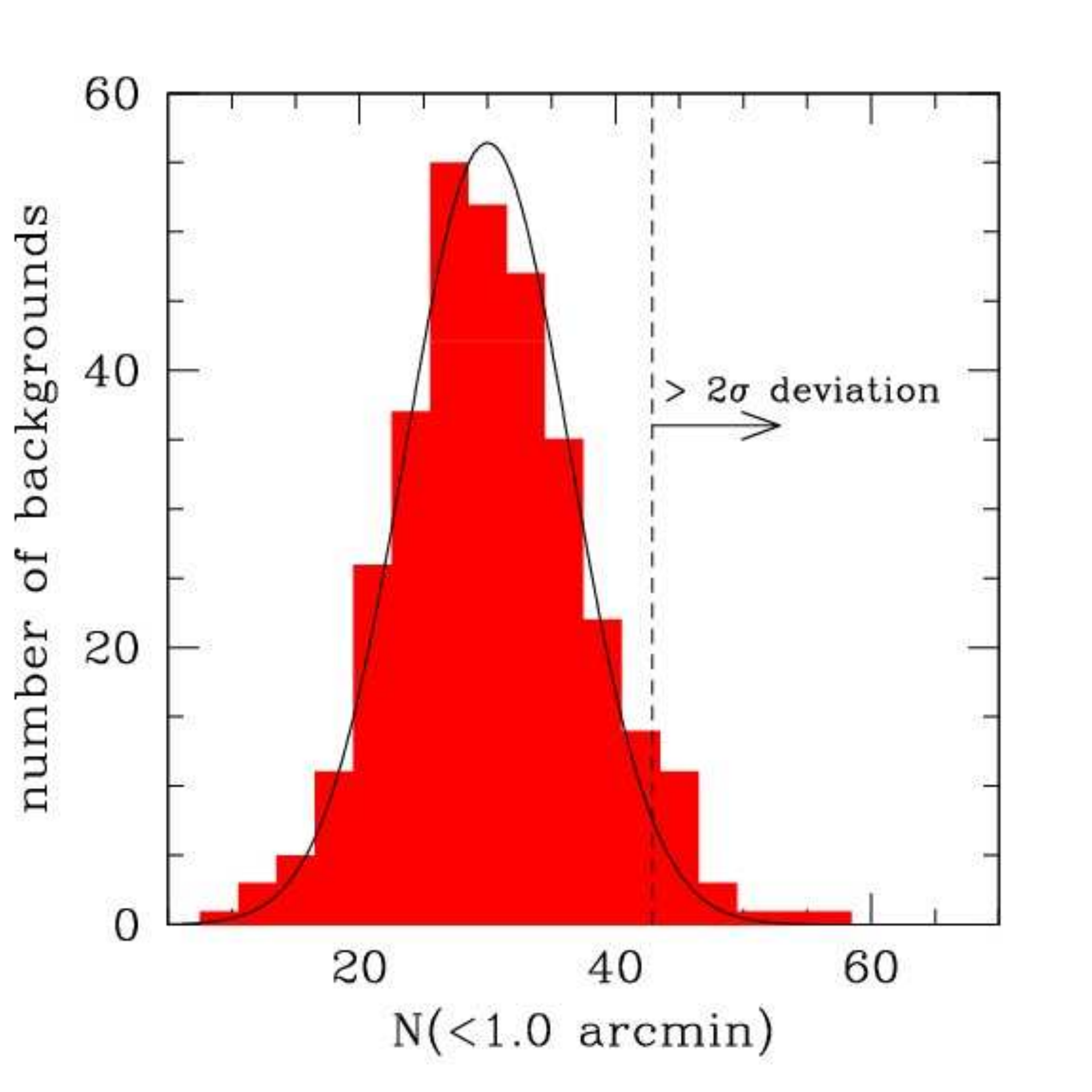}
\end{center}
\caption{Results of the \citet{papovich2008} method. Red histogram: distribution of sources within $\sim$300
randomly selected non-overlapping circular fields of 1 arcmin radius selected from the COSMOS area.
The solid line represents the Gaussian best fit curve obtained iteratively clipping at 2$\sigma$ above the best fit average.
The vertical dashed line is located at the 2$\sigma$ deviation from the best fit average.}
\label{fig:papovich_test}
\end{figure}

In Figure~\ref{fig:papovich_test} we plot
the number count distribution for the $\sim300$ fields
as a function of the number of sources in each field that satisfy the P08 criterion.

Similarly to what done in \citet{mayo2012} and \citet{galametz2012}, we fit such a
distribution with a Gaussian function, iteratively
clipping at 2-$\sigma$ above the best fit average.
This is done in order to exclude from the fit the high number count tail of the distribution.
In fact, it  might be contaminated by those fields that are populated by a significant high number of red objects.
They might be associated with Mpc-scale overdensities and therefore, not
representative of the overall number count distribution in the COSMOS survey.

We estimate the average number of sources
per field which satisfy the P08 criterion.
It is equal to $N = 30.0 \pm 6.4$
where the average and the reported uncertainty are the
mean value and square root of the variance of the
best fit Gaussian function, respectively.

For each 1~arcmin radius field centered around the galaxies in our sample
we count the sources in the S-COSMOS catalog that satisfy the P08 criterion, analogously
to what done for each of the $\sim300$ randomly selected fields.
Then, we estimate the overdensity significance level as the ratio of the number excess with respect the average $N = 30.0$
and the 1-$\sigma$ dispersion ($=6.4$) associated with $N$.

 The P08 method is expected to be effective 
at redshifts $z\gtrsim1.3$ \citep[see e.g.][]{galametz2012,mayo2012}.
As further discussed in \cite{galametz2012}, this is due to the fact that the specific color selection criterion
detects the rest-frame $1.6\mu$m bump in the SED of the galaxies, that is originated by
a minimum in the opacity of the H$^{-}$ ion in the
atmospheres of cool stars \citep{john1988}. Such a feature is redshifted 
out of the Spitzer filters at $3.6\mu$m and $4.5\mu$m,
in the case of lower redshift ($z\lesssim1.3$) sources.


Note that, even if the radio galaxy is at a redshift $z<1.3$, 
the P08 method might detect those overdensities in the field that are not
associated with the radio galaxy, but are at $z\geq1.3$.
As discussed in Sect.~\ref{sec:other_cluster_candidates} and as it is 
clear from visual inspection of the PPM plots in Figure~\ref{fig:ppm_plots}, overdensities not associated with
the radio galaxy are also found by the PPM in the fields of the radio sources, at different redshifts.

 The results of the P08 method are shown in Table~\ref{tab:papovich_test}, where we report the number counts
and the associated significance levels of the overdensities in the fields of the
sources in our sample. In the Table we only report two objects at $z<1.3$, namely 13 and 39. 
This is because these are the only two fields at $z<1.3$ in which 
overdensities are detected by such a method. 
For all other objects that are not reported in the Table the P08 method does not find any overdensity.

Negative significances correspond to underdense fields.
Similarly to what done in \citet{galametz2012} and \citet{mayo2012}, we consider as dense Mpc-scale
environments only the regions with an overdensity detected at
a level $>2\sigma$, i.e. sources with more than  42 counts within 1~arcmin radius.

According to the P08 method,  six sources are found to be in a $\geq2$-$\sigma$ dense Mpc-scale
environment. The source for which the highest significance is observed is object  03 with a photometric redshift of  2.2.
 Note also that the field of 28, that has a photometric redshift $z = 2.9$,
 is detected with a $\sim2.6\sigma$ significance.
While this object is formally beyond the redshift range for which
this sample has been built it is still an interesting case worth mentioning.
This is because   such an overdensity might be a $z\sim3$ ${\rm (proto-)cluster}$ around a $\sim2$ order of magnitude
lower power radio galaxy than those commonly found in clusters or protoclusters at similar redshifts \citep[][]{miley_debreuck2008,galametz2013}.

\begin{table}[htbc]
\caption{\citet{papovich2008} method results.}
\label{tab:papovich_test} \centering
\begin{tabular}{ccc|ccc}
\hline\hline
ID  & n. of sources & $\sigma$ & ID &  n. of sources & $\sigma$ \\
\hline
      02  &    36  &   0.93  & 32  &    33  &   0.47\\
      03  &    51  &    3.26 & 34  &    31  &   0.16\\
      04  &    47  &    2.64 & 37  &    38  &    1.24\\
      05  &    28  &  -0.31  & 38  &    37  &    1.09\\
      11  &   24  &  -0.93   & 39$^{\ast}$  &    47  &    2.64\\
      13$^{\ast}$  &    49  &    2.95 & 70  &    33 &    0.47\\
      22  &    40 &     1.56 & 202  &    34  &   0.62\\
      25  &    30 &    0.00  & 226  &    34  &   0.62\\
      28  &    47  &    2.64 & 228  &    33 &    0.47\\
      29  &   49  &    2.95  &      &       &        \\

\end{tabular}
\tablecomments{Column description: (1) ID number of the
radio galaxy,  radio galaxies 13 and 39 have photometric redshift $z<1.3$ and are marked with an asterisk; 
(2) number of sources within 1~arcmin radius with flux $>$1~$\mu$Jy and S/N$>$10 at both 
3.6 and 4.5~$\mu$m, as well $([3.6]-[4.5])_{AB}>-0.1$~mag; (3) overdensity significance  (in units of $\sigma$).
 Negative values refer to underdense regions.}
\end{table}

 In the following sections we discuss the results obtained by the P08 method and we compare them with those of the PPM.

\subsection{Comparison with the results of the \cite{papovich2008} method}\label{sec:results_papovich_test}
We compare our results with those obtained independently by using the P08 method, as described in Sect.~\ref{sec:papovich_test}.
All the  six cluster candidates found with the P08 method are
also detected by the PPM.  Five of them are associated with radio galaxies in the sample, according to the PPM procedure.
The  sixth overdensity is the cluster candidate found in the field of 13 by both the PPM and the P08 method.
However, according to the method procedure, such an overdensity is not associated with the radio galaxy by the PPM
(see Sect.~\ref{sec:other_cluster_candidates}).
 Note that all of the six overdensities detected by both the P08 method and the PPM are at
redshift $z\gtrsim1.3$ (within the 
corresponding uncertainties), as estimated by
the PPM procedure. This is also true for the overdensities in the fields of 13 and 39. Even if the radio sources are at 
redshift  $z= 1.19\pm^{0.08}_{0.11}$ and $z=1.10\pm^{0.05}_{0.05}$, 
the PPM detects overdensities in their fields at $z=1.42\pm0.06$ and $z=1.27\pm0.06$, respectively.
These results are not surprising since the P08 method is effective to find clusters at $z>1.3$.

 Excluding the overdensity in the field of 13 that is not associated with source 13,  
only  five out of the  12 cluster candidates at  $z\gtrsim1.3$ in our catalog are also found with the P08 method.
 Among the 12 clusters we conservatively do not consider the overdensities in the fields of the 
sources 38 and 228. 
Even if these sources have photometric redshifts $z=1.30\pm^{0.17}_{0.28}$ and $z= 1.31\pm^{0.05}_{0.07}$, respectively, the 
PPM detects clusters in their fields at redshifts below $z =1.3$.

 Two out of the  five clusters, namely 29 and 39, that are associated with the radio galaxies and
detected by both the P08 and the PPM,
are around LLRGs, the other three  (namely source 03,  04, and  28) 
 are around HLRGs.
 As discussed above, source 39 is the only source out of those five that has a photometric redshift below $z=1.3$.

 If we consider our  seven cluster candidates at $z\gtrsim1.3$  in our catalog that are not detected by the P08 method
we find that  three of them are associated with LLRGs (i.e. sources  2, 22, and 25).
The remaining  four out of the  seven are associated with $z\gtrsim1.3$ HLRGs
(i.e. 05, 34,  37, and 226).
Since  the P08 method was primarily designed to search Mpc-scale overdensities at these redshifts,
it is interesting that many of our $z\gtrsim1.3$ cluster candidates are not detected by  such a method.
It is therefore worth reconsidering in more details our cluster candidates found around our $z\gtrsim1.3$ sources.

Three of our cluster candidates are at $z\simeq2$. These are
the overdensities associated with sources 03, 05, and 226.
As mentioned before, the presence of Mpc-scale overdensities around those sources were previously suggested in C10.
Interestingly, the P08 method finds the overdensity in the field of 03 only.

If we focus on the nine $1.3\lesssim z\lesssim2$ sources that the PPM finds to be in dense Mpc-scale environments,
(i.e. sources 02, 04, 22, 25, 29, 34, 37, 38 and 228) we find that only  two out of the nine  are found in dense environments by
the P08 method (i.e. sources  04 and 29).
However, among them, the estimated redshifts of the cluster candidates associated with the sources 37 and 38
are only marginally consistent within the redshift uncertainties of the two sources.
These two cases  could correspond to false positive overdensity PPM detections.
 Furthermore, the P08 method  should not be able to detect the  $z=0.88$ overdensity associated with the source 38, 
since such a redshift is well below the redshift range where the method is effective.
The case of 37 is different, this is because the overdensity associated with this source
has an estimated redshift $z=1.95$. Therefore it falls within the
redshift range allowed by the method.

 Excluding source 38, the results reported above imply that $75\%\pm15\%$ 
of our $1.3\lesssim z\lesssim2$ cluster candidates are not detected by the P08
 method (we have conservatively excluded the above mentioned source 39 that is at redshift formally below $z=1.3$).
Such a percentage  decreases down  to $71\%\pm17\%$ if  also the source  37 is not considered.

We consider apart the high redshift $z\sim3$ source 28 that is detected to be in a dense
environment at $\sim 2.6\sigma$ and $\sim2.5\sigma$ significance levels by the P08 method and by the PPM method, respectively.
Even if such a redshift is formally
beyond the redshift interval ($z\sim1-2$) of our interest, we do not reject the source.

These results suggest that the great majority ($\gtrsim70\%$) of our $z\gtrsim1.3$
cluster candidates are not detected by the P08 method, while all the
seven cluster candidates found with  such a method are also detected by the PPM.
This suggests that our method might be more effective to find cluster candidates, at least limited to
our sample and dataset used.
We will further discuss these results in  the following section.

\vspace{0.1cm}
\subsection{Do we find blue or still forming clusters?}\label{sec:discussion_papovich_test}

 In the previous section we found that the great majority (i.e. $\sim70\%$)
of our $z\gtrsim1.3$ cluster candidates are not detected with the P08 method,  while all of the cluster candidates 
detected by such a method are also found with the PPM.
This is interesting, since such redshifts correspond to the range within which the P08 method is effective \citep{galametz2012}.
 Although we cannot fully understand the details for such a discrepancy
we believe that the method might miss those overdensities that do not fulfill the specific P08 color selection.

This result could also have physical implications.
The P08 method searches for segregations of red  $([3.6]-[4.5])_{AB}$
galaxies. In principle,  it is sensitive to both passively evolving and star-forming galaxies.
However, the method might miss overdensities that are populated by a great amount of bluer galaxies
than those required in order to detect the overdensity.

 {As argued by \citet{muzzin2013}, foreground galaxies at redshift $0.2<z<0.4$ have colors similar to those at redshift 
$z>1.0$ and might add noise, thus affecting the detections.

Furthermore, we also found that the majority of the objects that are used for the PPM and 
are selected within the I09 catalog 
are not included in the S-COSMOS survey and, therefore, they are not used by the P08 method.
Hence, a mismatch between the P08 method and the PPM is not surprising.}




 Note that we applied the P08 method performing a counts-in-cell analysis, i.e. we counted 
objects within a fixed circle centered at a given position in the sky, as done in previous work 
\citep[e.g.][]{galametz2012,mayo2012,wylezalek2013}.
 
On the contrary, the search for cluster candidates performed in this work by adopting the PPM 
is based on number counts and does not rely on peculiar and specific properties (e.g. colors of the sources) and
a specific segregation of the galaxies within the cluster core (see also Sect.~\ref{sec:disc_FRIlocation}).

 Since the P08 method is applied performing a counts-in-cell analysis, some of the clusters that are not detected by 
such a method might be populated by galaxies that are 
not completely segregated in the cluster core.

 Interestingly, C10 suggested the presence of a high fraction of star forming galaxies 
in the $z\sim2$ cluster candidates associated with
sources 03, 05, and 226, on the basis of the visual inspection of the RGB images of their fields.

In a forthcoming paper we will perform the color magnitude diagrams to study the
star formation activity of the galaxies in our clusters and address the problems 
of detecting and studying the red sequence, as well as understanding where 
star forming and quiescent galaxies are located within the cluster.

 The evidence for star formation activity in some of our clusters is not surprising,
 especially at $z\gtrsim1.5$, where
cluster galaxies are expected to have ongoing or increasing star formation \citep{zeimann2012}.
In fact, in some of these high redshift clusters, a significant fraction of the cluster galaxy population
is constituted by
highly dust reddened sources \citep{strazzullo2013} or by blue and irregular galaxies \citep{tozzi2013}.

From a theoretical point of view, previous studies
made predictions for the mass function of galaxy clusters \cite[e.g.][]{bode2001,tinker2008}.
However, since the cluster/group population at redshift $z\gtrsim1.5$ is limited to a few known
 spectroscopically confirmed clusters, observational studies are limited to single high redshift clusters.
This implies that the mass function is only poorly determined by observations.

The spectroscopic confirmation of our $z\gtrsim1.5$ cluster candidates
would increase the number count statistics.
This will help constraining the cluster mass function and will
support previous cluster studies
from both a theoretical and observational point of view.

\section{Discussion}\label{sec:Discussion}
The main goal of this project is to confirm that FR~I radio galaxies at redshifts
$z\sim1-2$ are
preferentially found in rich groups or clusters,
as already proved for local objects, at variance with what found for local powerful FR~II sources
\citep{hill1991,zirbel1997,wing2011}.
For this reason we selected a subsample of  {\it bona fide}
Low Luminosity Radio Galaxies (LLRGs) from the original C09 sample.
This was done to derive a sample of sources with radio powers compatible with those of FR~Is at low redshifts.

We also examine the properties of the subsample of relatively high radio power objects (HLRGs) with respect to the LLRGs.
In the following we discuss the implications of our results for these two groups of objects.

\subsection{Mpc--scale environments of the C09 sample}
As reported in Sect.~\ref{sec:PPMresults} both the LLRGs and HLRGs are found in dense environments. The fraction
of galaxies in groups or clusters is about $\sim70\%$ for both subsamples, consistently within the
1-$\sigma$ uncertainties.
We also found that the detected overdensities have comparable (within a factor of $\sim$2-3) estimated sizes,
independently of both the subsample and the redshift considered (we will discuss this in detail in Sect.~\ref{sec:cluster_prop}).
Therefore, {\it a posteriori}, this result strongly suggests that, on a statistical basis,
the two subsamples constitute a single
population of radio galaxies with similar Mpc--scale environments and similar properties.

\subsection{Comparison with low-redshift radio galaxy environments}
We found  that the majority  (69$\%\pm$8$\%$) of the radio  galaxies in
our  sample  reside in  dense  environments.   Here we  quantitatively
compare  our results  with  the  results obtained  for  samples of  low
redshift FR~Is.

Note that it is difficult to compare the estimated cluster richness of
our candidates  with that  of other samples  of low  redshift clusters
associated  with  radio  galaxies.   This  is mainly  because  of  the
different datasets  used and of  the different techniques  employed in
measuring the cluster richness.

\cite{zirbel1997}      found     that  $70\%$ (with an estimated uncertainty of
$11\%$)\footnote{We estimated  the error on  the percentage by
  adopting   $1\sigma$  uncertainties   according   to  the   binomial
  statistics,  for consistency  with  our results.}   of low  redshift
(i.e. $z<0.25$) FR~Is  in their sample reside in  intermediate or rich
groups  (i.e.   structures with  10  or  more  members). In  terms  of
richness, these  groups could roughly correspond  to the overdensities
detected  by  the  PPM  around  the  radio  galaxies  in  our  sample.

Instead, only $(24\pm8)\%$ of  the low redshift (i.e. $z<0.25$) FR~IIs
in  the \cite{zirbel1997}  sample reside  in  intermediate  or  rich groups.   Such  a
percentage  increases  up  to  $(41\pm8)\%$  if  high  redshift
(i.e.     $0.25\lesssim z\lesssim0.5$)      FR~IIs     are
considered.  The  results obtained by \cite{zirbel1997} are  also  in  agreement  with what independently found
for FR~IIs at $z<0.3$ by \cite{smithheckaman1990} and what found by \cite{ramosalmeida2013}
for a $z\leq0.7$ sample of luminous radio galaxies, mainly comprised of FR~IIs.

Interestingly,  the fraction  we found for the $z \gtrsim 1$
sources in our sample is fully consistent with the percentage (i.e. $70\%$) found
by \cite{zirbel1997} for their sample  of low  redshift
(i.e. $z<0.25$) FR~Is. Note that this holds not only for the LLRGs but also
for the HLRGs.
This implies that  the  environments of FR~Is
and FR~IIs are  different and that they also evolve differently with
redshift.  While the majority of FR~Is seem to be found in rich groups
or  clusters  at  all  redshifts,  the FR~IIs  seem  to  inhabit  rich
environments only at $z>0.25$.  However, as discussed in the following
section, the fraction of FR~IIs that reside in rich groups or clusters
is significantly lower than that of FR~Is even at higher redshifts.

\subsection{Comparison with high-z FR~IIs}
In  this  section  we  compare our  results  with  the
environment properties found for high redshift FR~IIs.
Note  that, thanks  to  the analysis  of the  C09
sample,  this is the  first time  that the  environments of  FR~Is and
FR~IIs can be directly compared at such high redshifts.

High redshift  ($z\sim1-2$) low power radio galaxies  (i.e. FR~Is) are
found in rich  environments more frequently than high  power FR~IIs at
similar  redshifts.   In fact,  if  we  consider  the sample  of  high
redshift     $(z\gtrsim1.3)$     powerful     FR~IIs    studied     by
\cite{galametz2012}, 11  out of 48 objects  (i.e. $23\%\pm7\%$) reside
in Mpc scale environments that  are at least $2\sigma$ denser than the
field.

However,  \cite{wylezalek2013}  extended  this  analysis to  a  larger
sample of 387 radio galaxies  at $1.3<z<3.2$.  They found evidence for
dense environments  for 55$\%$ of these  sources.  Interestingly, this
percentage is consistent  with what found for FR~II  radio galaxies at
redshifts $z\sim0.5$ \citep[$\sim 50$\%, ][]{hill1991}.

Note that the radio powers that characterize the objects in all of the
samples cited above
(L$_{1.4}  \gtrsim 10^{34}$~erg~s$^{-1}$~Hz$^{-1}$)
are about 2
order of  magnitudes higher  than those of all of the radio galaxies  in our
sample, including the HLRGs.  Hence, they undoubtedly represent a different class of radio
galaxies.

The  comparison  between our results and those cited above for powerful high-z FR~IIs confirms that
the environment of high redshift FR~Is and FR~IIs
is different.

This implies that the  Mpc--scale environments of FR~Is and FR~IIs
undergo a different evolution. If we adopt a $\sim50$\% level of FR~IIs
in clusters at  high redshifts as a fiducial  value, we could conclude
that at $z>0.5$ the environments of FR~Is  and FRII~s are similar
(but not identical!).  However, as we already discussed above, this is
clearly  not  true  at  lower redshifts.   Furthermore,  the  values
reported in \cite{galametz2012} and \cite{wylezalek2013} are
not  consistent with each  other  within the  number count  uncertainties.
\cite{wylezalek2013} suggested that  this may be due to the small size of
the \cite{galametz2012}  sample.  It might be interesting  to study in
more detail  the selection criteria of  these two samples  in order to
test whether  the differences are due to  significant discrepancies in
the two sample selections.

Therefore, in light of the results presented here, we confirm that the
connection between the active  nucleus and its large scale environment
could play  a fundamental role in determining  the specific properties
of each radio galaxy.
Clearly, it would be interesting to study X-ray or optically selected samples
of clusters of galaxies at redshifts $z\gtrsim1$ to investigate how the cluster properties
(e.g. richness, halo mass, gas content, and X-ray luminosities) are related to those of the hosted radio galaxies
(e.g. their radio power, their number within the cluster sample, and the mass and size of the host galaxy) and more in
general, to those of the entire cluster galaxy population.
However, these studies require complete and well studied samples of clusters.
Therefore, previous work has been so far limited
to low or intermediate redshifts \citep[e.g.][]{ledlow_owen1996}.

%
%
%
\subsection{Intermediate redshift cluster samples}
We here focus on previous studies on intermediate ($0.3\lesssim z\lesssim1$) redshift cluster samples.
Radio sources with radio power
L$_{1.4}\simeq10^{32-33}$~erg~s$^{-1}$~Hz$^{-1}$ which is typical of those
of the objects in our sample,  are found in $10\%-20\%$ of the X-ray and optically selected clusters
\citep{branchesi2006,gralla2011}.

However, such a percentage rapidly increases up to $\gtrsim90\%$ if lower power radio sources are included
\citep[L$_{1.4}\simeq10^{30}$~erg~s$^{-1}$~Hz$^{-1}$,][]{branchesi2006}.
This is in agreement with previous studies on local Abell clusters \citep{ledlow_owen1995,ledlow_owen1996}.

The fact that such a fraction increases for low power sources
might be explained as a straightforward consequence of the steepness of the radio luminosity function of the radio
galaxies in clusters \citep{branchesi2006}.
This strongly confirms that low power radio galaxies
can be more successfully used to search for  clusters of galaxies than radio galaxies with higher power.

\subsection{Detection efficiency}\label{sec:discussion_sig_redshifts}
The number density per unit redshift (${\rm dn}/{\rm d}z/{\rm d}\Omega$)
in the COSMOS survey is low and it is equal to $\simeq25$, 10, and 3~arcmin$^{-2}$
 at redshift $z\simeq1$, 1.5, and 2.0, respectively \citep[][]{ilbert2009}.
The steep decrease of the number counts for increasing redshifts is a strong constraint for all of the
methods (including the PPM) that search for Mpc-scale overdensities on the basis of number counts \citep{scoville2013}.

In addition, photometric and spectroscopic redshifts cannot
be easily obtained within $z\sim1-2$, where
most of the relevant spectral features fall
outside of the instrumental wavelength bands \citep{steidel2004,banerji2011}.

Therefore, methods that are based on number counts and redshift information and
that are used to  search for clusters and groups in the COSMOS survey
are usually applied up to redshifts $z\lesssim1$ \citep[e.g.][]{knobel2009,george2011,knobel2012}, or
at redshifts higher than $z\simeq2$ \citep[e.g.][]{diener2013}.
Note also that such methods commonly use spectroscopic redshifts so that a small number (i.e. $\lesssim5$)
of cluster galaxies is sufficient to establish the presence of a cluster or group candidate.

The clusters in our sample are detected within the entire redshift range $z\sim1-2$ of our interest.
For each overdensity we estimate detection significance, redshift and size.
The overdensities are detected up to 5.6$\sigma$ significance.
All these results are ultimately due to the flexibility of the PPM to obtain robust results in presence of low number counts.
The overdensities are detected with median significances of 3.3$\sigma$ and 2.5$\sigma$ for the LLRGs and the HLRGs, respectively.
Since the cluster candidates around the LLRGs and the HLRGs have a median redshift $z=1.17$ and $z=1.97$, respectively,
we suggest that the discrepancy between the detection significances of the clusters associated with the two different
subsamples is due to the decreasing number counts in the COSMOS survey for increasing redshifts.
However, such discrepancy is relatively small considering that the number density in the COSMOS field dramatically
drops down by a factor of $\sim$8 from $z=1$ to $z=2$ \citep{ilbert2009}.

In Paper~I we tested the ability of the PPM to detect overdensities at different redshifts,
with richness and size spanned within the ranges found for the cluster candidates in our sample.
Interestingly, we found that our method is able to efficiently detect clusters within our redshift interval,
despite the wide range allowed for the cluster richness and size.

Therefore, we are confident that the detection efficiency (i.e. the number of clusters with homogeneous properties
that are potentially detectable per unit redshift by the PPM) is fairly constant with redshift.
The fact that the detection rate is about 70$\%$ for both our subsamples confirms it, {\it a posteriori}.
Conversely, if the detection efficiency dramatically decreased for increasing redshifts, we would
significantly underestimate the fraction of HLRGs in clusters.

\subsection{The $z\gtrsim1.5$ cluster candidates}\label{sec:z1.5_clusters}
Six overdensities in our sample are found at redshift $z>1.5$.
These correspond to the sources 03, 04, 05, 28, 37, and 226.
All of them are HLRGs.
The fact that we find 6 overdensities at such a high redshift, despite the small area of the COSMOS survey,
further suggests that
these might be  clusters with a low or intermediate mass (i.e. M$\simeq10^{13-14}$M$_\odot$).

Furthermore, the number density of clusters of
higher mass (i.e. ${\rm M}\gtrsim10^{14}$~${\rm M}_\sun$)
is expected to drop down by more than an order of magnitude
between $z=1$ and $z=2$, according to the current $\Lambda$CDM scenario \citep[e.g.][]{bode2001,tinker2008}.
In fact, clusters with masses ${\rm M}\gtrsim10^{14}$~${\rm M}_\sun$, at redshift $z\sim2$, are most likely 
the progenitors of massive ${\rm M}\gtrsim10^{15}{\rm M}_\sun$ clusters at $z=0$ \citep{chiang2013}.
Conversely, assuming hierarchical clustering \citep{cooray_sheth2002}, at $z\sim2$, groups
of lower mass could represent a larger fraction
of the group/cluster population than at lower redshifts.

Furthermore, by definition, groups have a lower richness than clusters, they exhibit
fainter X-ray emission, and they have lower mass content
in terms both of dark matter and gas than clusters of galaxies.
They are therefore more difficult to find with the conventional techniques adopted for clusters.
High redshift groups are in fact usually identified up to $z\lesssim1$ with methods
such as those based on number counts \citep{knobel2012,more2012},
or searching for strong lensing signatures originated from Mpc-scale dark matter halos
\citep[][see also Sect.~\ref{sec:bright_arc_01}]{cabanac2007,limousin2009,more2012}.
Interestingly, if our cluster candidates were confirmed to be rich groups 
 (see Sect.~\ref{sec:masses_sizes_other_cat}), they would constitute a high redshift sample.

\cite{diener2013} obtained a number of 42 candidate groups at $z\gtrsim2$ in the COSMOS field.
They used spectroscopic redshifts, so that a small number (i.e. $\lesssim5$) of members is effective to
establish the detection of a cluster candidate.
Impressively, for the only object in common with our list (i.e. their cluster candidate
22 corresponds to our cluster candidate 03)
the redshift and the size of the cluster estimated by the PPM fully agree with the spectroscopic
measurement and the cluster size estimated in \cite{diener2013}.\footnote{The redshift and the size
estimated by the PPM for one of the two overdensities
associated with the source 03 are  $z=2.39\pm0.09$
and $617\pm57$~kpc, respectively.
\cite{diener2013} found a spectroscopic redshift $z=2.440$ and estimated a size
of 412~kpc for their group candidate 22.}
Note that this cluster candidate was suggested by previous work \citep{chiaberge2010}.
With its five spectroscopically selected cluster members, this is the richest among the groups in the
\cite{diener2013} catalog.

On the basis of the redshift information, the authors also estimated the velocity dispersion
of the cluster members (526~km~s$^{-1}$) which is significantly higher than the average $\sim300$~km~s$^{-1}$ among the
group candidates in their sample.
 This might suggest that the
cluster members are still encompassing a spatial segregation and that the cluster
is still forming, as also discussed for other cluster candidates in our sample (see also Sect.~\ref{sec:discussion_papovich_test}).

\subsection{Cluster properties}\label{sec:cluster_prop}

The general relationship among richness, size of the cluster, and the cluster mass is quite complex
(i.e. it depends on the depth of the photometric 
catalog, the redshifts, the evolution of luminosity function), especially at the redshifts of our interest ($z\sim1-2$),
where the properties of the cluster galaxy population in terms of luminosity and segregation within the cluster
are expected to evolve and are not fully understood. 
In the following sections we discuss size, mass, and richness estimates for
the clusters we find in COSMOS.

\subsubsection{Size and mass estimates for the $z\sim 1$ clusters}\label{sec:masses_sizes_other_cat}
In this section we compare our size estimates with those obtained by previous work
for our $z\sim1$ cluster candidates that are also found in the \citet{finoguenov2007,knobel2009,george2011,knobel2012} catalogs,
namely the clusters in the fields of 01, 16, 18, and 20.
Interestingly, all  of the cluster mass estimates in these catalogs are consistent with each other
and the reported cluster sizes are in good agreement with ours.

In particular, for the cluster candidate associated with our source 01 we roughly estimate a core size of $\sim$71~arcsec (i.e. $\sim500$~kpc).
On the basis of Newton-XMM data, \citet{finoguenov2007} estimated the virial core mass
and the size for the same cluster candidate.
They reported  r$_{500}$~=~48~arcsec and $M_{500}=5.65\times10^{13}M_{\odot}$
\citep[see Table~1 in][for further properties]{finoguenov2007}\footnote{Here r$_{500}$ (r$_{200}$) is the radius
encompassing the matter density 500 (200) times the
critical one and M$_{500}$ (M$_{200}$) is the mass enclosed within such radius.}.

By assuming spherical symmetry and a $\beta$-model density profile for the cluster matter distribution \citep{cavaliere_fusco1978}
we estimate r$_{200}$~=76~arcsec\footnote{In estimating r$_{200}$
we also assume hydrostatic equilibrium. We use Equation~(3)
of \citet{reiprich_bohringer99} and the core radius estimates as in Equation~(4) of  \citet{finoguenov2007}.}.
\citet{george2011} estimated for the same cluster candidate a core size  r$_{200}$~=~73~arcsec,
and a core mass  $M_{200}=5.25\times10^{13}M_{\odot}$, on the basis of  the mass vs. X-ray luminosity relation
given in \citet{leauthaud2010}.
Note that the \citet{george2011} group catalog was obtained by using photometric redshifts and previous
X-ray selected group catalogs.
Both the \citet{knobel2009,knobel2012} group catalogs were instead obtained by using spectroscopic redshifts.
They reported fiducial mass estimates ($M\sim6-9\times10^{13}M_{\odot}$) for
the Mpc-scale overdensity associated with the source 01.
They were obtained by using spectroscopic redshift information.
\citet{knobel2012} also estimated a size of 659~kpc for this cluster candidate.

Concerning the cluster candidates in the fields of 16, 18, and 20,
\citet{knobel2009,knobel2012} reported masses ($M\simeq1.4-2.2\times10^{13}M_{\odot}$)
and sizes \citep[$\sim327-378$~kpc,][]{knobel2012}.
These sizes are roughly consistent even if lower than those
estimated by the PPM for these three groups ($\sim600-800$~kpc).

These results suggest that
the $z\sim1$ cluster candidates associated with sources 01, 16, 18, and 20  are all groups of intermediate/small size,
even if that in the field of 01 is likely more massive than the others.
(see also Sect.~\ref{sec:bright_arc_01} for further discussion).
Interestingly, this result seems to  be independent of the cluster selection (i.e. optical or based on X-ray data).
This is also consistent with previous work by \citet[][see their Table~1]{bahcall2003}, who
found that the clustering lengths for optical selected clusters are comparable with (even if preferentially smaller than)
those obtained for X-ray selected clusters.

We nevertheless note that our cluster sizes are only rough estimates   or upper limits of the cluster core in the optical
 bands  (see also Sect.~\ref{sec:PPMresults_sizes})
and,  therefore, a robust comparison with previous X-ray cluster sizes is beyond the purposes of our work.
 In particular, the core size might be overestimated by at most a factor
of $\sim2$ if the radio galaxy is located in the outskirts
of the cluster. This possibility is further discussed and tested in Paper~I.
Despite this, our estimates are reasonable and typical of rich groups and clusters for all of the clusters candidates in our sample.
Furthermore, the sizes estimated in this work for each of the two subsamples (i.e. the LLRGs and the HLRGs) are consistent with
each other within the uncertainties. On average, comoving and physical sizes for the cluster 
candidates in our sample are about 1.8 and 0.8~Mpc,
respectively. Therefore, all these results allow us to draw general considerations on our cluster candidates,
as shown in the following sections.


\subsubsection{Cluster richness and mass}\label{sec:results_richness}
According to the PPM procedure, we count the galaxies
within a redshift bin $\Delta z=0.28$ centered at the estimated redshift of the cluster and within the projected
area enclosed between the median values of angular separations ${\rm r_{min}}$
 and ${\rm r_{max}}$ from the coordinates of the radio galaxy (see Table~\ref{tab:PPM_results}).
This is not the number of cluster members, but simply the number of sources in the I09 catalog that are found in the field
of each overdensity, around the estimated redshift of the cluster.
Such a number can be considered as a rough estimate of the richness of the cluster, because of
both the instrumental and the PPM limitations.

In detail, the overdensities in the fields of 18 and 26 are those that have the highest number of fiducial cluster members
(i.e. $\sim200$). They are also detected at high significances ($5.6\sigma$ and $3.9\sigma$, respectively).
About $\sim$100 galaxies are instead associated with the overdensities
in the fields of 01, 02, 16, and 20, which are detected at significances of
3.5$\sigma$, 4.3$\sigma$, 3.5$\sigma$, and 3.9$\sigma$, respectively.
About $\sim50$ sources are selected as cluster members of the overdensities associated with the sources 39 and 228, which are detected
at lower significance levels of 3.5$\sigma$ and 3.2$\sigma$, respectively.
At the high redshift end of our sample (i.e. $z\simeq2$) the overdensities are instead defined by only
$\sim10$ galaxies, as it is e.g. for the sources 03 and 05, that are detected at 2.6$\sigma$ and 2.2$\sigma$, respectively.

Therefore, the estimated number of the fiducial cluster members
varies with the cluster detection significance from $\sim10$ for our cluster candidates at the highest
redshifts ($z\sim2$) to more than $\sim200$
for our $z\sim1$ clusters candidates.
 This is most likely because of the overall decrease in the number count density of the COSMOS survey
for increasing redshifts.

High-z faint cluster galaxies (i.e. ${\rm I}\geq25$) are not included in the I09 catalog and therefore we might
miss a significant part of the cluster galaxy population.
However, as discussed in Sect.~\ref{sec:discussion_sig_redshifts}, this does not affect much the detection efficiency of the PPM.

Also note that our method is not highly biased
towards large scale structures with specific characteristics.
Previous work found that there is no clear correlation between cluster richness and mass and the radio power
of the source up
to intermediate redshifts ($z\lesssim0.95$) for radio galaxies with
radio power L$_{1.4}\simeq10^{32}$~erg~s$^{-1}$~Hz$^{-1}$ or even lower \citep{ledlow_owen1995,gralla2011}.
However, \cite{magliocchetti_bruggen2007} found contrasting results based on a small sample of 12 X-ray selected clusters
at low-intermediate redshift ($z<0.3$). In particular, they suggested that low power radio
sources  (down to L$_{1.4}\simeq10^{28}$~erg~s$^{-1}$~Hz$^{-1}$) are
preferentially hosted by low-mass clusters.

However, irrespectively of the number of the fiducial cluster members estimated by the PPM, we expect that, on average,
our group/cluster candidates have a low or intermediate mass (i.e. ${\rm M}\simeq10^{13-14}{\rm M}_\odot$).
  The fact that our size estimates are consistent 
with those found in previous work and are typical of those of rich groups and clusters strengthens such a scenario.
 Furthermore, as pointed out in Paper~I, we stress that the PPM effectively finds systems whose masses 
are typical of rich groups, i.e. are below the typical cluster mass cutoff $\sim1\times10^{14}~{\rm M}_\odot$.
In particular, this is the case of our $z\sim1$ cluster candidates that are found in previous catalogs of groups in the COSMOS
field (see Sect.~\ref{sec:PPMresults_sizes}).
This is clearly due to the small area of the COSMOS survey and the steepness of cluster mass function more than any detection biases
of our method. 
Hence, we will extend our work to wider surveys (e.g. stripe 82 of the SDSS), where
we expect to have a higher chance to find more massive structures.

\subsection{The location of the FR~I within the cluster}\label{sec:disc_FRIlocation}
Previous work investigated the position of BCGs and radio galaxies in clusters.
\cite{ledlow_owen1995} found that about $90\%$ of the radio galaxies  hosted in local ($z<0.09$)
Abell clusters are located
within 200~kpc from the cluster center.
Furthermore, the great majority of such local radio galaxies are FR~Is.
Similarly, \cite{smolcic2011} studied a sample of X-ray selected groups up to $z\simeq1.3$.\
They found that low power radio galaxies (L$_{1.4}\simeq10^{30.6-32.0}$~erg~s$^{-1}$~Hz$^{-1}$)
are preferentially found within  $0.2\times{\rm r}_{200}$ from the group center (i.e. about $\lesssim60$~kpc).

This could also be true at our redshifts.
In fact, for the six cluster candidates that are found by other authors
in the fields and at the redshifts of our sources (namely 01, 03, 16, 18, 20, and 31)
using different techniques \citep[i.e. X-ray emission
and overdensities based on redshift information,][]{finoguenov2007,knobel2009,george2011,knobel2012,diener2013}
we can compare the locations of our FR~I beacons with the coordinates of the cluster centers, as estimated by these authors.
We find that in the cases of 01, 03 and 31 the offset is less than $\sim14$~arcsec. They correspond
to $\lesssim120$~kpc at the redshifts of the overdensities.
In the cases of sources 16, 18 and 20 the association between our FR~I beacons
and the cluster candidates found in other catalogs \citep{knobel2009,knobel2012} is less certain.
This is because the offset is higher than the cases outlined above.
It is about 40~arcsec for sources 18 and 20 (i.e. $\sim$300~kpc at their redshifts)
and it is $\sim1$~arcmin (i.e. $\sim500$~kpc) for source 16.
All these values statistically agree, on average, with the result reported by \cite{ledlow_owen1995}.

This is also consistent with the offset of $\sim$100~kpc, typically found
between the optical and the X-ray cluster centroids  \citep{dai2007}.
Furthermore (as pointed out in Sect.~\ref{sec:intro}),
at variance with FR~II radio galaxies or other types of AGNs,
low-redshift FR~Is are typically hosted by undisturbed ellipticals or cD galaxies \citep{zirbel96},
which  are often associated with the BCGs \citep{vonderlinden2007}.
To the best of our knowledge, the bright BCG discovered by \citet{liu2013}
at $z=1.1$ is the most distant cD galaxy confirmed to date.
Therefore, in light of the results presented here,
the hosts of our FR~Is could also constitute a sample of high-z
cD galaxy candidates.

Concerning the BCGs, previous work found that they preferentially reside within $\lesssim41$~kpc from the X-ray cluster center
up to $z\simeq1$ \citep{semler2012}.
However, \citet{zitrin2012} found that the offset, if estimated from the optical cluster centroid,
increases for increasing redshifts (i.e. up to $\sim14$~kpc at $0.52<z<0.55$).
A similar trend is not excluded for our cluster candidates.
In fact, we find that
six of our cluster candidates are detected within an annulus centered at the coordinates of the radio galaxy
and an internal radius of $\gtrsim50$~arcsec
(see also Table~\ref{tab:PPM_results} and related discussion in Sect.~\ref{sec:results_multiple_associations}).
Note that 50~arcsec correspond to  427~kpc at redshift $z=1.5$.
These six overdensities correspond to $32\%\pm11\%$ of our 19 cluster
candidates.\footnote{Note that for this case we consider 19 clusters because for the purpose of estimating
sizes of clusters and locations of the FR~I beacons we exclude multiple overdensities
within the same field (see Sect.~\ref{sec:PPMresults_sizes}).}

The six sources are the LLRGs 26, 29, and 285  and the HLRGs 34, 37, and 226.
Although the statistics is extremely poor, this result implies that half of the sample of the HLRGs show significant offsets (i.e. $\geq50$~arcsec),
while a non-null offset occurs for only $\sim20\%$ of the LLRGs.
However, based on such a small sample we do not draw firm conclusions.

In order to investigate the marginal discrepancy found between the two subsamples,
it would be interesting (i) to look for FR~I radio galaxies
in COSMOS at redshifts similar to those of the HLRGs, but with radio powers comparable with those of the LLRGs, and
(ii) to search for radio galaxies with redshifts similar to those of LLRGs and radio powers comparable with those of the HLRGs.
This will improve the sample statistics
and will allow us to understand if the trend is due either to evolutionary properties (being the LLRGs, on average,
at lower redshifts than the HLRGs) or to the difference in radio power between the LLRGs and the HLRGs.

A possibility is that such radio galaxies are hosted in underdense regions within their
cluster environment.
To further investigate the above scenario we visually inspected the fields of the six sources.
We did not find any evidence that the non-null offsets are present because of an artificiality or a
technical bias of the I09 catalog (e.g. that some sources at the redshift of the cluster candidate and in the field of
the corresponding FR~I are not included in the I09 catalog  or that their redshifts are erroneously estimated).
We also found that the galaxies in each of these fields at redshifts around that of the corresponding FR~I are
homogeneously distributed around the position of the radio galaxy.
This means that, altough these overdensities are detected with significant offsets from the location of the corresponding FR~I,
each radio source is still likely located around the barycentric center of the galaxies in the field, in the projected sky,
and not in the outskirts of the cluster candidate.

Furthermore, our results could also imply that our cluster candidates are still encompassing
a strong evolution in terms of the spatial segregation of the galaxies within the core
\citep[see e.g.][for a very detailed study about a $z\sim1.6$ forming cluster]{bassett2013}.

\subsection{A bright arc in the field of 01}\label{sec:bright_arc_01}

\begin{figure}[htbc]
\begin{center}
\includegraphics[width=0.5\textwidth,natwidth=610,natheight=642,natwidth=610,natheight=642]{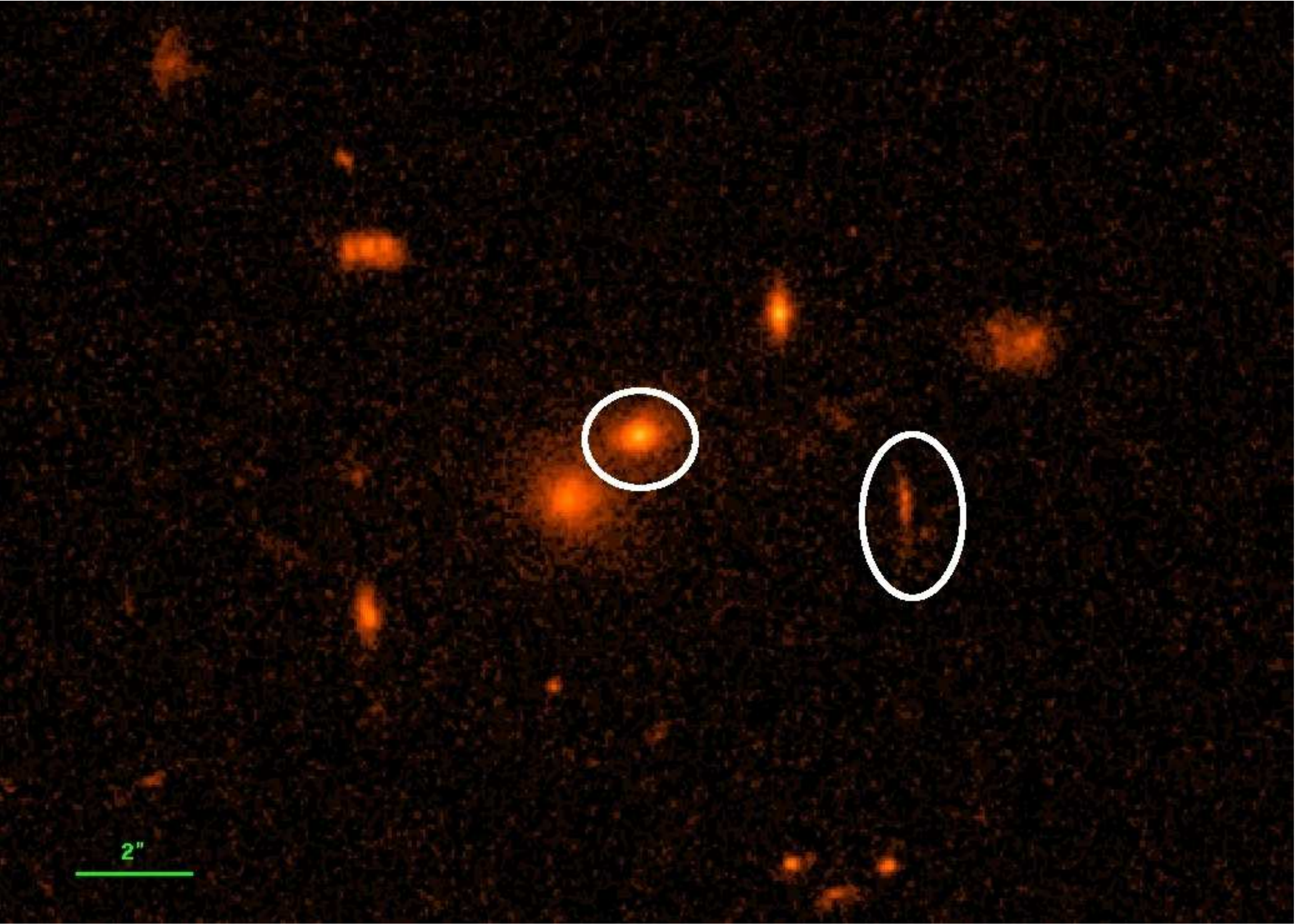}
\end{center}
\caption{Field ($22''\times16''$ dimensions) of source 01 as observed by ACS on board of HST \citep{koekemoer07}.
The galaxy host of the source 01 and the bright arc are marked with the left and right ellipses, respectively.}
\label{fig:acs_01_zoom}
\end{figure}

In this section we discuss the serendipitous discovery of a bright arc detected with the ACS camera on board of HST
in the in the field of the source 01, at $z_{\rm spec}=0.88$.
In Figure~\ref{fig:acs_01_zoom} we report the ACS image \citep[][]{koekemoer07} of the field of 01.
The source 01 and the arc are marked in Figure with the left and the right ellipses, respectively.

The arc is clearly visible about $\sim5$~arcsec westward of the pair formed
by the radio galaxy host and a larger elliptical companion.
Such a projected angular separation corresponds to $\sim$39~kpc at the redshift of the source.
The arc is very close to the radio galaxy, and it resides within the core of the Mpc--scale overdensity
associated with the source 01.

Strong lensing phenomena are expected to be originated close to the densest
regions of dark matter halos.
Since such a projected separation is consistent with
the typical size \citep[i.e. $\sim60$~kpc,][]{halkola2007} of the dark matter halos of BCGs, it is likely
that the arc is originated by the dark matter halo of the galaxy pair.

An alternative scenario is motivated by the fact that the
overdensity associated with the source 01 is a relatively compact rich group with
an estimated core size of about 70~arcsec \citep[as suggested by][and in this work]{finoguenov2007,george2011,knobel2012}.
Therefore, it is also possible that the group halo itself is responsible for the observed effect.
In fact, groups with intermediate masses in the range $10^{12}-10^{14}$~M$_{\odot}$
are usually more massive than galactic halos
and concentrated enough to act as lenses \citep{more2012}.

\begin{figure*}[htbc]
\begin{center}
\includegraphics[width=0.8\textwidth,natwidth=610,natheight=642,natwidth=610,natheight=642]{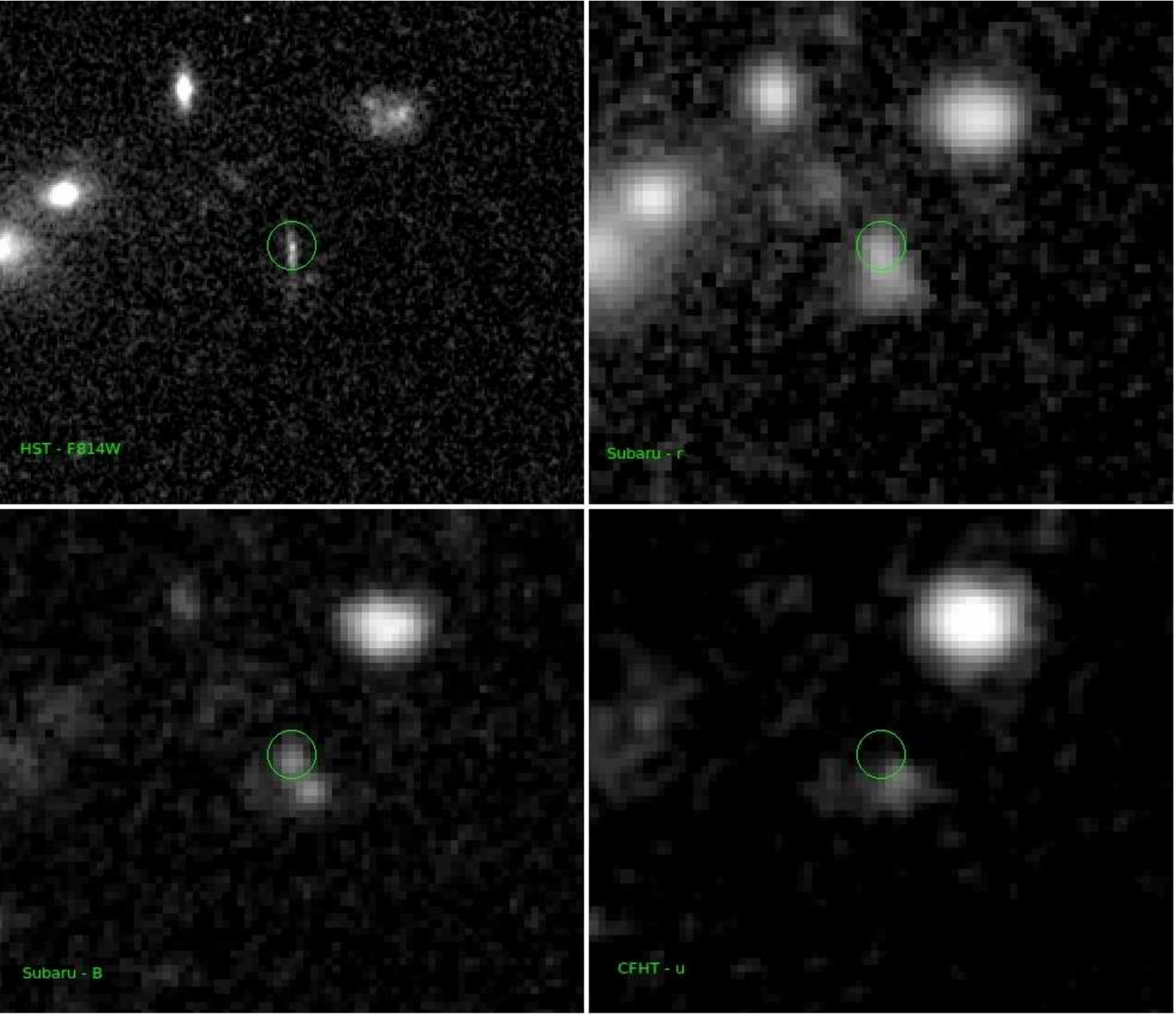}
\end{center}
\caption{Images ($10''\times10''$ dimensions) of arc located in the field source 01 approximately from i- to u-bands.
The arc is marked with a green circle in the center of each image.
Top left: HST/ACS image \citep[F814W filter,][]{koekemoer07}. Top right: {\it Subaru} r$^{+}-$band.
Bottom left: {\it Subaru} B-band \citep{taniguchi2007}. Bottom right: u$^{\ast}$ CFHT image \citep{capak2007}. }
\label{fig:arc_01_4images}
\end{figure*}

%
The I09 catalog reports a photometric redshift $z=0.715$ for the arc.
However such a redshift is significantly lower than that of 01.
This is unexpected, since the dark matter halo should be located between the observer and the lensed  object.
In order to understand the discrepancy we visually inspected the COSMOS archival images of the field
at different wavelengths, roughly between the i- and the u-bands.
In Figure~\ref{fig:arc_01_4images} we report four images ($10''\times10''$ each)
of the field of the arc, that is clearly marked with a green circle in each of them.

We find that the arc is very bright from the F814W filter to the B-band,
but it completely disappears in the u$^{\ast}$-band.
Therefore, we suspect that this is a u-band drop out and that the source associated with the arc
is located at redshift $z\simeq2.3$ or even higher.

While the arc clearly disappears in the u-band image, a close companion SW of the arc is clearly visible
in all the four images.
We suspect that, during their automatic procedure, I09 erroneously associated with the bright arc
the u$^{\ast}$-band flux measurement that corresponds to this companion.
This likely lead to an incorrect photometric redshift estimate.

Hence, our serendipitous discovery suggests that this project
might also be promising for systematic studies of (strong) lensing
features observed in rich groups or clusters.
Our method
might be complementary and would extend to higher redshifts
projects that find rich groups on
the basis of strong lensing signatures \citep[e.g.][]{cabanac2007,limousin2009,more2012}.

One limitation of such searches is that lensing features
are less likely at increasing redshifts.
This is mainly because the projected number density
of background objects decreases as the redshift of the lens increases.
This has so far limited the number of high redshift groups detected
by means of strong lensing phenomena to $z\lesssim1.2$.
Similarly, we expect to have a better chance to observe possible occurrence of lensing phenomena for
our $z\simeq1$ cluster candidates than at higher redshifts.
Therefore, our sample might not include a large number of strongly lensed objects while it includes an
extremely useful number of high redshift
groups.
\subsection{The nature of the HLRGs}

The HLRGs represent the
class of relatively higher power radio galaxies in our sample. As discussed in Sect.~\ref{par:sample_redef} and clearly shown in
Fig.~\ref{fig:scatterplot} such sources have radio power slightly above the formal FR~I/FR~II radio power divide.
Furthermore, the possible presence of bimodality in the radio power distribution of the
FR~Is in our sample suggests that the HLRGs might be drawn from a different parent population
(see Sect.~\ref{sec:stat_prop}).
In this section we will discuss the properties of the HLRGs with respect to their radio properties.

Radio galaxies with clear FR~II morphology (i.e. that showed evidence of clearly separated hot spots)
 were rejected during the C09 sample selection procedure.
This immediately excludes the possibility that the HLRGs might be classical FR~IIs radio sources, on the basis of their radio morphology.

\subsubsection{Radio galaxies of transitional type}

A possible scenario is that the HLRGs are radio galaxies
of transitional type, i.e. with radio morphology typical of FR~I sources and radio power typical of
the local faint FR~II radio galaxy population.
This is not surprising, because the high power tail of the FR~I radio power distribution partially
overlaps with the low luminosity tail of the FR~IIs, at least at low redshifts.

Furthermore, it has been proposed that the classical FR~I/FR~II radio luminosity divide undergos a positive evolution
with increasing redshift \citep[e.g.][]{heywood2007}.
In such a scenario, radio galaxies with radio morphology typical of FR~I sources and radio power typical of local
FR~IIs would be more common at the redshifts of our interest than at low-intermediate redshifts.

\subsubsection{Compact radio sources}

As discussed in C09, the rejection of radio galaxies with clear FR~II morphology was performed firstly on the basis of the
FIRST survey \citep{becker95}, and then by using the  VLA-COSMOS survey \citep{VLA_COSMOS}.
Their radio maps have a typical resolution of $\sim$5~arcsec (FIRST) and $\sim1.5$~arcsec (VLA-COSMOS), that correspond to 43~kpc and
13~kpc, at redshift $z=1.5$, respectively.
This selection excludes the presence of classical FR~IIs in the sample, since the radio jets of these sources
typically extend to distances larger than $\sim$a few tens of kpc, up to Mpc scale.

Almost all of the LLRGs and all of the HLRGs are observed as compact radio sources in both the
FIRST and the  VLA-COSMOS surveys.
As pointed out in C09, there are two possible scenario. i) While the core has a flat radio spectrum,
the extended emission of radio sources has a steep spectrum.
Because of the light redshifting, the extended emission is therefore increasingly more difficult
to detect at increasing redshifts.
Therefore it might be that both the FIRST and the VLA-COSMOS surveys detect the core emission only.
ii) Alternatively, the radio galaxies in our sample are intrinsically small.
The first scenario was discussed in C09. Therefore, we limit our discussion to the second possibility.

If the sources in our sample are intrinsically compact, they are entirely
contained within a few $\sim10$~kpc scale. They might show a radio morphology somehow different from that of classical FR~Is.
If this is the case we suggest that the HLRGs might be Compact Steep Spectrum sources
\citep[CSS, e.g.][]{saikia1988,fanti1990,fanti_fanti1994,dallacasa1993,fanti_spencer1995}
or GHz Peaked Sources \citep[GPS, e.g.][]{odea1991}.

The GPS are commonly contained within the Narrow Line Region at $\lesssim1$~kpc scale,
while the CSS sources are usually contained within the host galaxy (i.e. $\lesssim15$~kpc).
They would not be resolved at redshift $z\gtrsim1$
by using the VLA-COSMOS and the FIRST surveys.
Therefore, the possibility that some of the HLRGs are GPS or CSS cannot be excluded.

The GPS and CSS sources show a complex multiple radio morphology
\citep[see also][and references therein for a review]{odea1998}.
They are preferentially found at lower redshifts \citep[$z\lesssim1$][]{fanti1990,odea1991},
and have higher radio powers \citep[i.e. $\sim2$ orders of magnitude brighter,][]{odea_baum1997}
 than those of HLRGs.
This also implies that the presence of GPS or CSS sources within the HLRGs is more likely
than for the LLRGs.


However, the radio powers of the FR~Is in our sample (including both the LLRGs and the HLRGs)
are fully consistent with those of local faint radio sources studied by \cite{drake2004}.
Most of the galaxies in their sample are compact and therefore resemble
CSS or GPS sources.
They have redshifts and low frequency radio luminosities between $z\simeq0.05-0.35$ and
L$_{1.4}\simeq31.0-34.2$~erg~s$^{-1}$~Hz$^{-1}$, respectively.
Interestingly, this suggests that all of the radio galaxies in our sample might
be similar to the local radio sources  in the \cite{drake2004} catalog.

If some of our sources were confirmed to be CSSs or GPSs, they would constitute
a population of compact radio sources with higher redshifts and lower radio power than
those included in previous samples of intermediate redshift
objects of these two classes
\citep[e.g.][]{dallacasa1995,dallacasa1998,dallacasa2013}.

It would be interesting to study the spectral properties of the HLRGs in our sample
with multiwavelength radio observations, to see if they are consistent with the steep spectra typical of CSS
or if the SEDs are instead consistent with those of GPS sources that show a peak at high radio frequencies.
High angular resolution ($\lesssim0.1$~arcsec) radio observations with the 
Very-long-baseline interferometry (VLBI) network
may allow us to investigate
in detail the radio morphology of these sources.

According to the theoretical evolutionary scenario suggested for CSSs and GPSs by \cite{snellen2000},
if the radio galaxies in our sample are compact $\gtrsim1$~kpc sources, they will evolve into classical FR~Is
increasing their size and decreasing their radio luminosity.
Alternatively, if our sources are $\lesssim1$~kpc GPSs,
they will increase their luminosities and sizes, until
they reach a $\sim1$~kpc size. Then, they will decrease their radio power evolving into CSSs and
finally into radio galaxies.

Conversely, \citet{tinti_dezotti2006} found observational evidence
that GPS sources always evolve decreasing their luminosity and increasing their size.
This is in agreement with the theoretical model suggested by \citet{begelman1996}.

Therefore, it might be that, during their evolution,
some of our sources will reach a higher radio power.
However, it is unlikely that they will
increase their radio luminosities enough to evolve into radio galaxies with a radio morphology
typical of classical FR~IIs,
as also suggested by \cite{drake2004} for their sample of lower redshifts compact sources.


\section{Conclusions and future work}\label{sec:conclusions}
We applied a newly developed method
to search for overdensities around the $z\sim1-2$ FR~Is of the C09 sample,
which has been accurately redefined in this work.
We found that the great majority of the FR~Is in the sample reside
in Mpc-scale rich groups or clusters. We estimated, for each cluster candidate: (i) detection significance,
(ii) redshift, (iii) size, and (iv) richness.

We also compared our results with those obtained by previous work on the environments of low redshift radio galaxies,
high redshift FR~IIs and cluster samples at intermediate redshifts.
The fraction of FR~Is that are associated with cluster environments in our redshift range is consistent with what
found for low redshift (i.e. $z\leq0.25$) FR~Is.
However, it is significantly higher than what found
for both local and high redshift FR~IIs.

Moreover, we applied an independent method based on IR colors to search for high redshift overdensities \citep[][P08]{papovich2008}
 performing a counts-in-cell analysis.
Interestingly, all of the  six cluster candidates that are found with such a method, are also detected by the PPM.
Vice-versa, the great majority (i.e. $\sim70\%$) of our $z\gtrsim1.3$ cluster candidates
are not found by the P08 method.
 Since the P08 method is applied performing a counts-in-cell analysis, some of the clusters that are not detected 
by the P08 method might be populated by galaxies that are 
not completely segregated in the cluster core.

Spectroscopic confirmations and detailed multiwavelength observations of our cluster candidates
are nevertheless required to study them in more detail,
to confirm the results obtained in this work.
This is especially important for our high redshift ($z\gtrsim1.5$) cluster candidates.
These would significantly increase the statistics of cluster samples
at such high redshifts and might
allow a more complete understanding of the ongoing processes involved in the formation and the evolution of
these structures.

In more details, it would be interesting to observe the cluster candidates with deeper IR
and optical observations, to look for any evidence (or absence) of the red sequence or a segregation
of faint red objects in the fields that we
might be missing by using the COSMOS catalog \citep{ilbert2009}.
Rest frame  ultra-violet (UV) observations might also help to search for the possible presence of Lyman-$\alpha$ emitting regions that
are commonly found in $z\gtrsim2$ protoclusters.
X-ray observations deeper than those available within the COSMOS survey will allow to search for signatures of
hot plasma within the Intra Cluster Medium \citep{tundo2012}.
All of these observations will help
establishing if our clusters are
still evolving. Alternatively, they might
exhibit transitional properties
between those typical of high redshift ($z>2$) Lyman-$\alpha$ emitter protoclusters and those associated with
low redshift clusters, that show
common features  such as X-ray emission, red-sequence, and segregation of red objects within the core.

More in general, our results suggest that the Mpc--scale overdensities associated with
the radio galaxies in our sample are similar, independently of the two subclasses considered
throughout this work (i.e. the LLRGs and the HLRGs), in terms of estimated richness, mass, and size.
Interestingly, on the basis of their multi-component SED fitting, \cite{baldi2013}
 found that also the host galaxies
of both low and high power radio galaxies in the C09 sample have homogeneous properties,
in terms of UV, IR luminosities, stellar mass content, and dust temperature, independently of the subsample considered.
Therefore, we can conclude that the radio galaxies in the C09 sample constitute a homogeneous population.

Furthermore, we reported the serendipitous discovery of a bright arc in the field of 01, that is at $z_{\rm spec}=0.88$.
This might suggest that the cluster associated with that source is rich and compact
\citep[as suggested by][and in this work]{finoguenov2007,george2011,knobel2012}.
The presence of strong and weak lensing features in our sample might be present for some of our cluster candidates.
We will investigate this scenario in a forthcoming paper.

The above results, combined with the steepness of the radio luminosity function of the radio galaxies,
suggest that low power FR~Is are more effective than
FR~IIs as beacons to search for groups and clusters at high redshifts.

Radio sources with radio powers typical of those of our FR~Is are found only in $10-20\%$ of
X-ray and optically selected clusters at $z\lesssim1$ \citep{branchesi2006,gralla2011}.
Therefore, unless this percentage dramatically changed at $z\geq1$ we might still be
missing $80-90\%$ of the entire cluster population at the redshifts of our interest.
It would be interesting to blindly apply the PPM to the entire COSMOS field to robustly estimate
such a total number of overdensities.
This will allow us to compare that with
the number counts predicted by the $\Lambda$CDM model.
We will investigate these aspects in a future work.

Interestingly, our
cluster candidates might be also studied by using the next generation telescopes such as JWST.
Although the PPM is primarily introduced for the COSMOS survey, it may be applied to
wide field surveys to blindly search for cluster candidates by using accurate photometric redshift information.
In particular, we will also extend our work to wider surveys (e.g. stripe 82 of the SDSS), where
we expect to find a higher number of both FR~Is ($\sim3000$) and cluster candidates ($\sim2100$).
Furthermore, we will have a higher chance to find more massive
structures and lensing phenomena.
Two possible limitations are that the FR~Is are difficult to find and that the PPM requires good photometric redshifts.
Moreover, our method will be less effective for those surveys that will provide sufficient spectroscopic high redshift information,
where standard 3-D methods (e.g. correlation functions) might be more successfully applied.

Conversely, the PPM might be also applied
to future wide field surveys such as LSST that will provide good photometric redshift information.
Another possible use of the PPM is a search for (proto-)clusters at $z\gtrsim 2$, by
adopting radio galaxies or other sources (e.g. Lyman break galaxies) as beacons.

The careful selection of our FR I sample and the accurate redshift estimates have also allowed us to estimate the comoving space density of sources with $L_{1.4}\simeq 10^{32.3}\,\hbox{erg}\,\hbox{s}^{-1}\,\hbox{Hz}^{-1}$ at $z\simeq 1.1$. Previous direct observational estimates and model predictions span a quite broad range. Our result is consistent with the upper values and
strengthens the case for a strong cosmological evolution of these sources.



\begin{acknowledgements}
 We thank the anonymous referee for helpful comments.
We  thank the Space Telescope Science Institute, where part of this work was developed.
We also thank Roberto Gilli, Piero Rosati, and Paolo Tozzi for fruitful discussion.
This work was partially supported by the STScI JDF account D0101.90157, and (G.C.) by both the Internship Program ISSNAF-INAF 2010 and
 one of the Foundation Angelo Della Riccia fellowships both in 2012 and in 2013.
This research has made use of the NASA/IPAC Extragalactic Database (NED) which is operated by
 the Jet Propulsion Laboratory, 
California Institute of Technology, under contract with the National Aeronautics and Space 
Administration.
\end{acknowledgements}


\begin{thebibliography}{161}
\bibitem[Abell(1958)]{abell1958} Abell, G. O., 1958, ApJS, 3, 211
\bibitem[Adami et al.(2010)]{adami2010} Adami, C., Durret, F., Benoist, C., Coupon, J., et al., 2010, A\&A, 509A, 81	
\bibitem[Adami et al.(2011)]{adami2011} Adami, C., Mazure, A., Pierre, M., et al., 2011, A$\&$A, 526, 18
\bibitem[Allen et al.(2011)]{allen2011} Allen, S. W., Evrard, A. E., Mantz, A. B., 2011, ARAA, 49:409-470
\bibitem[Adami et al.(2013)]{adami2013} Adami, C., Durret, F., et al., 2013, A\&A, 551, A20
\bibitem[Alberts et al.(2013)]{alberts2013} Alberts, S., Pope, A., Brodwin, M., et al., 2013, MNRAS, tmp, 2563
\bibitem[Ashman et al.(1994)]{ashman1994} Ashman, K. M., Bird, C. M., Zepf, S. E., 1994, AJ, 108, 2348
\bibitem[Bahcall et al.(2003)]{bahcall2003} Bahcall, N. A., Dong, F., et al., 2003, ApJ, 599, 814
\bibitem[Baldi et al.(2013)]{baldi2013} Baldi, T., Chiaberge, M., et al., 2013, ApJ, 762, 30B (B13)
\bibitem[Banerji et al.(2011)]{banerji2011} Banerji, M., Chapman, S. C., Smail, I., et al., 2011, MNRAS, 418, 1071
\bibitem[Barthel \& Arnaud(1996)]{barthel_arnaud1996} Barthel, P. D., \& Arnaud, K. A. 1996, MNRAS, 283, 45
\bibitem[Barone-Nugent et al.(2013)]{barone_nugent2013}Barone-Nugent, R. L.,  Wyithe, J. S. B., et al.,  arXiv1303.6109
\bibitem[Bassett et al.(2013)]{bassett2013} Bassett, R., Papovich, C., et al., 2013, arXiv1305.0607
\bibitem[Bayliss et al.(2013)]{bayliss2013} Bayliss, M. B., Ashby, M. L. N., Ruel, J., et al., 2013, arXiv1307.2903
\bibitem[Becker et al.(1995)]{becker95} Becker, R. H., White, R. L., \& Helfand, D. J. 1995, ApJ, 450, 559
\bibitem[Begelman(1996)]{begelman1996} Begelman, M. C., 1996, in Cygnus A: Study of a Radio Galaxy, eds C. L. Carilli \& D. A. Harris (Cambridge: Cambridge University Press), 209
\bibitem[Benson et al.(2013)]{benson2013} Benson, B. A., de Haan, T., Dudley, J. P., 2013, ApJ, 763, 147
\bibitem[Blandford \& Narayan(1986)]{blandford_narayan1986} Blandford, R., \& Narayan, R. 1986, ApJ, 310, 568
\bibitem[Blandford \& Narayan(1992)]{blandford_narayan1992} Blandford, R. D., \& Narayan, R. 1992, ARA\&A, 30, 311
\bibitem[Bode et al.(2001)]{bode2001} Bode, P., Bahcall, N. A., et al., 2001, ApJ, 551, 15
\bibitem[B\"{o}hringer et al.(2004)]{bohringer2004} B\"{o}hringer, H., Schuecker, P., Guzzo, L., et al., 2004 A\&A, 425, 367
\bibitem[Branchesi et al.(2006)]{branchesi2006} Branchesi, M., Gioia, I. M., Fanti, C., et al., 2006 A\&A, 446, 97
\bibitem[Brodwin et al.(2011)]{brodwin2011} Brodwin, M., Stern, D., Vikhlinin, A., et al., 2011, ApJ, 732, 33
\bibitem[Brodwin et al.(2012)]{brodwin2012} Brodwin, M., Gonzalez, A. H., Stanford, S. A., et al., 2012, ApJ, 753, 162
\bibitem[Brodwin et al.(2013)]{brodwin2013} Brodwin, M., Stanford, S. A., Gonzalez, A. H., et al., 2013, arXiv1310.6039
\bibitem[Cabanac et al.(2007)]{cabanac2007} Cabanac, R. A., Alard, C., Dantel-Fort, M., et al. 2007, A\&A, 461, 813
\bibitem[Capak et al.(2007)]{capak2007} Capak, P., et al. 2007, ApJS, 172, 99
\bibitem[Capak et al.(2011)]{capak2011} Capak, P. L., Riechers, D., Scoville, N. Z., et al. 2011, Nature, 470, 233
\bibitem[Cappi \& Maurogordato(1995)]{cappi1995} Cappi, Alberto, \& Maurogordato, Sophie, 1995, ApJ, 438, 507
\bibitem[Casasola et al.(2013)]{casasola2013} Casasola, V., Magrini, L., Combes, F., et al., 2013, A\&A, 558, 60
\bibitem[Castignani et al.(2014)]{PPMmethod} Castignani, G., Chiaberge, M., Celotti, A., and Norman, C., 2014, arXiv:1405.7973, ApJ in press (Paper~I)
\bibitem[Cavaliere \& Fusco-Fermiano(1978)]{cavaliere_fusco1978} Cavaliere, A. \& Fusco-Fermiano, R., 1978, A\&A, 70, 677
\bibitem[Chiaberge et al.(1999)]{chiaberge1999} Chiaberge, M., Capetti, A., \& Celotti, A., 1999, AA, 349, 77
\bibitem[Chiaberge et al.(2009)]{chiaberge2009} Chiaberge, M., Tremblay, G., Capetti, A., et al., 2009, ApJ, 696, 1103 (C09)
\bibitem[Chiaberge et al.(2010)]{chiaberge2010} Chiaberge, M., Capetti, A., Macchetto, F. D., et al., 2010, ApJ, 710L, 107 (C10)
\bibitem[Chiang et al.(2013)]{chiang2013} Chiang, Y.-K., Overzier, R., \& Gebhardt, K., arXiv:1310.2938
\bibitem[Collister et al.(2007)]{collister2007} Collister A. et al., 2007, MNRAS, 375, 68
\bibitem[Condon et al.(1998)]{condon1998} Condon, J. J., Cotton, W. D., Greisen, E. W., et al., 1998, AJ, 115, 1693
\bibitem[Cooray \& Sheth(2002)]{cooray_sheth2002} Cooray, A. \& Sheth, R., PhR, 372, 1
\bibitem[Cruddace et al.(2002)]{cruddace2002} 	Cruddace, R., Voges, W., B\"{o}hringer, H., et al., 2002,  ApJS, 140, 239
\bibitem[Dai et al.(2007)]{dai2007} Dai, X., Kochanek, C. S., \& Morgan, N. D., 2007, ApJ, 658, 917
\bibitem[Dalal et al.(2008)]{dalal2008} Dalal, N., White, M., Bond, J. R., Shirokov, A., 2008, ApJ, 687, 12D
\bibitem[Dallacasa et al.(1993)]{dallacasa1993} Dallacasa, D., Fanti, C., \& Fanti, R. 1993, in Jets in Extragalactic Radio Sources, ed. H.-J. Roser \& K. Meisenheimer(Heidelberg: Springer), 27
\bibitem[Dallacasa et al.(1995)]{dallacasa1995} Dallacasa D., Fanti C., Fanti R., Schilizzi R. T., Spencer R. E., 1995, A\&A, 295, 27
\bibitem[Dallacasa et al.(1998)]{dallacasa1998} Dallacasa, D., Bondi, M., Alef, W., \& Mantovani, M., 1998, A\&ASS, 129, 219
\bibitem[Dallacasa et al.(2013)]{dallacasa2013} Dallacasa, D., Orienti, M., Fanti, C., Fanti, R., Stanghellini, C., 2013, MNRAS, 433, 147
\bibitem[Davis \& Mushotzky(1993)]{davis1993} Davis, D. S. \& Mushotzky, R. F., 1993, AJ, 105, 409
\bibitem[Diener et al.(2013)]{diener2013} Diener, C., Lilly, S. J., Knobel, C., Zamorani, G., et al., 2013, ApJ, 765, 109
\bibitem[Donoso et al.(2009)]{donoso2009} Donoso, E., Best, P. N., \& Kauffmann, G., 2009, MNRAS, 392, 617
\bibitem[Donoso et al.(2010)]{donoso2010} Donoso, E., Li, C., Kauffmann, G., et al., 2010,  MNRAS, 407, 1078
\bibitem[Donzelli et al., 2007]{donzelli07} Donzelli, C. J., Chiaberge, M., Macchetto, F. D., Madrid, J. P., Capetti, A., \& Marchesini, D.s, 2007, ApJ, 667, 780
\bibitem[Drake et al.(2004)]{drake2004} Drake, C. L., Bicknell, G. V., McGregor, P. J., \& Dopita, M. A. 2004, AJ, 128, 969
\bibitem[Durret et al.(2011)]{durret2011} Durret, F., Adami, C., Cappi, A., et al., 2011, A\&A, 535A, 65
\bibitem[Ebeling et al.(1993)]{ebeling1993} Ebeling H., Wiedenmann G., 1993, Phys. Rev. E, 47, 704
\bibitem[Eisenhardt et al.(2008)]{eisenhardt2008} Eisenhardt, P. R. M., Brodwin, M., et al., 2008, ApJ, 684, 905
\bibitem[Falder et al.(2010)]{falder2010} Falder, J. T., Stevens, J. A., Jarvis, M. J., et al. 2010, MNRAS, 405, 347
\bibitem[Fanaroff \& Riley(1974)]{fr74} Fanaroff, B. L., \& Riley, J. M. 1974, MNRAS, 167, 31
\bibitem[Fanti et al.(1990)]{fanti1990} Fanti, R., Fanti, C., Schilizzi, R. T., Spencer, R. E., Rendong, N., Parma, P., van Breugel, W. J. M., \& Venturi, T., 1990, A\&A, 231, 333
\bibitem[Fanti \& Fanti(1994)]{fanti_fanti1994} Fanti, C., \& Fanti, R., 1994, in ASP Conf. Ser. 54, The Physics of Active Galaxies, ed. G. V. Bicknell, M. A. Dopita, \& P. J. Quinn(San Francisco: ASP), 341
\bibitem[Fanti \& Spencer(1995)]{fanti_spencer1995} Fanti, R., \& Spencer, R. E., 1995, in IAU Symp. 175, Extragalactic Radio Sources, ed. R. Ekers, C. Fanti, \& L. Padrielli(Dordrecht: Kluwer), 63
\bibitem[Fassbender et al.(2011)]{fassbender2011} Fassbender, R., Nastasi, A., Böhringer, H., et al. 2011, A\&A, 527, 10
\bibitem[Finoguenov et al.(2007)]{finoguenov2007} Finoguenov, A., Guzzo, L., Hasinger, G., et al., 2007, ApJS, 172, 182F
\bibitem[Galametz et al.(2012)]{galametz2012} Galametz, A., Stern, D., De Breuck, C., et al., 2012, ApJ, 749, 169
\bibitem[Galametz et al.(2013)]{galametz2013} Galametz, A., Stern, D., Pentericci, L., et al., 2013,arXiv:1309.6645
\bibitem[George et al.(2011)]{george2011} George, M. R., Leauthaud, A., Bundy, K., Finoguenov, A., et al., 2011, ApJ, 742, 125
\bibitem[George et al.(2012)]{george2012} George, M. R., Leauthaud, A., Bundy, K., Finoguenov, A., et al., 2012, ApJ, 757, 2
\bibitem[Giavalisco(2002)]{giavalisco2002} Giavalisco, M. 2002, ARA\&A, 40, 579
\bibitem[Gladders \& Yee(2005)]{gladders2005} Gladders, M. D., \& Yee, H. K. C., 2005, ApJS, 157, 1	
\bibitem[Gobat et al.(2011)]{gobat2011} Gobat, R., Daddi, E., Onodera, M., et al., 2011, A\&A, 526, 133
\bibitem[Gobat et al.(2013)]{gobat2013} Gobat, R., Strazzullo, V., Daddi, E., et al., 2013, arXiv1305.3576
\bibitem[Gomez et al.(1997)]{gomez1997} Gomez, P. L., Pinkney, J., Burns, J. O., et al., 1997, ApJ, 474, 580
\bibitem[Gralla et al.(2011)]{gralla2011} Gralla, M. B., Gladders, M. D., et al., 2011, ApJ, 734, 103
\bibitem[Halkola et al.(2007)]{halkola2007} Halkola, A., Seitz, S., Pannella, M., 2007, ApJ, 656, 739
\bibitem[Harris(2012)]{harris2012} Harris, Kathryn A., 2012, arXiv1201.5746
\bibitem[Hasselfield et al.(2013)]{hasselfield2013} Hasselfield, M., Hilton, M., Marriage, T. A., et al. 2013, arXiv:1301.0816
\bibitem[Henry et al.(2006)]{henry2006} Henry, J. P., Mullis, C. R., Voges, W., et al.,  2006, ApJS, 162, 304	
\bibitem[Heywood et al.(2007)]{heywood2007} Heywood, I., Blundell, K. M., and Rawlings, S., 2007, MNRAS, 381, 1093	
\bibitem[Hickox et al.(2009)]{hickox2009} Hickox, R. C., Jones, C., Forman, W. R., et al. 2009, ApJ, 696, 891
\bibitem[Hill \& Lilly(1991)]{hill1991} Hill, G. J., \& Lilly, S. J., 1991, ApJ, 367, 1
\bibitem[Hilton et al.(2010)]{hilton2010} Hilton, M., Lloyd-Davies, E., Stanford, S. A., et al., 2010, ApJ, 718, 133
\bibitem[Hinshaw et al.(2009)]{Hinshaw2009} Hinshaw, G., et al. 2009, ApJS, 180, 225
\bibitem[Ilbert et al.(2009)]{ilbert2009} Ilbert, O., Capak, P., Salvato, M., et al. 2009, ApJ, 690, 1236(I09)
\bibitem[Jee et al.(2011)]{jee2011} Jee, M. J., Dawson, K. S., Hoekstra, H., Perlmutter, S., Rosati, P., et al., 2011, ApJ, 737:59
\bibitem[Jian et al.(2013)]{jian2013} Jian, H.-Y., Lin, L., Chiueh, T., 2013, arXiv1305.1891
\bibitem[John(1988)]{john1988} John, T. L., 1988, A\&A, 193, 189
\bibitem[Jones et al.(2004)]{jones2004} Jones, D.~H., et al., 2004, MNRAS, 355, 747
\bibitem[Klypin \& Kopylov(1983)]{klypin1983} Klypin, A. A., \& Kopylov, A. I., 1983, SvAL, 9, 41
\bibitem[Knobel et al.(2009)]{knobel2009} Knobel, C., Lilly, S. J., Iovino, A., Porciani, C., et al., 2009, ApJ, 697, 1842
\bibitem[Knobel et al.(2012)]{knobel2012} Knobel, C., Lilly, S. J., Iovino, A., et al. 2012, ApJ, 753, 121
\bibitem[Kochanek(2006)]{kochanek2006} Kochanek, C. S. 2006, in Saas-Fee Advanced Course 33: Gravitational Lensing:Strong, Weak and Micro, ed. G. Meylan, P. North, \& P. Jetzer(Berlin: Springer), 91
\bibitem[Koekemoer et al.(2007)]{koekemoer07} Koekemoer, A. M., et al. 2007, ApJS, 172, 196
\bibitem[Koyama et al.(2014)]{koyama2014} Koyama, Y., Kodama, T., Tadaki, K., et al., 2014, arXiv:1405.4165K
\bibitem[Kravtsov \& Borgani(2012)]{kravtsov_borgani2012} Kravtsov, Andrey V., Borgani, Stefano, 2012, ARA\&A, 50, 353
\bibitem[Landy \& Szalay(1993)]{landy1993} Landy, S. D., \& Szalay, A. S., 1993, ApJ, 412, 64
\bibitem[Leauthaud et al.(2010)]{leauthaud2010} Leauthaud, A., Finoguenov, A., Kneib, J.-P., et al. 2010, ApJ, 709, 97
\bibitem[Ledlow \& Owen(1995)]{ledlow_owen1995} Ledlow, M. J. \& Owen, F. N., 1995, AJ, 109, 853
\bibitem[Ledlow \& Owen(1996)]{ledlow_owen1996} Ledlow, M. J. \& Owen, F. N., 1996, AJ, 112, 9
\bibitem[Lilly et al.(2007)]{lilly2007} Lilly, S. J., Le F\`evre, O., Renzini, A., et al. 2007, ApJS, 172, 70
\bibitem[Limousin et al.(2009)]{limousin2009} Limousin, M., Cabanac, R., Gavazzi, R., et al. 2009, A\&A, 502, 445
\bibitem[Liu et al.(2013)]{liu2013} Liu, F. S., Guo, Y., et al., 2013, ApJ, 769, 147
\bibitem[Magliocchetti \& Br\"{u}ggen(2007)]{magliocchetti_bruggen2007} Magliocchetti, M. \& Br\"{u}ggen, M., 2007, MNRAS, 379, 260
\bibitem[Mantz et al.(2010)]{mantz2010} Mantz, A., Allen, S. W., Rapetti, D., Ebeling, H., 2010, MNRAS, 406:1759-1772
\bibitem[Marriage et al.(2011)]{marriage2011} Marriage, T. A., Acquaviva, V., et al., 2011, ApJ, 737, 61
\bibitem[Massardi et al.(2010)]{massardi2010} Massardi, M., Bonaldi, A., et al., 2010,  MNRAS, 404, 532
\bibitem[Mauch \& Sadler(2007)]{mauch_sadler2007} Mauch, T., Sadler, E.~M., 2007, MNRAS, 375, 931
\bibitem[Mayo et al.(2012)]{mayo2012} Mayo, J. H., Vernet, J., De Breuck, C., et al., 2012, A\&A, 539A, 33
\bibitem[McAlpine et al.(2013)]{mcalpine2013} McAlpine, K., Jarvis, M.~J., \& Bonfield, D. G., 2013, MNRAS, 436, 1084
\bibitem[McIntosh et al.(2013)]{mcintosh2013} McIntosh, D. H., Wagner, C., Cooper, A., et al., 2013,  arXiv:1308.0054
\bibitem[Mei et al.(2009)]{mei2009} Mei, S., Holden, B. P., Blakeslee, J. P., et al., 2009, ApJ, 690, 42
\bibitem[Menci et al.(2008)]{menci2008} Menci, N., Rosati, P., Gobat, R., et al., 2008, ApJ, 685, 863
\bibitem[Miley \& De~Breuck(2008)]{miley_debreuck2008} Miley, G., \& De Breuck, C., 2008, A\&ARv, 15, 67
\bibitem[Mobasher et al.(2007)]{mobasher2007} Mobasher, B., Capak, P., Scoville, N. Z., et al. 2007, ApJS, 172, 117
\bibitem[Mohr(2005)]{mohr2005} Mohr, J.~J., 2005, ASPC, 339, 140
\bibitem[More et al.(2012)]{more2012} More, A., Cabanac, R., More, S., et al. 2012, ApJ, 749, 38
\bibitem[Mortonson et al.(2011)]{mortonson2011} Mortonson, M. J., Hu, W., Huterer, D., 2011, Phys. Rev. D, 83:023015
\bibitem[Muzzin et al.(2013)]{muzzin2013} Muzzin, A., Wilson, G., et al., 2013, ApJ, 767, 39
\bibitem[Nastasi et al.(2011)]{nastasi2011} Nastasi, A., Fassbender, R., B\"{o}hringer, H., et al., 2011, A\&A, 532, 6
\bibitem[Newman et al.(2013)]{newman2013} Newman, A. B., Ellis, R. S., Andreon, S., et al., 2009, arXiv1310.6754
\bibitem[Noble et al.(2013)]{noble2013} Noble, A. G., Geach, J. E.,  van Engelen, A. J., et al., 2013, MNRAS, tmpL156
\bibitem[O'Dea(1998)]{odea1998} O'Dea, C. P., 1998, PASP, 110, 493O
\bibitem[O'Dea \& Owen(1986)]{Odea_Owen1986} O'Dea C. P. \& Owen, F. N., 1986, ApJ, 301, 841O
\bibitem[O'Dea et al.(1991)]{odea1991} O'Dea, C. P., Baum, S. A., \& Stanghellini, C., 1991, ApJ, 380, 66
\bibitem[O'Dea \& Baum(1997)]{odea_baum1997} O’Dea, C. P., \& Baum, S. A., 1997, AJ, 113, 148
\bibitem[Papovich(2008)]{papovich2008} Papovich, C., 2008, ApJ, 676, 206 (P08)
\bibitem[Papovich et al.(2010)]{papovich2010} Papovich, C., Momcheva, I., Willmer, C. N. A., et al., 2010, ApJ, 716, 1503
\bibitem[Peacock(1999)]{peacock1999} Peacock, J. A., 1999, Cosmological Physics, Cambridge, UK, Cambridge Univ. Press
\bibitem[Peebles(1980)]{peebles1980} Peebles P. J. E., 1980, The Large-Scale Structure of the Universe. Princeton Univ. Press, Princeton
\bibitem[Peebles(1993)]{peebles1993} Peebles, P. J. E., 1993, Physical Cosmology, Princeton, NJ, Princeton Univ. Press
\bibitem[Planck Collaboration XX(2013)]{planckXX_2013} Planck Collaboration XX, 2013, arXiv1303.5080	
\bibitem[Planck Collaboration XXIX(2013)]{planckXXIX_2013} Planck Collaboration XXIX, 2013, arXiv1303.5089
\bibitem[Prescott et al.(2006)]{prescott06} Prescott, M. K. M., Impey, C. D., Cool, R. J., \& Scoville, N. Z. 2006, ApJ, 644, 100
\bibitem[Ramos~Almeida et al.(2013)]{ramosalmeida2013} Ramos Almeida, C., Bessiere, P. S., Tadhunter, C., et al., arXiv1308.4725
\bibitem[Reichardt et al.(2013)]{reichardt2013} Reichardt, C. L., Stalder, B., et al., 2013, ApJ, 763, 127
\bibitem[Reiprich \& B\"{o}hringer(1999)]{reiprich_bohringer99} Reiprich, T. H., B\"{o}hringer, H., 1999, dtrp conf, 157
\bibitem[Rigby et al.(2008)]{rigby2008} Rigby, E.~E., Best, P.~N., \& Snellen, I.~A.~G., 2008, MNRAS, 385, 310
\bibitem[Rigby et al.(2013)]{rigby2013} Rigby, E. E., Hatch, N. A., et al., 2013, arXiv1310.5710
\bibitem[Rosati et al.(2002)]{rosati2002} Rosati, P.,  Borgani, S., Norman, C., 2002, Annu. Rev. Astron. Astrophys.,  40, 539-77
\bibitem[Rosati et al.(2009)]{rosati2009} Rosati, P., Tozzi, P., Gobat, R., et al., 2009, A\&A, 508, 583
\bibitem[Rozo et al.(2010)]{rozo2010}	Rozo, E., Wechsler, R. H., Rykoff, E. S., et al., 2010, ApJ, 708:645–660
\bibitem[Sadler et al.(2007)]{sadler2007} Sadler, E. M., Cannon, R. D., Mauch, T., Hancock, P. J., 2007, MNRAS, 381, 211
\bibitem[Saikia(1988)]{saikia1988} Saikia, D. J. 1988, in Active Galactic Nuclei, ed. H. R. Miller \& P. J. Wiita(Berlin: Springer), 317
\bibitem[Sanders et al.(2007)]{sanders2007} Sanders, D.~B., Salvato, M., Aussel, H., et al., 2007, ApJS, 172, 86
\bibitem[Santos et al.(2009)]{santos2009} Santos, J. S., Rosati, P., Gobat, R., et al., 2009, A\&A, 501, 49S
\bibitem[Santos et al.(2011)]{santos2011} Santos, J. S., Fassbender, R., Nastasi, A., et al. 2011, A\&A, 531, 15
\bibitem[Santos et al.(2013)]{santos2013} Santos, J. S., Altieri, B., Popesso, 2013, arXiv1305.1938S
\bibitem[Schinnerer et al(2007)]{VLA_COSMOS} Schinnerer, E., et al. 2007, ApJS, 172, 46
\bibitem[Schmidt(1968)]{schmidt1968} Schmidt, M., 1968, ApJ, 151, 393
\bibitem[Scoville et al.(2007a)]{scoville07} Scoville, N., et al. 2007a, ApJS, 172, 1
\bibitem[Scoville et al.(2007b)]{scoville07b} Scoville, N., et al. 2007b, ApJS, 172, 150
\bibitem[Scoville et al.(2013)]{scoville2013} Scoville, N.; Arnouts, S.; Aussel, H. et al., 2013, arXiv1303.6689
\bibitem[Semler et al.(2012)]{semler2012} Semler, D. R., et al., 2012, ApJ, 761, 183
\bibitem[Sheth \& Tormen(2004)]{sheth2004} Sheth, R. K., Tormen, G., 2004, MNRAS, 350, 1385
\bibitem[Sikora et al.(2007)]{sikora2007} Sikora, M., Stawarz, \L, \& Lasota, J.-P. 2007, ApJ, 658, 815
\bibitem[Smith \& Heckman(1990)]{smithheckaman1990} Smith, E. P., \& Heckman, T. M. 1990, ApJ, 348, 38
\bibitem[Smol\v{c}i\'{c} et al.(2009)]{smolcic2009} Smol\v{c}i\'{c}, V., Zamorani, G., Schinnerer, E., et al., 2009, ApJ, 696, 24
\bibitem[Smol\v{c}i\'{c} et al.(2011)]{smolcic2011} Smol\v{c}i\'{c}, V., Finoguenov, A., et al., MNRAS,  416, 31
\bibitem[Snellen \& Best(2001)]{snellen2001} Snellen, I. A. G. \& Best, P. N., 2001, MNRAS, 328, 897
\bibitem[Snellen et al.(2000)]{snellen2000} Snellen, I. A. G., Schilizzi, R. T., Miley, G. K., de Bruyn, A. G., et al., 2000, MNRAS, 319, 445
\bibitem[Song et al.(2012)]{song2012} Song, J., Zenteno, A., et al., 2012, ApJ, 761, 22
\bibitem[Spitler et al.(2012)]{spitler2012} Spitler, L. R., et al., 2012, ApJ, 748L, 21
\bibitem[Stanford et al.(2012)]{stanford2012} Stanford, S. A., Brodwin, M., Gonzalez, A. H., et al., 2012, ApJ, 753, 164
\bibitem[Steidel et al.(2000)]{steidel2000} Steidel, C. C., Adelberger, K. L., Shapley, A. E., et al., 2000, ApJ, 532, 170
\bibitem[Steidel et al.(2004)]{steidel2004} Steidel, C. C.; Shapley, A. E., 2004, ApJ, 604, 534
\bibitem[Strazzullo et al.(2010)]{strazzullo2010} Strazzullo, V., Rosati, P., Pannella, M., et al., 2010, A\&A, 524, A1
\bibitem[Strazzullo et al.(2013)]{strazzullo2013}  Strazzullo, V., Gobat, R., Daddi, E., et al., 2013, arXiv1305.3577	
\bibitem[Sunyaev \& Zel'dovich(1972)]{sunyaev_zeldovich1972} Sunyaev \& Zel'dovich, 1972, CoASP, 4, 173
\bibitem[\v{S}uhada et al.(2012)]{suhada2012} \v{S}uhada, R., Song, J., B\"{o}hringer, H., et al., 2012, A\&A, 537A, 39
\bibitem[Tanaka et al.(2013)]{tanaka2013} Tanaka, M., Finoguenov, A., Mirkazemi, M., et al., 2013, PASJ, 65, 17T
\bibitem[Taniguchi et al.(2007)]{taniguchi2007} Taniguchi, Y., et al. 2007, ApJS, 172, 9
\bibitem[Tinker et al.(2008)]{tinker2008} Tinker, J., Kravtsov, A. V., Klypin, A., et al., 2008, ApJ, 688, 709
\bibitem[Tinti \& De Zotti(2006)]{tinti_dezotti2006} Tinti, S., \& De Zotti, G., 2006, A\&A, 445, 889–899
\bibitem[Tozzi et al.(2013)]{tozzi2013} Tozzi, P., Santos, J. S., Nonino, M., Rosati, P., 2013, A\&A, 551A, 45
\bibitem[Trump et al.(2007)]{trump2007} Trump, J. R., Impey, C. D., McCarthy, P. J., et al. 2007, ApJS, 172, 383
\bibitem[Tundo et al.(2012)]{tundo2012} Tundo, E., Tozzi, P., \& Chiaberge, M. 2012, MNRAS, 420, 187
\bibitem[Venemans et al.(2007)]{venemans2007} Venemans, B. P., R\"{o}ttgering, H. J. A., Miley, G. K., et al. 2007, A\&A, 461, 823
\bibitem[Venturi et al.(1995)]{venturi1995} Venturi, T., Castaldini, C., Cotton, W. D., et al., 1995, ApJ, 454, 735
\bibitem[Vikhlinin et al.(2009)]{vikhlinin2009} Vikhlinin, A., Kravtsov, A. V., Burenin, R. A., Ebeling, H., Forman, W. R., et al., 2009, ApJ, 692:1060–1074
\bibitem[von der Linden et al.(2007)]{vonderlinden2007} von der Linden, A., Best, P. N.; Kauffmann, G., 2007, MNRAS, 379, 867
\bibitem[Weinberg et al.(2012)]{weinberg2012} Weinberg, D. H., Mortonson, M., Einsenstein, D., Hirata, C., et al., 2012, Phys. Reports in press(arXiv/1201.2434)
\bibitem[Wen \& Han(2011)]{wen_han2011} Wen, Z. L. \& Han, J. L., 2011, ApJ, 734, 68
\bibitem[Willott et al.(2001)]{willot2001} Willott, C.~J., Rawlings, S., et al., 2001, MNRAS, 322, 536W (W01)
\bibitem[Wilson \& Colbert(1995)]{wilson_colbert1995} Wilson, A. S., \& Colbert, E. J. M. 1995, ApJ, 438, 62
\bibitem[Wing \& Blanton(2011)]{wing2011} Wing, J. D. ,\& Blanton, E. L., 2011, AJ, 141, 88	
\bibitem[Wylezalek et al.(2013)]{wylezalek2013} Wylezalek, D., Galametz, A., Stern, D., 2013, ApJ, 769, 79
\bibitem[Worpel et al.(2013)]{worpel2013} Worpel, H.,  Brown, M. J. I., et al., 2013, ApJ, 772, 64
\bibitem[Zatloukal et al.(2007)]{zatloukal2007} Zatloukal, M., R\"{o}ser, H.-J., Wolf, C., Hippelein, H., \& Falter, S., 2007, A$\&$A, 474, 5
\bibitem[Zeimann et al.(2012)]{zeimann2012} Zeimann, G. R., Stanford, S. A., et al., 2012, ApJ, 756, 115Z
\bibitem[Zeimann et al.(2013)]{zeimann2013} Zeimann, G., Stanford, S. A., Brodwin, M., et al., 2013, arXiv1310.6037
\bibitem[Zirbel(1996)]{zirbel96} Zirbel, E. L. 1996, ApJ, 473, 713
\bibitem[Zirbel(1997)]{zirbel1997} Zirbel, E. L., 1997, ApJ, 476, 489
\bibitem[Zitrin et al.(2012)]{zitrin2012} Zitrin, A., Bartelmann, M., et al., 2012, MNRAS, 426, 2944
\end{thebibliography}
\end{document}